\documentclass{article}

\usepackage{arxiv}
\usepackage{placeins}
\usepackage[utf8]{inputenc} 
\usepackage[T1]{fontenc}    
\usepackage{xr-hyper}
\usepackage{hyperref}       
\usepackage{url}            
\usepackage{booktabs}       
\usepackage{amsfonts}       
\usepackage{nicefrac}       
\usepackage[protrusion = true, final]{microtype}      
\usepackage{lipsum}		
\usepackage{graphicx}
\usepackage[numbers, sort, compress]{natbib}
\usepackage{doi}
\usepackage{hyphenat}
\usepackage{amsmath, amssymb}
\usepackage{authblk}
\usepackage[symbol]{footmisc}
\usepackage{subfiles}
\usepackage{array}
\graphicspath{{Figures/}}

\title{Quantifying the Spatiotemporal Dynamics of Engineered Cardiac Microbundles} 

\author[1,2]{Hiba~Kobeissi}
\author[3]{Samuel~J.~DePalma}
\author[3]{Javiera~Jilberto}
\author[3,4]{David~Nordsletten}
\author[3]{Brendon~M.~Baker}
\author[1,2, *]{Emma~Lejeune}

\affil[1]{Department of Mechanical Engineering, Boston University, Boston, MA 02215, USA}
\affil[2]{Center for Multiscale and Translational Mechanobiology, Boston University, Boston, MA 02215, USA}
\affil[3]{Department of Biomedical Engineering, University of Michigan, Ann Arbor, MI 48109, USA}
\affil[4]{Department of Cardiac Surgery, University of Michigan, MI 48109, USA}

\date{}

\renewcommand{\shorttitle}{Microbundle Analysis}

\hypersetup{
pdftitle={Microbundle Analysis},
pdfsubject={q-bio.NC, q-bio.QM},
pdfauthor={Hiba~Kobeissi,Emma~Lejeune},
pdfkeywords={First keyword, Second keyword, More},
}

\begin{document}
\maketitle

\begin{abstract}
Brightfield time-lapse imaging is widely used in cardiac tissue engineering, yet the absence of standardized, interpretable analytical frameworks limits reproducibility and cross-platform comparison. We present an open, scalable computational pipeline for quantifying spatiotemporal contractile dynamics in microscopy videos of human induced pluripotent stem cell–derived cardiac microbundles. Building on our open-source tools ``MicroBundleCompute'' and ``MicroBundlePillarTrack,'' we define a suite of $16$ interpretable structural, functional, and spatiotemporal metrics that capture tissue deformation, synchrony, and heterogeneity. The framework integrates full-field displacement tracking, strain reconstruction, spatial registration, dimensionality reduction, and topology-based vector-field analysis within a unified workflow. Applied to a dataset of $670$ cardiac microbundles spanning $20$ experimental conditions, the pipeline reveals continuous variation in contractile phenotypes rather than discrete condition-specific clustering, with intra-condition variability often exceeding inter-condition differences. Redundancy analysis identifies a reduced core set of $10$ metrics that retain most informational content while minimizing multicollinearity. Analysis of denoised displacement fields shows that contraction is dominated by a global isotropic mode, with localized saddle-type deformation patterns present in approximately half of the samples. All software and workflows are released openly to enable reproducible, scalable analysis of dynamic tissue mechanics.
\end{abstract}


\footnotetext{* Corresponding author: elejeune@bu.edu}

\section*{Author Summary} 
Stem cells can be guided to form many types of cells, including cardiomyocytes, offering new ways to repair damaged tissue and to model disease. In cardiac tissue engineering, human induced pluripotent stem cell-derived cardiomyocytes (hiPSC-CMs) are grown in two- and three-dimensional systems to create functional heart tissues. These constructs, currently valuable for drug testing and heart disease modeling, are promising as future implantable patches. However, the field lacks consistent protocols and quantitative metrics that are both standardized and reproducible to evaluate tissue function. We address this need by building on our open-source tools, ``MicroBundleCompute'' and ``MicroBundlePillarTrack,'' and a public dataset of videos of beating engineered tissues. We introduce quantitative metrics that describe tissue behavior across space and time, including motion patterns, beat timing, and the coordination and propagation of contraction. Using these metrics, we apply statistical analyses and machine learning approaches to identify distinct contraction phenotypes and to compare performance across samples. We also demonstrate how the choice of metrics has the potential to influence scientific conclusions. All code, documentation, and analysis workflows are openly available. By sharing these methods and a reproducible computational pipeline, we aim to support transparent benchmarking, improve cross-lab comparisons, and accelerate the development of reliable cardiac tissue models.

\section{Introduction} 
\label{sec:intro}
Tissue engineering increasingly serves as an experimental platform for studying human tissue function by enabling the controlled fabrication of three-dimensional constructs with tunable structural and mechanical properties \citep{de2024tissue, hoang2025tissue}. Advances in stem cell technologies \citep{lian2012robust, burridge2014chemically}, biomaterials \citep{eldeeb2022biomaterials, ketabat2024cardiac}, and microfabrication \citep{de2021scaffold, cho2022challenges} have pushed this capability further, yielding engineered tissues of increasing complexity that more closely recapitulate native architecture and function \citep{lian2012robust, burridge2014chemically, eldeeb2022biomaterials, ketabat2024cardiac, de2021scaffold, cho2022challenges}. In parallel, improvements in live-cell imaging and integrated culture platforms have facilitated longitudinal monitoring of these constructs, yielding rich, time-resolved datasets that capture tissue-level dynamics such as contraction, remodeling, or failure \citep{dou2022microengineered, ouyang2017imaging}. As a result, the overall \textit{scale} of tissue engineering research has increased substantially, and high-throughput platforms now routinely generate large imaging datasets across many samples and conditions \citep{ashammakhi2022highlights, kalkunte2024review, zhuang2022opportunities, ewoldt2025induced}. 

Despite this growth, quantitative methods capable of fully leveraging these data remain limited \citep{luo2024current}. In contrast to transcriptomics, where standardized analytical frameworks and metrics have rapidly matured to support large-scale, reproducible analysis \citep{du2023advances, grases2025practical}, methods for extracting and interpreting spatiotemporal mechanical information from engineered tissue datasets are still emerging. This gap is particularly acute in cardiac tissue engineering, a technically demanding domain where analysis remains highly fragmented \citep{zhuang2022opportunities,ewoldt2025induced}. Specifically, most studies rely on custom scripts and laboratory-specific workflows, limiting reproducibility, cross-study comparison, and the establishment of shared benchmarks \citep{cho2022challenges,hirt2014cardiac,zhuang2022opportunities,ketabat2024cardiac,ewoldt2025induced}. Recent advances, including shared differentiation protocols \citep{lian2012robust, burridge2014chemically}, emerging data and metadata standards \citep{sarkans2021rembi}, publicly available datasets \citep{kobeissi2024fibroTUG, kobeissi2024strain, iudin2023empiar, Mohammadzadeh2025dataset}, and accessible analysis tools \citep{kobeissi2024microbundlecompute, kobeissi2024microbundlepillartrack, huebsch2015automated, sala2018musclemotion, ronaldson2019engineering, toepfer2019sarctrack, psaras2021caltrack, tsan2021physiologic, tamargo2021millipillar, rivera2023automated, rivera2025forcetracker, mohammadzadeh2025quantifying, mohammadzadeh2025sarcgraph}, have begun to address these challenges. However, unified, scalable analytical workflows and standardized, interpretable metrics for heterogeneous tissue dynamics remain largely absent.

\begin{figure}[h]
\begin{center}
\includegraphics[width=0.98\textwidth, keepaspectratio]{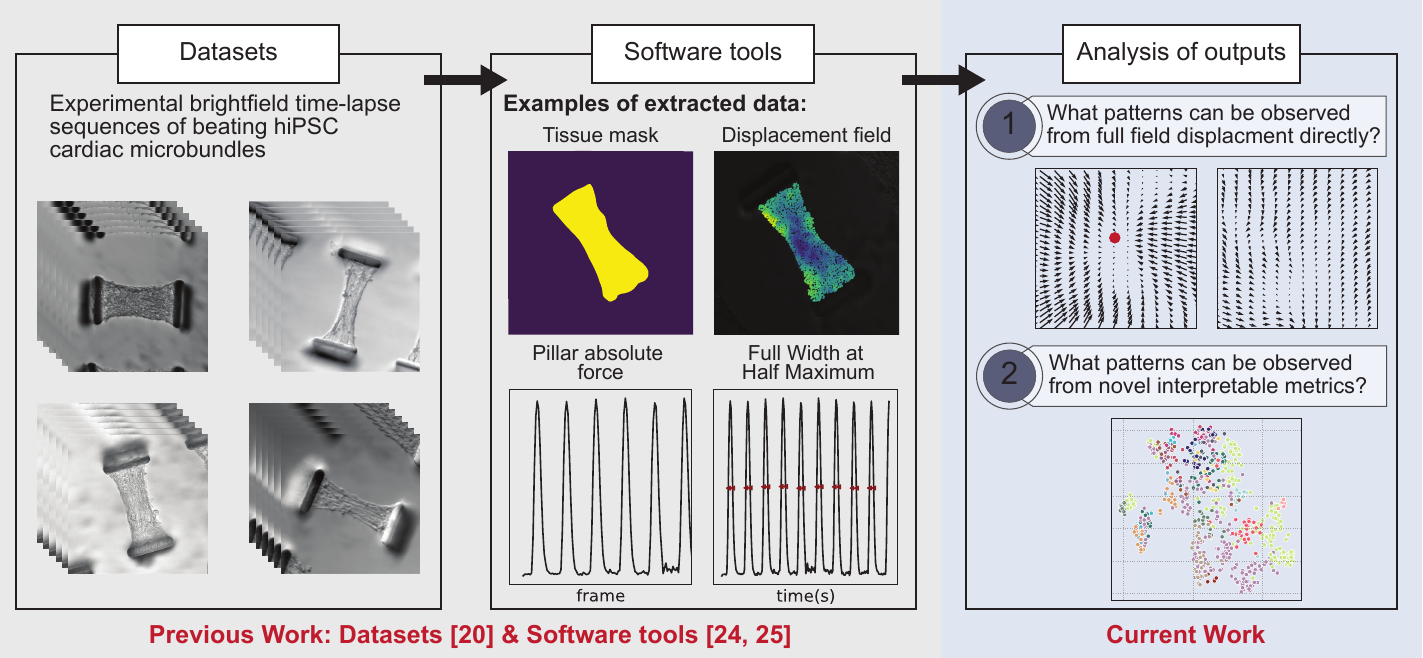}
\caption{\label{fig:overview}Overview of the study scope for mechanical analysis of engineered cardiac microbundle microscopy data. Previous work established a dataset of $808$ brightfield time-lapse image sequences of contracting hiPSC-based cardiac microbundles on FibroTUG platforms \citep{kobeissi2024fibroTUG} and open-source tools \citep{kobeissi2024microbundlecompute, kobeissi2024microbundlepillartrack} for high-throughput extraction of tissue masks, displacement fields, and temporal contractility measures. The present work extends these efforts through analysis of the extracted outputs by (1) examining patterns in full-field displacement data and (2) defining interpretable metrics to characterize tissue behavior and dataset-level organization.} 
\end{center}
\end{figure}

Addressing this gap requires overcoming two key challenges: (1) the absence of standardized, openly accessible computational tools for extracting quantitative information from imaging data, and (2) the lack of interpretable metrics capable of capturing spatiotemporally resolved mechanical and functional tissue behavior. Ideally, such tools would enable reproducible metric computation, while the resulting metrics would support both robust cross-sample comparison and meaningful biological interpretation. To address the first gap, we have previously developed and openly released a comprehensive computational pipeline for analyzing brightfield microscopy videos of human induced pluripotent stem cell derived cardiac microbundles (Fig \ref{fig:overview}). Specifically, our open-source tools, ``MicroBundleCompute'' \citep{kobeissi2024microbundlecompute} and ``MicroBundlePillarTrack'' \citep{kobeissi2024microbundlepillartrack}, enable the robust extraction of full field tissue displacement and contractile force measurements, respectively. 

In this paper, we focus on the second gap and build on these tools to define a validated suite of $16$ interpretable structural, functional, and spatiotemporal metrics. We showcase these metrics by applying them to a previously published dataset of $808$ cardiac tissues generated using the fibroTUG platform \citep{kobeissi2024fibroTUG,depalma2024matrix} (Fig \ref{fig:overview}). Using integrated statistical and machine learning based analyses, we evaluate metric informativeness and redundancy, and compare the efficacy of different metrics at assessing tissue contractile behavior. All software, analysis scripts, and implementation details are openly available to support transparency, reproducibility, and ultimately broad adoption of this approach (\href{https://github.com/HibaKob/MicroBundleAnalysis}{https://github.com/HibaKob/MicroBundleAnalysis}).

The remainder of this paper is organized as follows. In Section \ref{sec:materials&methods}, we detail our methodology for extracting contractile dynamics and computing interpretable metrics from microscopy videos of cardiac microbundles. We begin by introducing the experimental dataset, then describe how we use custom software to extract full field tissue displacement and pillar force measurements. We also outline post-processing steps to address sample variability, calculate strain fields, and reduce the dimensionality of displacement data. We then present a diverse suite of structural, functional, and spatiotemporal metrics that capture the heterogeneity of tissue contractions. In Section \ref{sec:results}, we present a systematic assessment of these metrics, leverage machine learning methods to quantify informational overlap, and demonstrate how the choice of metrics influences statistical outcomes. Finally, in Section \ref{sec:concl}, we discuss the advances enabled by our pipeline, its strengths and limitations, and new opportunities for the field. Through this work, we establish a broadly applicable computational framework that advances interpretable metric extraction in cardiac tissue engineering, laying a foundation for rigorous quantitative analysis and meaningful biological interpretation.

\section{Methods} 
\label{sec:materials&methods}
In this Section, we present our methodology for computing the set of $16$ interpretable metrics that capture the complex and heterogeneous spatiotemporal behavior of cardiac microbundles. 
Section~\ref{subsec:dataset} introduces the experimental dataset of cardiac microbundles, followed by Section~\ref{subsec:data_prep}, which details the implementation of custom software tools for extracting displacement fields and quantifying pillar forces from the recorded movies. We further describe post-processing procedures that address geometric variability across samples, compute strain fields, perform dimensionality reduction on the displacement vector fields, and analyze the resulting low-dimensional representations to identify characteristic flow and deformation patterns.

Section~\ref{subsec:metric_ex} introduces a suite of structural, functional, and spatiotemporal metrics derived from these data, enabling a comprehensive and interpretable assessment of microbundle contractility. Although the approaches described here are demonstrated using the cardiac microbundle dataset, these methods are broadly applicable to other time-lapse image-based datasets.

\subsection{Dataset} 
\label{subsec:dataset}
Access to openly available datasets is essential for reproducible analysis, benchmarking, and cumulative progress, particularly as tissue engineering studies generate increasingly large and complex imaging data. Consistent with open science and FAIR (Findable, Accessible, Interoperable, Reusable) principles \citep{wilkinson2016fair,bertram2023open}, biomedical image analysis, including applications in tissue engineering, is increasingly adopting community‑driven standards for data and metadata deposition and public release, strengthening the foundations for transparent and reproducible quantitative research \citep{national2018open,kemmer2023building,sarkans2021rembi,schapiro2022miti,hosseini2023fair}. Within this framework, we leverage our recently published open‑access dataset to evaluate the performance and relevance of our metrics.

Specifically, we use our previously published dataset of cardiac microbundle time-lapse movies  (\href{10.5061/dryad.3r2280gqd}{10.5061/dryad.3r2280gqd}) \citep{kobeissi2024fibroTUG} (see Fig \ref{fig:grid_reg}a for representative image examples). This dataset has been described in detail in previous studies \citep{ kobeissi2024microbundlecompute, kobeissi2024microbundlepillartrack, depalma2024matrix} and comprises a total of $808$ time-lapse images capturing the dynamic contractions of human induced pluripotent stem cell-derived cardiomyocyte (hiPSC-CM) microbundles on FibroTUG platforms. 

FibroTUG platforms are constructed from arrays of electrospun dextran vinyl sulfone (DVS) fiber matrices \citep{davidson2020myofibroblast}, which are suspended between pairs of poly(dimethylsiloxane) (PDMS) cantilevers. The matrices are first functionalized with cell adhesive cyclic RGD (cRGD) peptides and subsequently seeded with differentiated and purified hiPSC-CMs \citep{depalma2021microenvironmental}. Following a culture period of $3$–$21$ days, spontaneous microbundle contractions are recorded as time-lapse images at approximately $65~\text{Hz}$ on Zeiss LSM800 equipped with an Axiocam 503 camera, under controlled conditions of $37^\circ C$ and $5\%$ CO$_{2}$.

A key feature of the FibroTUG platforms is their versatility: $(1)$ the mechanical stiffness of both the fiber matrix and the PDMS cantilevers can be precisely tuned by modifying the photoinitiator concentration during matrix crosslinking and adjusting the cantilever height, respectively; and $(2)$ the alignment of the matrices can be controlled by the translation speed of the mandrel during fiber deposition. By systematically varying these parameters, the dataset encompasses a broad range of biomechanical and structural environments across $20$ distinct experimental conditions. 

Further details on experimental protocols, matrix fabrication, and metadata can be found in the published dataset and accompanying resources \citep{kobeissi2024fibroTUG,  kobeissi2024microbundlecompute, kobeissi2024microbundlepillartrack,depalma2024matrix, davidson2020myofibroblast, depalma2021microenvironmental}. Though this manuscript presents results on the FibroTUG microscopy movies exclusively, the methods presented here are extensible to all time-lapse images of cardiac tissue across different experimental platforms and imaging modalities where approximating full field deformation is feasible \citep{tsan2021physiologic, boudou2012microfabricated, ewoldt2024hypertrophic, zhao2019platform, karakan2024geometry}. 

\subsection{Time-lapse image data extraction} 
\label{subsec:data_prep}
For the dataset described in Section \ref{subsec:dataset}, we first performed detailed analysis of the microscopy movies using our previously developed tracking and quantification software, ``MicroBundlePillarTrack'' \citep{kobeissi2024microbundlepillartrack} and ``MicroBundleCompute'' \citep{kobeissi2024microbundlecompute}. Out of the initial $808$ samples, $670$ were processed successfully. Specifically, the software failed on $100$ examples due to data quality issues, as detailed in the \textit{Software failure and data exclusion} Section. Additional $38$ examples were excluded from further analysis due to structural issues with the tissue, such as being excessively thin, having extensions that extended beyond the width of the pillars causing unbalanced mechanical behavior, or showing partial detachment from one of the anchoring pillars.

\subsubsection{MicroBundlePillarTrack} 
\label{subsubsec:MBPT}
We developed ``MicroBundlePillarTrack'' \citep{kobeissi2024microbundlepillartrack} as a robust, open-source software tool for automated segmentation, tracking, and analysis of pillar deflection in beating microbundles imaged using brightfield microscopy. The software processes consecutive frames of a movie, automatically generating two distinct binary masks to delineate each pillar. After segmentation, fiducial markers identified using the Shi-Tomasi corner detection method \citep{shi1994good} on the first relaxed frame, are tracked throughout all subsequent frames via a pyramidal implementation of the Lucas-Kanade sparse optical flow algorithm \citep{lucas1981iterative, bouguet2001pyramidal}. This approach enables precise determination of pillar positions across time, from which both mean directional and absolute displacements are calculated. The software further derives quantitative outputs including microbundle twitch force and stress, as well as temporal metrics such as contraction and relaxation velocities, Full Width at Half Maximum (FWHM), and full width at 80\% maximum (FW80M).
Notably, ``MicroBundlePillarTrack'' \citep{kobeissi2024microbundlepillartrack} requires no parameter tuning, streamlining the high-throughput analysis of large datasets. Required user input is minimal and limited to frame rate (fps), length scale ($\mu$m/pixel), pillar stiffness ($\mu$N/$\mu$m), and tissue thickness ($\mu$m). A comprehensive description of the software’s methodology and capabilities can be found in its primary publication \citep{kobeissi2024microbundlepillartrack}.

For the present study, we primarily utilize outputs from ``MicroBundlePillarTrack''\citep{kobeissi2024microbundlepillartrack} related to pillar force and the temporal metric Full Width at Half Maximum (derived from the pillar mean absolute displacement time series), which become interpretable metrics detailed in Section \ref{subsec:metric_ex}.

\subsubsection{MicroBundleCompute} 
\label{subsubsec:MBC}
Our tool ``MicroBundleCompute'' \citep{kobeissi2024microbundlecompute} is also an optical flow-based tracking and analysis software, and is among the few specialized tools available for whole-tissue deformation analysis in microscopy movies of cardiac microbundles \citep{mery2023light, scalzo2021dense, huebsch2015automated}.``MicroBundleCompute'' is specifically designed for multi-purpose assessment of heterogeneous cardiac microbundle deformation and strain from brightfield and phase contrast videos, and the software has been extensively validated to ensure robust and reliable performance.

Analogous to ``MicroBundlePillarTrack'' \citep{kobeissi2024microbundlepillartrack}, ``MicroBundleCompute'' \citep{kobeissi2024microbundlecompute} automatically generates a binary mask of the tissue and identifies ``good features to track'' marker points using Shi-Tomasi corner detection \citep{shi1994good} within the masked region on the first relaxed frame. These points are then tracked across all frames, employing a sparse optical flow \citep{lucas1981iterative, bouguet2001pyramidal} approach and segmenting the analysis by individual contraction cycles to mitigate noise accumulation associated with extended temporal tracking. This process yields high-resolution vector maps of tissue displacements throughout the contraction-relaxation cycle and facilitates calculation of spatially-averaged Green-Lagrange strain within specified tissue subdomains. Post-processing modules further enhance analytical rigor by allowing for automated rotational alignment of image stacks and tracked data, as well as interpolation of displacement and strain fields onto query grids for spatiotemporal analyses. Importantly, the software is optimized for batch processing and requires minimal user intervention, with only the frame rate (fps) and pixel-to-micrometer conversion factor specified by the user. All other parameters are internally standardized to maintain analytic consistency across large datasets. For an in-depth description of the software’s methods and functionalities, as well as further details on its implementation and usage, please refer to its original publication \citep{kobeissi2024microbundlecompute}.

In the present study, we utilize the two-dimensional displacement fields generated by ``MicroBundleCompute'' \citep{kobeissi2024microbundlecompute} for direct computation of full-field Green-Lagrange strain, as outlined in the “Strain computation details” Section. These computational outputs serve as the starting point for a substantial subset of the interpretable metrics described in Section \ref{subsec:metric_ex}, supporting rigorous, spatially resolved characterization of contractile tissue mechanics.

\paragraph{Strain computation details}
\label{subsubsubsec:strain_comp}
As described in Section \ref{subsubsec:MBC}, ``MicroBundleCompute'' \citep{kobeissi2024microbundlecompute} is specifically designed to calculate average subdomain strain rather than full-field strain. This design choice was motivated by two main considerations: (1) minimizing the impact of imaging artifacts and noise, and (2) facilitating more consistent comparisons across different samples. While the subdomain averaging approach is practical and robust, it inherently restricts the spatial resolution of strain quantification, since the strain in each subdomain is estimated from the available fiducial marker points and requires a minimum number of points to ensure mathematical stability, particularly to avoid singular matrices during computations.

For our current study, we sought higher spatial resolution in strain mapping than what the subdomain-based method can provide. To achieve this, we adopted the approach described by Zimmerman et al. \citep{zimmerman2009deformation} and Benkley et al. \citep{benkley2023estimation}, which enables computation of a two-dimensional Green-Lagrange strain field. This method estimates the local deformation gradient tensor directly from the tracked positions of randomly distributed particles, utilizing a least-squares fitting procedure on nearest-neighbor vectors combined with a first-order finite difference approximation. In mathematical terms, the estimated deformation gradient $\mathbf{F}$ at a given point $\alpha$ can be expressed as follows \citep{zimmerman2009deformation}:

\begin{center}
\begin{equation}
\mathbf{F}^{\alpha} = \left[\sum_{\beta=1}^{n} \mathbf{x}^{\alpha \beta} \left(\mathbf{X}^{\alpha \beta}\right)^\top \right]\left[\sum_{\beta=1}^{n} \mathbf{X}^{\alpha \beta} \left(\mathbf{X}^{\alpha \beta}\right)^\top \right]^{-1}
\end{equation}
\end{center}

where $\mathbf{x}^{\alpha \beta}$ represents the vector connecting marker points $\alpha$ and $\beta$ in the current (deformed) configuration, while $\mathbf{X}^{\alpha \beta}$ denotes the corresponding vector in the reference (undeformed) configuration. The parameter $n$ specifies the total number of non-collinear neighboring marker points used in the estimation, which, for our analysis, was set to $8$ to ensure robust and accurate computation of the local deformation gradient.

Computing the Green-Lagrange strain ($\mathbf{E}$) tensor from the deformation gradient tensor is straightforward:  
\begin{center}
\begin{equation}
\mathbf{E}^{\alpha} = \frac{1}{2}\left[(\mathbf{F}^{\alpha})^\top \mathbf{F}^{\alpha} - \mathbf{I} \right]
\end{equation}
\end{center}
where $\mathbf{I}$ is a $2\times2$ identity matrix. 

We provide our python implementation of this strain computation approach, enabling users to derive strain fields from displacement data obtained via ``MicroBundleCompute'' \citep{kobeissi2024microbundlecompute}. This code, along with all of the scripts used to calculate the defined metrics in Section \ref{subsec:metric_ex}, are made available on GitHub (\href{https://github.com/HibaKob/MicroBundleAnalysis}{https://github.com/HibaKob/MicroBundleAnalysis}) to allow others to reproduce and build on our methods.

\paragraph{Software failure and data exclusion}
\label{subsubsubsec:quality}
The failure of $100$ microscopy movies to be processed successfully can be attributed to specific design requirements and stringent algorithmic constraints within ``MicroBundleCompute'' \citep{kobeissi2024microbundlecompute} and ``MicroBundlePillarTrack'' \citep{kobeissi2024microbundlepillartrack}. Specifically, two principal categories of movies were excluded: (1) those with blurred frames, which compromise image quality and hinder the accurate identification of fiducial markers essential for robust tracking; and (2) those comprising fewer than three contraction beats, which provide insufficient temporal information for meaningful characterization of contractile dynamics. By enforcing these exclusion criteria, the software ensures that only movies capable of yielding precise and interpretable results are included in subsequent analysis.

\subsubsection{Spatial registration of tissue domains} 
\label{subsubsec:unify_grid}
\begin{figure}[h!tp]
\begin{center}
\includegraphics[width=0.95\textwidth, keepaspectratio]{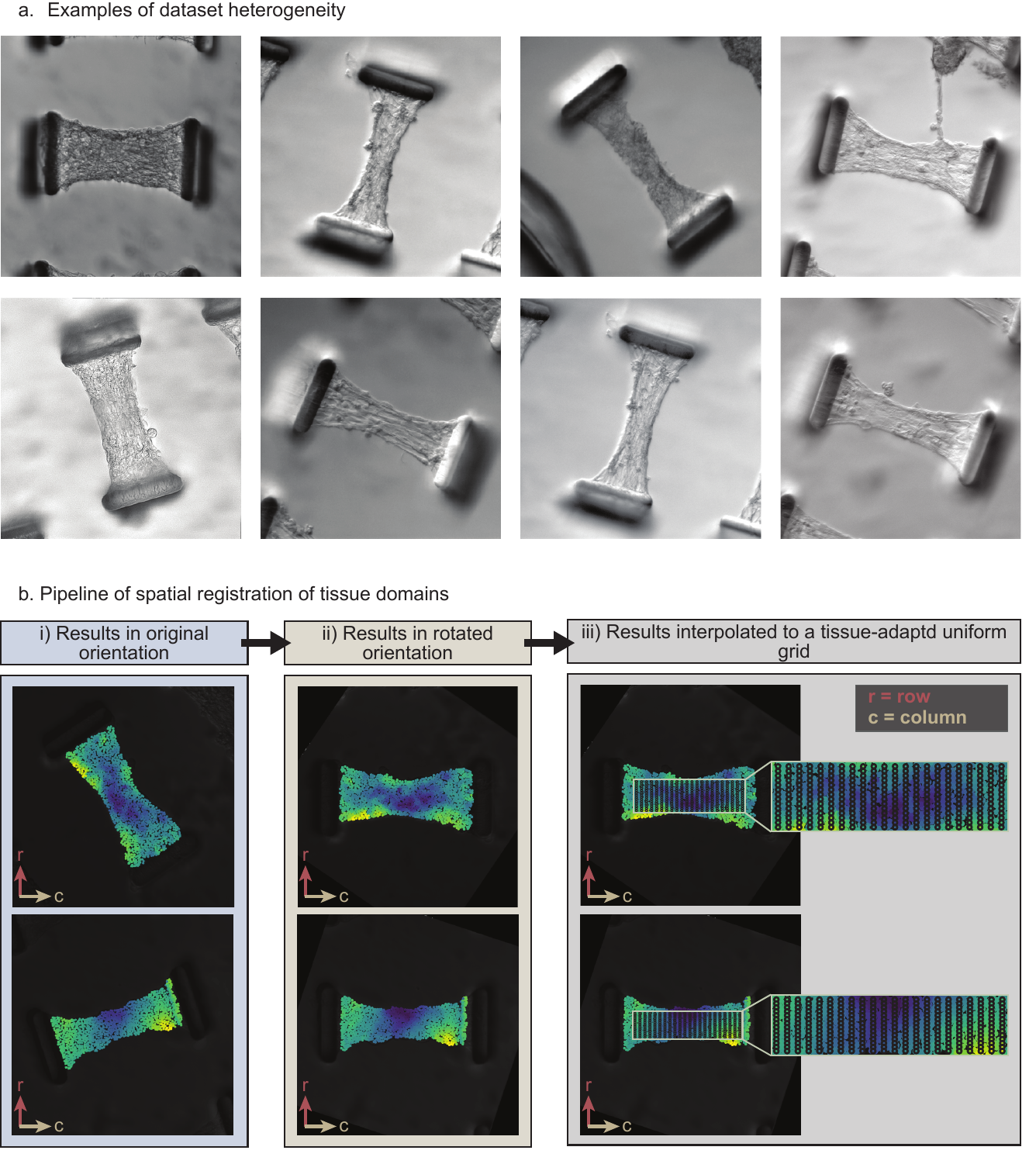}
\caption{\label{fig:grid_reg}Spatial manipulation to standardize tissue-domain data. (a) Representative raw images from the dataset illustrating spatial heterogeneity in tissue orientation, placement, and local geometry which complicates direct comparison across samples. (b) Workflow for spatial registration of tissue domains using the absolute displacement field as an example: (i) per-sample results in the original image orientation; (ii) domains rotated to a common axis to homogenize orientation across the dataset; (iii) rotated domains interpolated onto a tissue-adapted, uniform grid (illustrated inset) to produce consistently sampled measurements along the tissue axis. This two-step manipulation (rotation followed by interpolation) preserves local spatial features while enabling direct pixel-wise comparisons and downstream analyses.}
\end{center}
\end{figure}

The images present in this dataset are taken at multiple different angles, and individual tissues vary in size (Fig \ref{fig:grid_reg}a).
To make direct comparisons between these tissue, we began by performing a systematic registration of tissue domains using masks automatically generated by MicroBundleCompute (Fig \ref{fig:grid_reg}b) \citep{kobeissi2024microbundlecompute}. Each mask was rotated so that the tissue’s major axis aligns with the horizontal (column) axis, and a mean tissue mask was then computed from the set of aligned and rotated masks. To ensure standardized subsequent analyses, we performed a grid sensitivity analysis (Fig \ref{fig:grid_sens}) on the metrics defined in Section \ref{subsec:metric_ex}, which informed the selection of a regular grid with a resolution of $28 \times 33$, centered on the mean mask and spanning $80\%$ of its width and height. For each individual tissue, we calculated an affine transformation that maps the example rotated tissue mask onto the mean tissue mask. This transformation was then used to register the reference grid onto the individual tissue, thereby defining a tissue-specific grid. To enable direct comparisons across tissues, the field of interest (either displacement or strain) was interpolated onto each tissue-specific grid using SciPy’s Radial Basis Function (RBF) interpolation (v$1.13.1$) \citep{virtanen2020scipy,scipy_rbf_doc}. This workflow is critical for ensuring robust and spatially consistent comparisons across heterogeneous microbundle time-lapse images, enabling meaningful downstream analyses that would otherwise be confounded by variations in tissue orientation, size, and geometry.

It is worth noting that variability in tissue orientation and geometry is an inherent challenge across engineered heart tissue datasets, irrespective of experimental platform. The registration and grid optimization workflow detailed above is therefore not dataset-specific; rather, it is designed to be broadly applicable to alternative tissue geometries and experimental configurations. For datasets derived from distinct microbundle architectures or fabrication platforms, the grid sensitivity analysis outlined in this section would need to be repeated to identify a resolution that adequately captures spatially heterogeneous contractile behavior. The affine registration framework and RBF interpolation pipeline are similarly generalizable, provided that reliable tissue masks can be automatically or semi-automatically generated. Collectively, the methodology presented here establishes a systematic and extensible protocol for spatially consistent cross-tissue comparisons, accommodating the geometric diversity inherent to engineered cardiac tissue datasets produced across different experimental systems.

\subsubsection{Principal Component Analysis} 
\label{subsubsec:pca}
Principal component analysis (PCA) \citep{pearson1901onlines, hotelling1992relations} is a widely used unsupervised multivariate analysis technique that transforms complex datasets by projecting them onto a lower-dimensional orthogonal subspace. Typically, the number of retained principal components is far fewer than the original dimensions of the data, resulting in a more concise and informative representation. When applied to vector fields, PCA identifies the primary directions of variation, enabling significant reduction in dimensionality while efficiently filtering out noise and redundancy \citep{abney2011principal, grama2014computation, tyagi2017improving, arzani2021data}. Moreover, the dominant principal components not only capture the most meaningful patterns within the data, but also provide valuable physical insight into the underlying deformation modes and structures of the system under study \citep{abney2011principal, tyagi2017improving, arzani2021data}.

In this study, PCA was applied to the displacement vector fields within the beating cardiac microbundles, as extracted by ``MicroBundleCompute'' \citep{kobeissi2024microbundlecompute}. Our primary motivation for implementing principal component analysis in this study is twofold. First, we seek to uncover latent patterns within the displacement vector field that may reveal meaningful insights into the contractile behavior of cardiac microbundles. Second, we use PCA in this context to denoise the displacement data to enable additional analysis. By reconstructing each tissue’s displacement vector field using only the first $10$ principal components, which capture approximately $93\%$ of the total variance, we effectively remove noise and redundancy, producing cleaner datasets that are better suited for future metric extraction and quantitative analysis.

Given the inherent complexity and spatiotemporal variability of the raw displacement data across samples, two preprocessing steps are required to achieve both objectives. First, we performed temporal homogenization by selecting the displacement field at a single time point corresponding to peak tissue contraction. Next, as detailed in Section \ref{subsubsec:unify_grid}, we spatially homogenized the dataset by interpolating the displacement fields of all $670$ cardiac microbundle samples onto a unified $28\times33$ grid. This careful standardization of both temporal and spatial dimensions ensures consistency across examples.

To organize the data for PCA, we first retained the spatial structure of the grid points and constructed two three-dimensional matrices of size $28 \times 33 \times 670$, one for the horizontal (column) and one for the vertical (row) components of the displacement field. In each matrix, the first two dimensions represent the row and column positions on the spatial grid, while the third dimension indexes the $670$ tissue samples. We then transformed these $3$D matrices into a $2$D format suitable for PCA by performing mode-$3$ unfolding, as described in \citep{lu2008mpca, lu2011survey}. Specifically, each $28 \times 33 \times 670$ matrix was reshaped into a $670 \times 924$ matrix, where each row corresponds to a tissue sample and each column to a specific spatial displacement feature. We then concatenated the unfolded horizontal and vertical displacement matrices along the feature axis, resulting in a final data matrix ($\mathbf{Q}$) of size $670 \times 1848$. This organization allows each tissue sample to be represented as a single vector of $1,848$ displacement features, enabling a comprehensive and meaningful principal component analysis.

We performed PCA using the Python library scikit-learn (v$1.7.1$) \citep{scikit-learn} on the data matrix $\mathbf{Q}$. In accordance with the standard PCA theory, the process begins by centering the data, which involves subtracting the mean from each column to ensure that the analysis captures variance relative to the mean. The subsequent steps differ in implementation from the classical covariance-based derivation, but they yield mathematically equivalent results. Instead of explicitly forming the covariance matrix and performing an eigenvalue decomposition, scikit-learn performs a singular value decomposition (SVD) of the centered data matrix. Given a centered data matrix $\mathbf{Q_c}$, it computes $\mathbf{Q_c} = \mathbf{U\Sigma V^\top}$; the principal axes (directions) are the right singular vectors $\mathbf{V}$, and the explained variances are $\mathbf{\Sigma}^2/(n-1)$, which are mathematically equivalent to the eigenvectors and eigenvalues of the covariance matrix $\mathbf{C} = [1/(n-1)]\mathbf{Q_c}^\top\mathbf{Q_c}$ \citep{wall2003singular}. This approach avoids explicitly constructing the covariance matrix, which improves numerical stability and is more memory- and compute-efficient for large or high-dimensional datasets. Finally, the principal component scores, or in other words, the transformed variables summarizing the dominant patterns in the data, are obtained as $\mathrm{scores} = \mathbf{Q_c}\mathbf{V}$, that is by projecting the centered data onto the principal axes.

Finally, we make our complete implementation of PCA publicly available on GitHub (\href{https://github.com/HibaKob/MicroBundleAnalysis}{\nolinkurl{https://github.com/HibaKob/MicroBundleAnalysis}}), including detailed steps for matrix unfolding and the construction of the data matrix $Q$. We present the results of this analysis in Section \ref{res:pca}.

\subsubsection{Critical point analysis} 
\label{subsubsec:cpa}
Analysis of two-dimensional vector field topology has been foundational in fluid dynamics, where it is particularly useful for studying velocity fields in turbulent flows that exhibit intricate and dynamic patterns \citep{lee2001flow, adrian2000analysis}. This approach not only facilitates intuitive visual representation of complex datasets, making them more accessible for human interpretation, but also enables the identification and classification of a broad spectrum of wave phenomena \citep{lee2001flow, adrian2000analysis,helman1989representation, helman1991visualizing}. Central to this methodology is critical point analysis, which detects and categorizes local flow patterns and spatial features within vector fields at fixed time points, thereby enabling the identification of spatiotemporal wave phenomena through their evolution in time \citep{helman1989representation, helman1991visualizing, perry1975critical, perry1987description, shu1994vector, effenberger2010finding}. The versatility of vector field analysis and critical point identification has captured the interest of researchers across a broad range of disciplines, given its applicability to virtually any vector field. This has led to the adoption and extension of these methods in diverse fields such as neural circuit analysis and brain activity pattern detection \citep{townsend2018detection, hamid2021wave, xu2023interacting}, transcriptomics and cell fate mapping \citep{qiu2022mapping, sha2024reconstructing, zhu2025quantifying}, and the detection of irregularities in cardiac electrophysiology \citep{pancorbo2023vector, tonko2025vector}. 

Building on these advances, we further extend critical point analysis to the study of dynamic cardiac tissue behavior. By applying this technique to PCA-denoised displacement vector fields reconstructed for each tissue, we aim to identify and characterize significant localized spatial patterns underlying tissue contraction and coordination.

\begin{figure}[h]
\begin{center}
\includegraphics[width=0.8\textwidth, keepaspectratio]{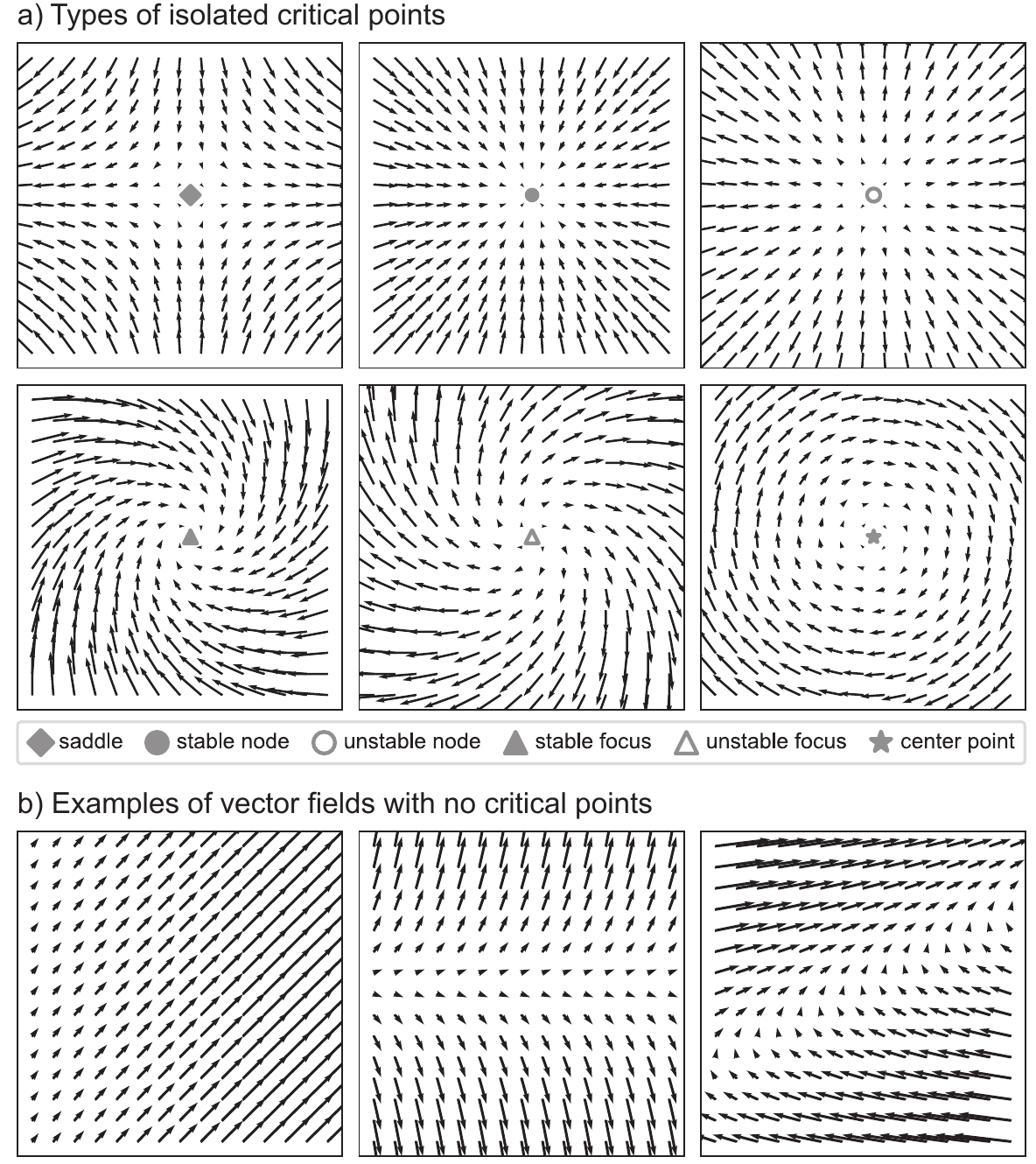}
\caption{\label{fig:critical_pts}Representative 2-dimensional vector fields depicting outcomes from critical point analysis: (a) isolated first-order critical points; (b) examples of vector fields in which no critical points are detected, demonstrating flows without singularities.}
\end{center}
\end{figure}

Critical points are locations where the vector magnitude is zero, and they are classified into six main types based on the behavior of nearby tangent curves (Fig \ref{fig:critical_pts}a), with nodes and foci each further distinguished into stable and unstable variants. These primary types include: (1) saddles, characterized by one stable and one unstable axis, often arising from interactions or collisions of different wavefronts; (2) nodes, which act as sinks (stable) or sources (unstable) and represent regions where flow contracts toward or expands from a central point; and (3) foci, around which flow spirals, corresponding to contracting (stable) or expanding (unstable) spiral waves \citep{helman1989representation, shu1994vector, theisel2005topological, liu2024two}. 

In our implementation, critical point analysis was performed on the reconstructed displacement vector field for each tissue, using only the first $10$ principal components (see Section \ref{subsubsec:pca}) to capture the most salient displacement patterns at peak contraction. The reconstructed field was then reshaped into horizontal and vertical displacement components on the unified $28\times 33$ spatial grid, yielding a denoised two-dimensional displacement vector field for each tissue. 

Critical points were subsequently identified and characterized in these reconstructed fields according to established methods from prior literature \citep{shu1994vector, perry1987description, effenberger2010finding}. Specifically, a grid-based Poincar'e Index approach was employed \citep{brasselet2009vector}, under the assumption that the vector field varies piecewise linearly within each grid cell. 

Given that our reconstructed displacement field per tissue is represented on a $28\times33$ grid, we partitioned the grid into an oriented triangular mesh, and each triangle was evaluated for the presence of a critical point. For each triangle, we calculated the winding number as follows \citep{townsend2018detection}:

\begin{center}
\begin{equation}
\text{winding number} = \frac{1}{2\pi}\sum_{k=1}^{n}\left(\theta_{k+1} - \theta_{k}\right),\qquad \left(\theta_{k+1} - \theta_{k}\right) \in \left[-\pi,\pi\right]
\end{equation}
\end{center}

where $n=3$ for a triangle, and $\theta_k$ denotes the angle of the displacement vector at the $k^{\text{th}}$ vertex along the triangle boundary. Here, $k$ indexes the ordered spatial vertices of the triangle, taken in counterclockwise order, with circular indexing such that $\theta_{n+1}=\theta_1$. The angular differences were wrapped to the interval $[-\pi,\pi]$. Based on the computed winding number for each triangle, we identified three possible scenarios:
\begin{itemize}
    \item winding number = 1: indicates the presence of a node or focus critical point within the triangle (Fig.\ref{fig:critical_pts}a);
    \item winding number = -1: indicates the presence of a saddle critical point (Fig.\ref{fig:critical_pts}a);
    \item winding number = 0: indicates no critical point within the triangle (Fig.\ref{fig:critical_pts}b).
\end{itemize}

When a nonzero winding number is detected, we used the piecewise linear approximation of the displacement field within the element to locate the critical point, solving for the position where the vector field vanishes in both horizontal (column) and vertical (row) directions.
To further characterize the nature of each critical point, we computed the Jacobian matrix $\mathbf{J}$ for the corresponding triangle, ensuring it is non-degenerate ($\det(\mathbf{J}) \neq 0$). We then solved the characteristic equation to obtain the eigenvalues:

\begin{center}
\begin{equation}
\lambda \mathbf{x} = \mathbf{J} \mathbf{x}, \qquad \lambda = \text{Re}_{1,2} + i\,\text{Im}_{1,2}
\end{equation}
\end{center}

We then classified the critical points based on the real and imaginary components of the eigenvalues:
\begin{itemize}
\item stable node: $\text{Re}_1, \text{Re}_2 < 0$, $\text{Im}_1 = \text{Im}_2 = 0$
\item unstable node: $\text{Re}_1, \text{Re}_2 > 0$, $\text{Im}_1 = \text{Im}_2 = 0$
\item stable focus: $\text{Re}_1, \text{Re}_2 < 0$, $\text{Im}_1 \times \text{Im}_2 < 0$
\item unstable focus: $\text{Re}_1, \text{Re}_2 > 0$, $\text{Im}_1 \times \text{Im}_2 < 0$
\item center point: $\text{Re}_1 = \text{Re}_2 = 0$, $\text{Im}_1 \neq \text{Im}_2 \neq 0$
\end{itemize}

Our complete Python implementation of this critical point analysis, including mesh generation and eigenvalue computation, is available on GitHub: \href{https://github.com/HibaKob/MicroBundleAnalysis}{https://github.com/HibaKob/MicroBundleAnalysis}. In Section \ref{res:cpa}, we show the corresponding findings.

\subsection{Interpretable metrics} 
\label{subsec:metric_ex}
While dimensionality reduction and topological analysis reveal dominant contraction modes and localized deformation patterns, they do not yield quantitative descriptors suitable for systematic comparison across tissues. In addition, conventional workflows that compare between tissue typically lack explicit measures for quantifying asynchrony and heterogeneity in contractile behavior across the tissue domain \citep{ewoldt2025induced}.

\begin{figure}[h!tp]
\begin{center}
\includegraphics[width=0.91\textwidth, keepaspectratio]{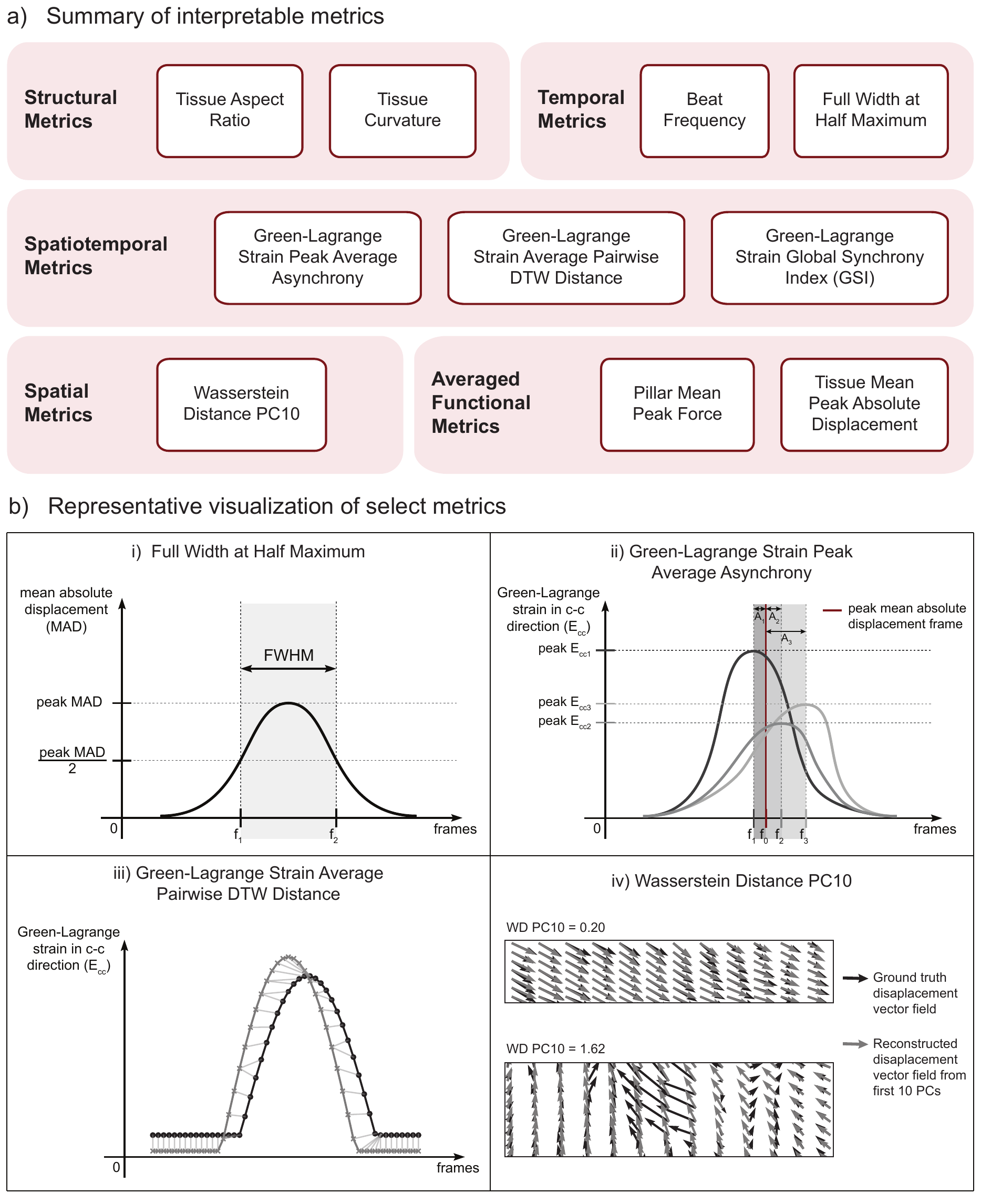}
\caption{\label{fig:met_summ}Summary of interpretable metrics and example computations. (a) Organized grouping of the metrics by the type of descriptive information they provide. (b) Representative visualizations showing how select, less‑intuitive metrics are computed: (i) Full Width at Half Maximum (FWHM)--the temporal width of the mean absolute displacement (MAD) curve at half its peak amplitude, obtained as $f_2 - f_1$; (ii) Green–Lagrange Strain Peak Average Asynchrony-- $A_1 \cdots An$, the differences between the timing of peak strain across spatial locations relative to the tissue’s peak mean absolute displacement are used to compute the average asynchrony; (iii) Green–Lagrange Strain Average Pairwise DTW Distance--example of pairwise dynamic time warping (DTW) distance between two strain time series from different locations within the tissue; (iv) Wasserstein Distance PC10--comparison of ground‑truth and reconstructed displacement vector fields using the first $10$ principal components, with shown Wasserstein distance values illustrating low (Wasserstein distance PC10 = $0.20$) versus high (Wasserstein distance PC10 = $1.62$) distance.}
\end{center}
\end{figure}

Here, we introduce a suite of structural, functional, and spatiotemporal metrics (Fig \ref{fig:met_summ}a), adapted from disciplines including optimal transport \citep{rubner1997earth}, neural activity analysis \citep{li2007synchronization}, and automatic speech recognition \citep{Shearme1968some}, specifically tailored to quantify the heterogeneous and dynamic behavior observed in cardiac tissues.

\subsubsection{Wasserstein distance} 
\label{subsubsec:wd}
The Wasserstein distance, also known as Earth Mover’s Distance, is a widely used metric for quantifying the difference between two probability distributions. Originally formulated by Rubner et al. \citep{rubner1997earth, rubner1998metric}, the Wasserstein distance measures the minimal ``work'' required to transform one distribution into another. In vector field analysis \citep{lavin1998feature, rimehaug2023uncovering}, it has been used to quantitatively compare computational and experimental flow fields under equivalent conditions, thereby providing an interpretable measure of differences between complex vector field patterns.

In this work, we use the Wasserstein distance to quantify the difference between the original displacement field at peak tissue contraction, interpolated onto a tissue-specific $28 \times 33$ grid (see Section \ref{subsubsec:unify_grid}), and its reconstruction from the first 10 principal components (see Section \ref{subsubsec:pca}). This approach  allows us to assess the overall similarity between the two vector-valued distributions, where a larger Wasserstein distance indicates greater dissimilarity in the displacement patterns and suggests that the original field contained greater irregularities not preserved in the PCA-based reconstruction (Fig. \ref{fig:met_summ}b-iv). By comparing the two displacement vector field distributions, the Wasserstein distance provides a measure of reconstruction fidelity beyond point-wise error metrics.

The first Wasserstein distance between two distributions, using the Euclidean norm as the ground metric, is defined as \citep{ramdas2017wasserstein}:
\begin{center}
\begin{equation}
l_1(u,v)=\inf_{\pi \in \Gamma(u,v)} \int_{\mathbb{R}^n \times \mathbb{R}^n} \|x-y\|_2 \, d\pi(x,y)
\end{equation}
\end{center}

where $\Gamma(u,v)$ denotes the set of distributions with marginals $u$ and $v$, $\pi$ is a transport plan, and $\|x-y\|_2$ is the Euclidean distance between $x$ and $y$ in $\mathbb{R}^n$. In the discrete setting, the Wasserstein distance can be interpreted as the cost of an optimal transport plan required to transform one distribution into the other, where the cost is given by the amount of probability mass moved multiplied by the distance over which it is transported. In this setting, the finite point sets $\{x_i\}$ and $\{y_j\}$ denote the support set of the probability mass functions $u$ and $v$, respectively.

To compute the Wasserstein distance in our analysis, the \verb|wasserstein_distance_nd| function \citep{lu2025multi, scipy_wd}, available in SciPy (v$1.13.1$) \citep{virtanen2020scipy}, was used. 
This function enables the efficient computation of the Wasserstein distance between $N$-dimensional discrete distributions. In this implementation, the inputs \verb|u_values| and \verb|v_values| correspond to the 2D displacement vectors from the original and reconstructed fields, respectively. The optional inputs \verb|u_weights| and \verb|v_weights| specify the associated nonnegative weights of the support points; however, these arguments were not provided in the present analysis, and equal weight ($1/M$) was therefore assigned to all support points, where $M=924$ denotes the number of grid points. For transparency and reproducibility, the complete script for implementing this function, as well as the procedure for reconstructing the displacement fields, is provided on GitHub: \href{https://github.com/HibaKob/MicroBundleAnalysis}{https://github.com/HibaKob/MicroBundleAnalysis}. We present the results of this analysis in Section \ref{subsec:correlations}. 

\subsubsection{Global Synchrony Index (GSI)} 
\label{subsubsec:gsi}
\begin{figure}[h]
\begin{center}
\includegraphics[width=1\textwidth]{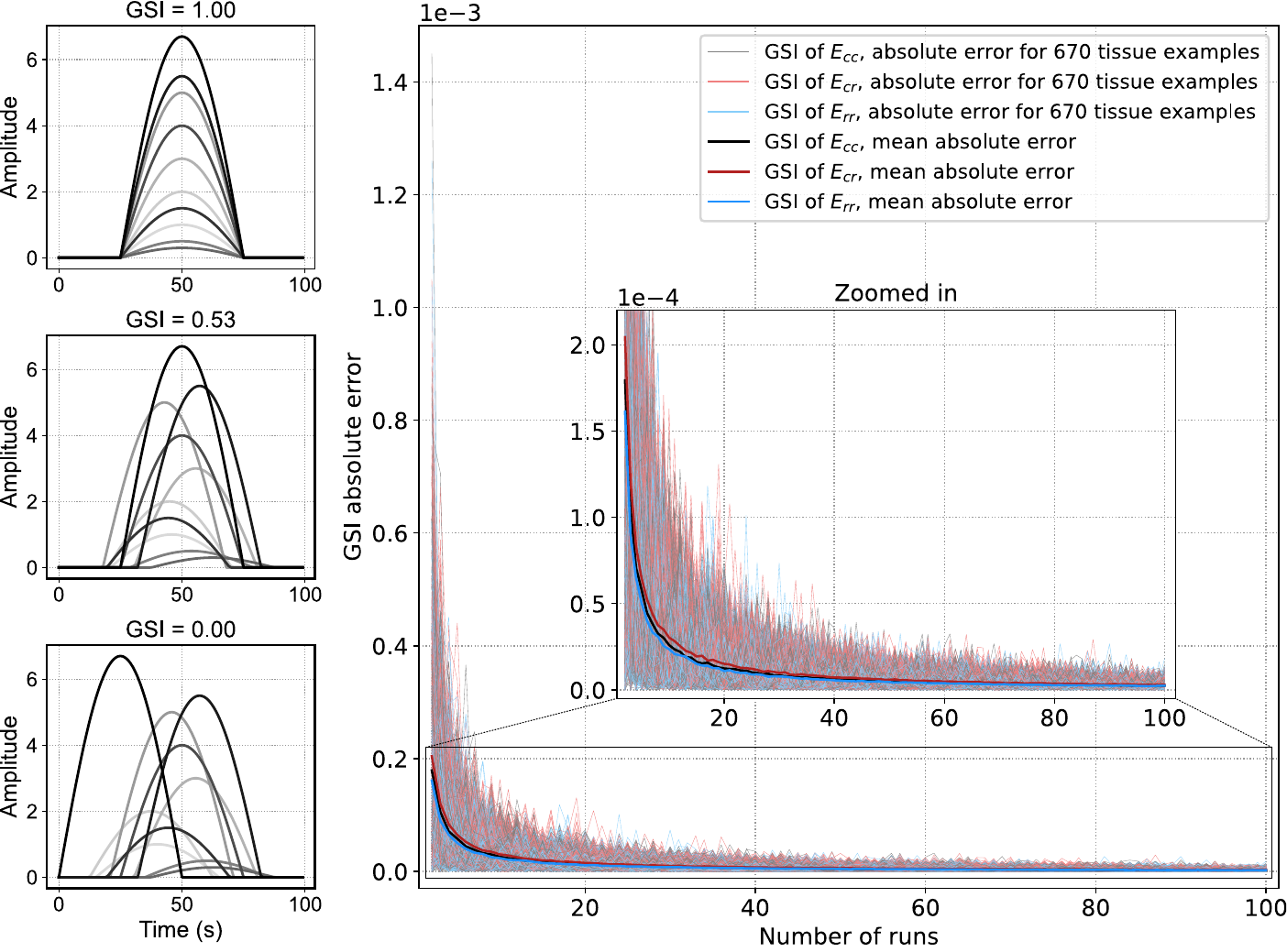}
\caption{\label{fig:GSI_conv}Convergence analysis of the normalized global synchrony index (GSI) for $670$ tissue examples over $100$ independent runs. The left panel presents representative strain time series for varying degrees of synchrony: perfect synchrony (GSI = $1.00$), moderate synchrony (GSI = $0.53$), and complete asynchrony (GSI = $0.00$), exemplifying how the GSI reflects regional temporal alignment in tissue deformation. The right panel illustrates the reduction in GSI mean absolute error as a function of the number of runs, demonstrating that the mean GSI values stabilize after approximately $80$ runs for all three strain components ($E_{cc}$, $E_{cr}$, and $E_{rr}$).}
\end{center}
\end{figure}

In order to quantitatively assess synchronization across multiple neuronal population time series, Li et al. \citep{li2007synchronization} introduced the normalized global synchrony index (GSI). This metric provides an interpretable scale where a value of $0$ denotes complete asynchrony among time series, and a value of $1$ represents perfect synchrony (Fig \ref{fig:GSI_conv}, left panel). In our dataset, the GSI is well-suited for the quantification of regional synchrony within tissue samples. In particular, the Green-Lagrange strain fields, evaluated on unified grids (see Section \ref{subsubsec:unify_grid}), display peak strain values that do not necessarily occur at the same time frame across all grid locations. 

To apply the global synchrony index to our tissue data, we computed GSI values for each tissue sample and for each of the three Green-Lagrange strain components: $E_{cc}$ (column-column based on row-column descriptions of image data, representing the main axis of tissue contraction), $E_{rr}$ (row-row, normal to the main contraction axis), and $E_{cr}$ (column-row). In order to make the synchrony metric comparable across all $670$ tissue examples, we homogenized the temporal profiles of the strain fields by rescaling each time series to span a normalized interval from $0$ to $1$ time units. For computational consistency, we sampled $25$ equally spaced time points within this interval, using steps of $0.04$ units. This sampling density was selected as a practical balance between preserving sufficient temporal resolution to capture waveform dynamics and avoiding unnecessary oversampling. At these standardized time points, strain values for each spatial grid location were interpolated using SciPy’s (v $1.13.1$) \citep{scipy_rbf_doc} CubicSpline implementation. This approach ensures that the GSI reflects true differences in strain synchrony across spatial regions and tissue samples, independent of variations in tissue beating frequency.

The global synchrony index (GSI) was implemented following the framework established by Li et al. \citep{li2007synchronization}, which leverages random matrix theory and equal-time correlation matrix analysis. For each tissue sample, the temporally homogenized strain data were organized as $M=924$ strain time series, one per spatial grid location, each sampled at $T=25$ normalized time points. The Pearson correlation matrix $\mathbf{R}\in\mathbb{R}^{M\times M}$ was then constructed, where
$R_{ij}=\mathrm{corr}(\mathbf{x}_i,\mathbf{x}_j)$, and $\mathbf{x}_i$ and $\mathbf{x}_j$ denote the strain time series at grid locations $i$ and $j$, respectively. Thus, each entry of $\mathbf{R}$ measures the equal-time temporal association between a pair of spatial locations within the tissue.

We then performed eigenvalue decomposition of $\mathbf{R}$ with the largest eigenvalue $\lambda$ serving as a robust measure of global synchronization across the tissue \citep{li2007synchronization, patel2012dynamic}. To ensure that the GSI values accurately reflect true synchronization rather than statistical artifacts, we normalized $\lambda$ using randomized surrogate datasets generated via amplitude-adjusted Fourier transform (AAFT). AAFT effectively preserves the amplitude distribution and approximate autocorrelation structure of each time series while removing genuine equal-time correlations. For each tissue sample, this randomization was repeated $100$ times, and the maximum eigenvalue $\lambda^{'}$ was recorded for each realization. We then calculated the mean ($\bar{\lambda}^{'}$) and the standard deviation ($SD$) of these surrogate eigenvalues to provide a robust estimate of expected synchronization under random conditions. 

The normalized GSI value was then calculated as follows:

\begin{center}
\begin{equation}
\text{GSI}_{norm} = \begin{cases}
\left( \lambda - \bar{\lambda}^{'} \right) / \left( M -  \bar{\lambda}^{'}\right) &\text{if } \lambda >  \bar{\lambda}^{'} + K \times SD \\ 
0
\end{cases}
\end{equation}
\end{center}
where $K$ is the Bonferroni-adjusted critical value from the standard normal distribution to control the overall probability of a Type I error (false positive) for multiple hypothesis tests \citep{dunn1961multiple}, set to $K = 4.42$ for an overall significance level of $\alpha = 0.01$ using Z-tests given $924$ comparisons.

Considering the inherent randomness in AAFT surrogate generation and to ensure reproducibility, we repeated the entire normalized GSI calculation $100$ times for each tissue sample and reported the average normalized GSI value. As demonstrated in Fig. \ref{fig:GSI_conv}, the mean measured GSI rapidly stabilizes, converging after approximately $80$ iterations.
The complete implementation, including scripts for GSI computation and temporal homogenization, is openly available on GitHub: \href{https://github.com/HibaKob/MicroBundleAnalysis}{https://github.com/HibaKob/MicroBundleAnalysis}. We leverage the results of this analysis in Section \ref{subsec:correlations}.

\subsubsection{Average pairwise Dynamic Time Warping (DTW) distance} 
\label{subsubsec:dtw}
Originally developed for speech recognition \citep{Shearme1968some, itakura1975minimum, sakoe1978dynamic} and later broadly adopted for pattern analysis in diverse time series applications \citep{berndt1994using, jeong2011weighted, olivares2019inferring, miralles2023forecasting}, Dynamic Time Warping (DTW) distance provides a robust measure of similarity between two temporal sequences that may differ in speed, phase, or timing. Unlike simple distance measures, DTW flexibly aligns sequences by dynamically ``warping'' the time axis, identifying the optimal path that minimizes the cumulative disparity between corresponding points in the sequences \citep{berndt1994using}. The resulting DTW distance reflects the total alignment cost, accounting for shifts and stretching along the time dimension.

A key advantage of DTW distance over standard Euclidean distance is its resilience to temporal misalignments. Euclidean-based comparisons are highly sensitive to even minor offsets: for example, if one time series is delayed or shifted relative to another, Euclidean distance will exaggerate their difference despite underlying similarity. In contrast, DTW distance accurately quantifies true similarity by aligning corresponding features, making it especially well-suited for biological time series data where phase variability is common.

In this study, we leverage the DTW distance (Fig \ref{fig:met_summ}b-iii) to quantitatively assess waveform heterogeneity within each tissue sample using an average pairwise DTW distance metric. For every tissue, we first normalized each of the three Green-Lagrange strain components ($E_{cc}$, $E_{cr}$, and $E_{rr}$) such that all strain time series amplitudes are scaled between $-1$ and $1$ at each spatial grid location. This normalization ensures meaningful cross-tissue comparisons by removing inter-tissue amplitude differences that could otherwise inflate the DTW metric, while preserving the intrinsic waveform differences among time series within a given tissue. We then computed the DTW distance for every possible pair among the $924$ normalized time series per tissue sample, utilizing the Python implementation provided by the DTAIDistance library \citep{meert2020DTAIDistance}. The average pairwise DTW distance was defined as follows:

\begin{center}
\begin{equation}
\text{average pairwise DTW distance} = \frac{\sum_{i=0}^{N}\sum_{j=0,j\neq i}^{N}\text{DTW}(\mathbf{x}_i, \mathbf{x}_j)}{N(N-1)}
\end{equation}
\end{center}

where $N$ is the total number of time series per tissue. Interpretively, a lower average pairwise DTW distance indicates that most time series within the tissue are temporally alike, with waveform patterns that align closely even if offset or stretched in time; a higher value signifies greater heterogeneity, capturing pronounced differences in strain dynamics, timing, or shape that persist despite optimal alignment. All code and workflow for calculating the average pairwise DTW distance are openly accessible on GitHub: \href{https://github.com/HibaKob/MicroBundleAnalysis}{https://github.com/HibaKob/MicroBundleAnalysis}. We present the results of this analysis in Section \ref{subsec:correlations}. 

\subsubsection{Additional metrics} 
\label{subsec:additional_metrics}
In addition to the metrics detailed in Sections \ref{subsubsec:wd}, \ref{subsubsec:gsi}, and \ref{subsubsec:dtw}, several informative measures are readily available through the core functionalities of the software tools ``MicroBundleCompute'' \citep{kobeissi2024microbundlecompute} and ``MicroBundlePillarTrack'' \citep{kobeissi2024microbundlepillartrack}. These common metrics offer complementary insights into fundamental aspects of tissue structure, function, and dynamics and include:

\begin{itemize}
\item \textbf{Tissue Aspect Ratio} – calculated as the ratio of tissue length to width, with both dimensions extracted from segmented tissue masks. Length is defined as the edge-to-edge distance along the tissue’s major axis, while width is measured at the midpoint perpendicular to the major axis.
\item \textbf{Tissue Curvature} ($1/\mu$m) – quantified as the mean curvature of the two free tissue edges. For each edge, curvature is determined by fitting a circle to the contour points and taking the reciprocal of its radius.
\item \textbf{Beat Frequency} (Hz) – determined by analyzing the mean absolute displacement time series and calculating the average number of beats per second. Individual beats are identified as intervals between two consecutive valleys (zero-crossings) in the displacement curve.
\item \textbf{Full Width at Half Maximum (FWHM)} (frames) – measured from the pillar mean absolute displacement time series as the number of frames between the points where displacement amplitude equals half of its peak value (Fig \ref{fig:met_summ}b-i).
\item \textbf{Pillar Mean Peak Force} ($\mu$N) – obtained by averaging the absolute peak contraction force measured at each pillar during tissue contraction events.
\item \textbf{Tissue Mean Peak Absolute Displacement} (pixels) – calculated by taking the mean of the absolute displacement values at maximal contraction across the tissue.
\item \textbf{Strain Peak Average Asynchrony} (frames) – computed as the mean difference between the frame at which peak mean absolute displacement occurs and the frame of peak Green-Lagrange strain ($E_{cc}$, $E_{cr}$, or $E_{rr}$) at each grid location, providing an aggregate measure of temporal offset across the tissue (Fig \ref{fig:met_summ}b-ii).
\end{itemize}

All code for extracting these direct metrics from our software tools, together with the complete post-processing workflow, is openly available on GitHub: \href{https://github.com/HibaKob/MicroBundleAnalysis}{https://github.com/HibaKob/MicroBundleAnalysis}. We present the results of these metrics in Section \ref{subsec:correlations}.

\section{Results and discussion} 
\label{sec:results}
We evaluate the full set of structural, functional, and spatiotemporal metrics defined in Section \ref{sec:materials&methods} to determine which features meaningfully characterize cardiac microbundle behavior and which add little additional insight. Rather than assuming all metrics are equally informative, our objective is to assess the choice of metrics itself and examine how different selections shape the interpretation of the dataset described in Section \ref{subsec:dataset}, which comprises $20$ experimental conditions. Consistent with this objective, we do not pursue mechanistic explanations for condition-dependent differences, but instead focus on an interpretation-agnostic analysis of how the metrics behave across conditions. 

We first examine how these metrics vary across the $20$ experimental conditions and identify features that remain consistent across fibroTUGs irrespective of condition (Section \ref{subsec:challenges}). We then analyze relationships among metrics to quantify correlations, multicollinearity, and informational overlap, using both statistical measures and machine learning approaches (Section \ref{subsec:correlations}). Building on this analysis, we assess how different feature-selection choices influence condition-level comparisons (Section \ref{subsec:statistics}) and show that multivariate analyses provide important context beyond univariate metrics by clarifying the underlying structure and dispersion of the data across conditions (Section \ref{subsec:multivariate}).

\subsection{Interpretable features show continuous variation across experimental conditions} 
\label{subsec:challenges}
As an initial step, we examined how the $16$ extracted interpretable metrics vary across the experimental conditions represented in the dataset. To this end, we applied three complementary dimensionality-reduction techniques: Principal Component Analysis (PCA) \citep{pearson1901onlines}, Uniform Manifold Approximation and Projection (UMAP) \citep{healy2024uniform}, and t-Distributed Stochastic Neighbor Embedding (t-SNE) \citep{maaten2008visualizing}. While PCA captures dominant linear variance and provides interpretable component loadings, UMAP and t-SNE generate embeddings that emphasize nonlinear and local relationships that may not be resolved by linear projection. Qualitative consistency across these complementary embeddings suggests that the observed patterns are not specific to a single dimensionality-reduction method and thus are more robust.

\begin{figure}[h]
\begin{center}
\includegraphics[width=1\textwidth]{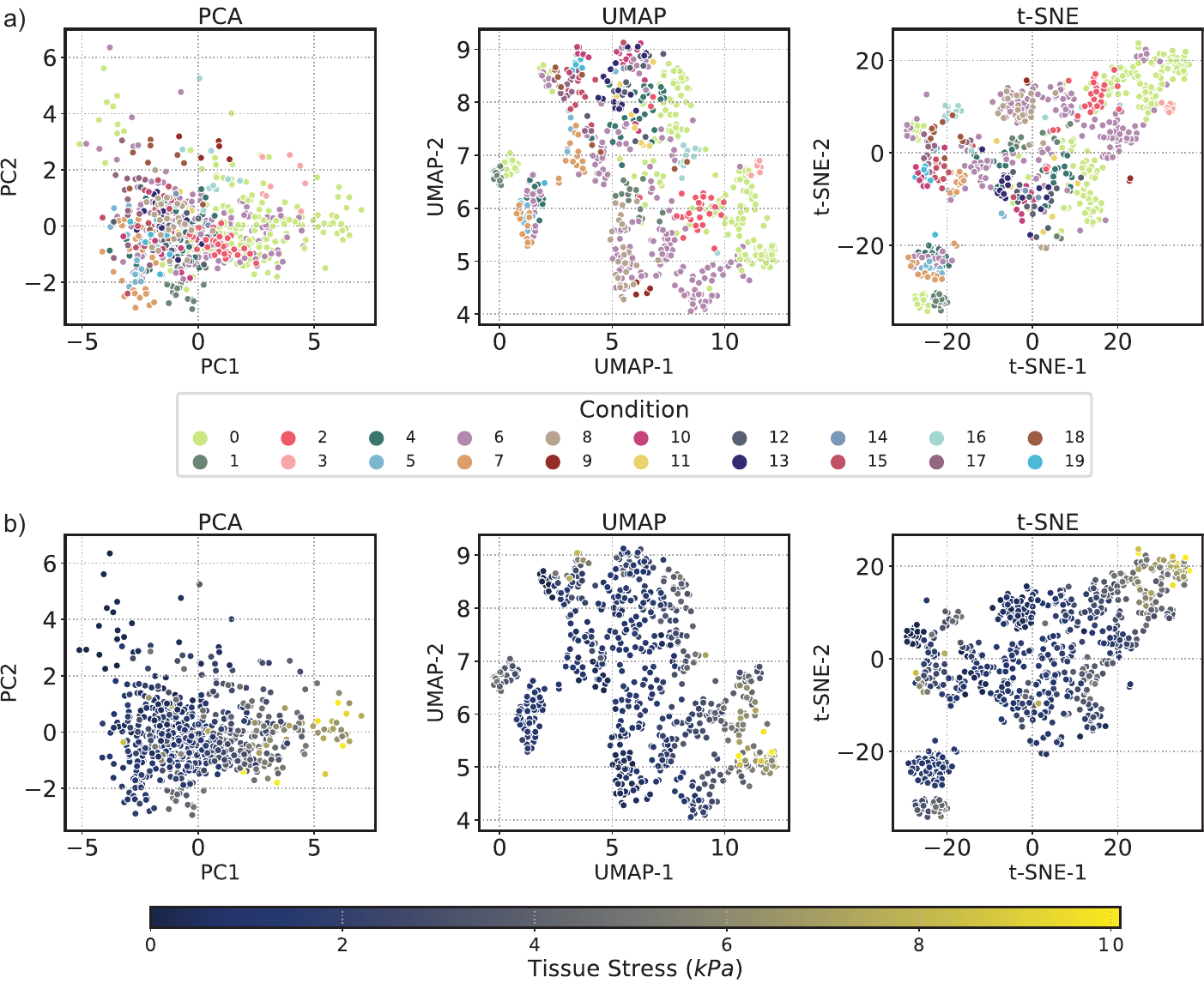}
\caption{\label{fig:pca_umap_tsne} Dimensionality reduction visualizations of the $16$ extracted features applied using three methods: PCA ({\fontfamily{pcr}\selectfont n\_components=2}), UMAP ({\fontfamily{pcr}\selectfont n\_components=2, n\_neighbors=30, min\_dist=0.2, spread=1.0, random\_state=42}), and t-SNE ({\fontfamily{pcr}\selectfont n\_components=2, perplexity=30, learning\_rate=100, early\_exaggeration=12, random\_state=42}). Each method is visualized with respect to: (a) the 20 distinct experimental conditions, and (b) peak tissue stress ($kPa$) at maximal contraction, calculated as the ratio of mean peak pillar force to the cross-sectional area at the tissue centroid, approximated as a rectangle. These representations highlight the clustering patterns and dispersion across conditions, as well as the nearly continuous variation of tissue stress across the multidimensional feature space.}
\end{center}
\end{figure}

The resulting embeddings are shown in Figure \ref{fig:pca_umap_tsne}. The condition-labeled projections (Fig. \ref{fig:pca_umap_tsne}a) reveal that no method yields perfectly separable groups. PCA shows substantial overlap among conditions, whereas UMAP and t-SNE produce more visually distinct cluster structures. However, these clusters are not exclusive: several conditions appear in multiple clusters, and considerable overlap persists across methods.

Notably, when the embeddings are visualized according to tissue stress at peak contraction (Fig. \ref{fig:pca_umap_tsne}b), a broad continuum becomes apparent, with samples graded from low to high stress along the dominant axes of each projection. Rather than forming condition-specific clusters, the data exhibit a smooth transition that organizes fibroTUGs into low-, medium-, and high-stress regimes. This pattern suggests that variation in tissue stress aligns more closely with the underlying structure captured by the metrics than do the discrete experimental labels. More broadly, these observations illustrate the inherent difficulty of analyzing biological systems: measurements are high-dimensional, interdependent, and often reflect continuous biological variation rather than sharply separated groups. As such, interpreting these datasets requires integrative approaches that can account for subtle gradients and shared features across conditions.

\subsection{Principal Component Analysis uncovers primary isotropic contraction mode in tissue displacement} 
\label{res:pca}

\begin{figure}[h]
\begin{center}
\includegraphics[angle=0,width=0.95\textwidth]{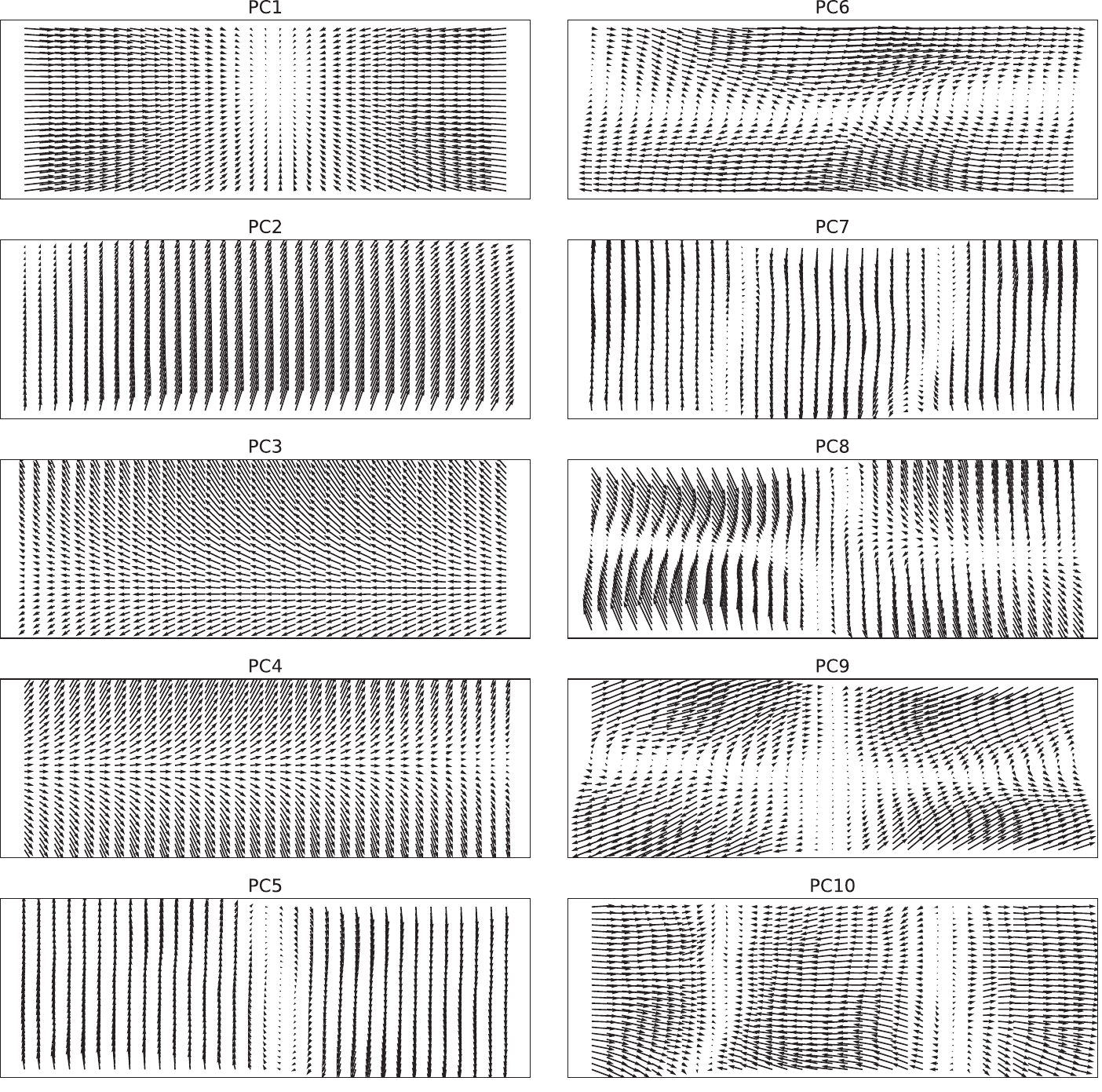}
\caption{\label{fig:pca_pcs}Vector plots illustrating the spatial displacement patterns corresponding to the first $10$ principal components derived from PCA of the tissue displacement fields at peak contraction. These visualizations highlight the dominant modes of motion present within the dataset.}
\end{center}
\end{figure}

\begin{figure}[p]
\begin{center}
\includegraphics[angle=0,width=0.95\textwidth]{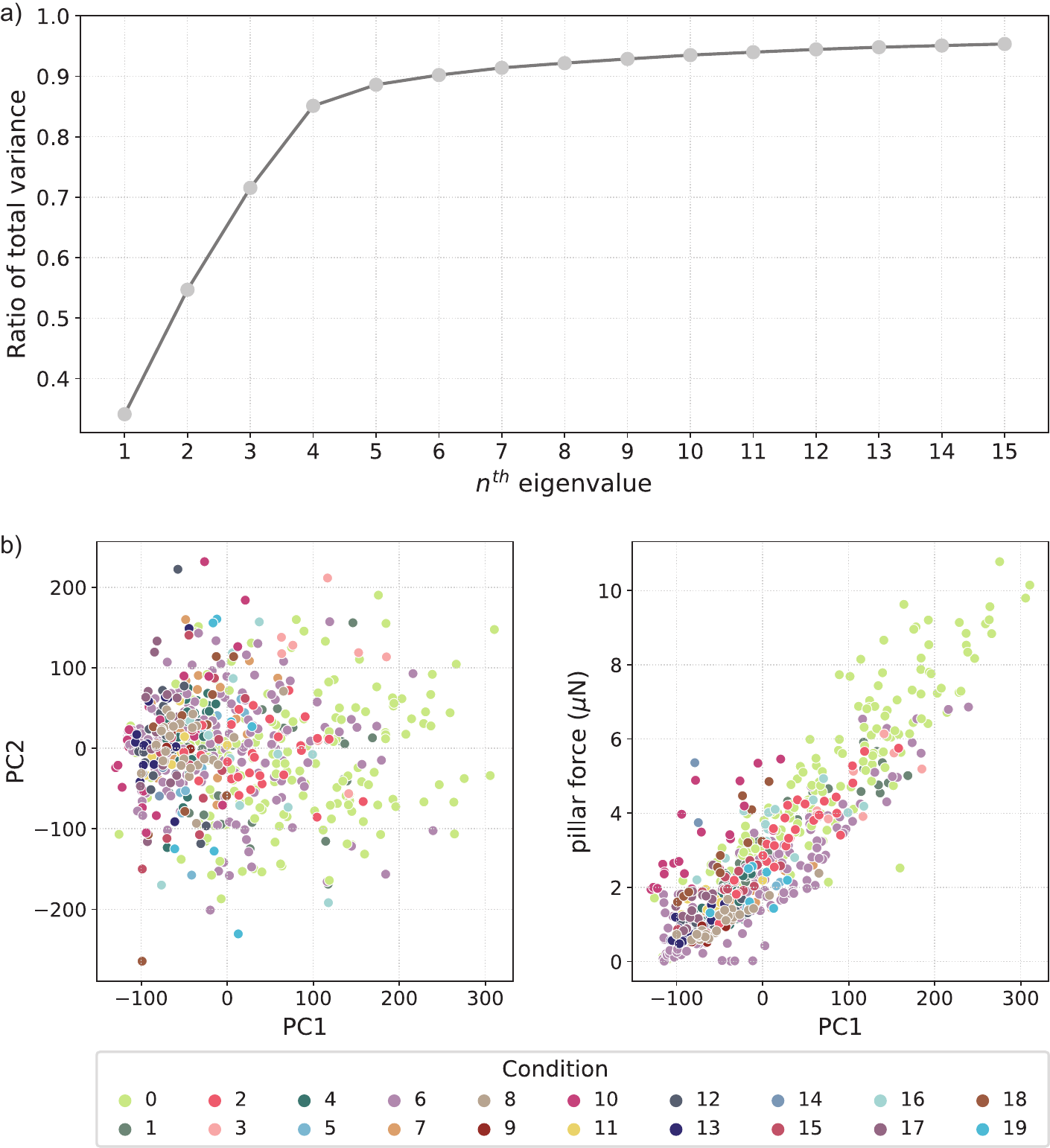}
\caption{\label{fig:pca_res}PCA provides additional insight into tissue displacement fields at peak contraction. (a) The cumulative variance plot demonstrates that the first $4$ principal components capture approximately $85\%$ of the total variance, indicating that most contraction patterns are efficiently described by a limited number of dominant modes. (b) Despite this, visualization in the PC1–PC2 space (left panel) does not reveal clear clustering by experimental condition, highlighting substantial overlap between groups. Notably, the projection onto PC1 exhibits a strong positive linear correlation with measured pillar force (right panel), confirming that the primary mode of displacement, corresponding to isotropic contraction, is highly correlated to increased tissue contraction force across all samples.}
\end{center}
\end{figure}

The PCA detailed in Section \ref{subsubsec:pca} offers novel perspectives on the collective contraction dynamics of the tissues. Examination of vector plots of the first $10$ principal components (PCs) (Fig \ref{fig:pca_pcs}) reveals distinct and interpretable spatial deformation patterns. For instance, PC1, which reflects the dominant contraction mode shared across all samples, is indicative of a  global isotropic contraction, aligned with the major axis of the tissue. On the other hand, PC2 is principally governed by lateral deformation, while PC3 and PC4 reflect a combination of lateral and anisotropic stretching. Notably, reconstruction using the first $4$ PCs accounts for approximately $85\%$ of the observed variance (Fig \ref{fig:pca_res}a), underscoring the predominance of linear, non-rotational contractile behavior in governing the motion of these tissues. 
 
Beyond the primary modes, higher order components (PC5 -- PC10) capture more intricate and localized displacement patterns, including clear evidence of vortex-like and rotational elements. These PCs introduce spatial heterogeneity, with local undulations, non-linear gradients, and twisting motions that are not embodied in the dominant contraction modes. Such features reflect finer-scale tissue mechanics, encompassing localized contraction and mechanical diversity that contribute incrementally to the overall displacement field.
This hierarchical progression, from simpler, coordinated global contraction to increasingly complex and localized dynamics, mirrors the organizational structure of tissue mechanics. Most of the tissue displacement can be explained by a handful of global, coordinated contraction modes; residual variance is distributed among less prominent, spatially complex processes.

Visualizing the dataset in the PC1–PC2 space (Fig \ref{fig:pca_res}b, left panel) reveals that dominant contraction patterns identified by PCA transcend experimental conditions, as substantial overlap and a lack of discrete clustering is prevalent. Crucially, projection onto PC1 demonstrates a strong positive linear correlation with measured pillar force (Fig \ref{fig:pca_res}b, right panel), confirming that global isotropic contraction along the tissue’s major axis acts as a principal driver of elevated contraction force across all samples.

With our analysis, we have only scraped the surface of what an in-depth PCA can reveal. As a starting point, one can apply unsupervised methods, such as clustering on PC scores, to identify new patterns across samples. Another immediate exploration is to extend PCA to dynamic analyses and apply temporal or spatiotemporal PCA to sequences of displacement fields over time, revealing how tissue contraction dynamics evolve. In future studies, which would require the generation of new experimental data, it would also be possible to specifically investigate the biological and experimental determinants of each PC. For instance, one can combine PCA insights with additional data modalities, such as gene expression, protein levels, or tissue composition to test if specific PCs are predictive of key outcomes of for example tissue viability, maturation, or pathological states.

Furthermore, the principal directions identified through PCA could serve as valuable benchmarks for both validating and informing computational models of fibroTUG tissues \citep{jilberto2023data, jilberto2025data}. Rather than relying solely on one-to-one comparisons of displacement field magnitudes, the dominant deformation vector patterns extracted via PCA offer a robust framework for assessing whether computational models accurately reproduce the essential modes of tissue behavior. Importantly, these principal patterns can reveal critical aspects such as global fiber orientations and the spatial distribution of contraction sites, allowing for targeted refinement of model parameters. By integrating PCA-derived insights, computational models can be iteratively calibrated to better capture the complex structural and functional characteristics observed in experimental data.

\subsection{Critical point analysis provides complementary insights to PCA}
\label{res:cpa}
\begin{figure}[p]
\begin{center}
\includegraphics[height=0.8\textheight, keepaspectratio]{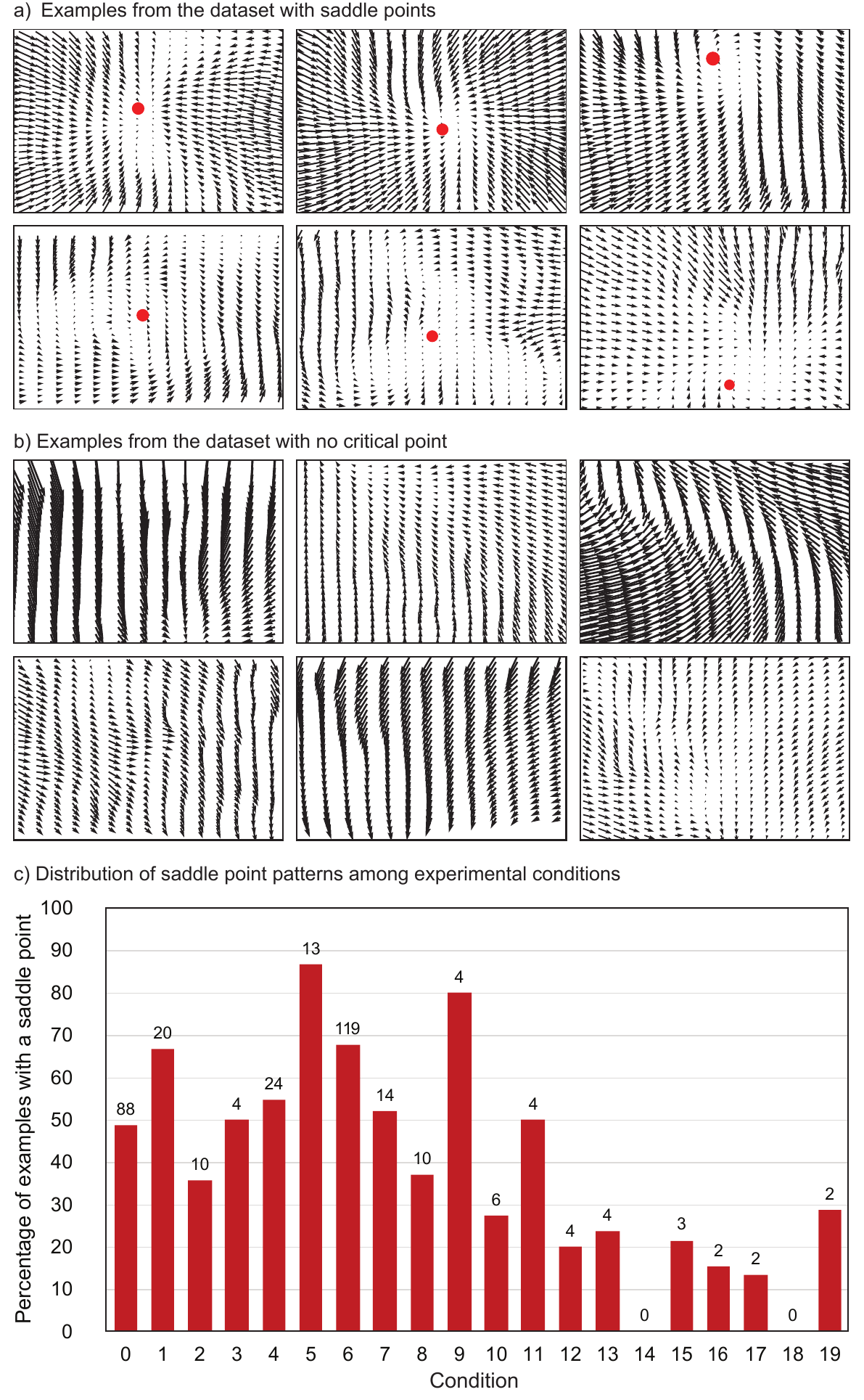}
\caption{\label{fig:critical_pts_res} Schematic overview of the critical point analysis results. (a) Representative examples from the dataset exhibiting saddle points in the displacement field, with saddle locations indicated by red dots. (b) Examples from the dataset that lack any identified critical points, illustrating alternative contraction or deformation patterns. (c) The occurrence of saddle point patterns varies across experimental conditions, and no experimental conditions exclusively yield saddle points. Numbers above each bar indicate the absolute count of examples featuring a saddle point within each condition. For completeness, we note that condition $14$ contains only $2$ examples whereas example $18$ contains $11$ examples.}
\end{center}
\end{figure}

As detailed in Section \ref{subsubsec:cpa}, we performed a critical point analysis on the reconstructed displacement fields of all $670$ tissue samples, using the first $10$ principal components. The results, summarized in Fig \ref{fig:critical_pts_res} and Table \ref{tbl:cpa_summ}, show that saddle point patterns are frequently observed but not ubiquitous. Of note, this dataset is evenly divided between examples exhibiting saddle points and those with no critical points at all. Representative vector field patterns illustrating both scenarios are shown in Fig \ref{fig:critical_pts_res} a and b.

A closer examination also reveals substantial variation in the prevalence of saddle points across experimental conditions. As visualized in Fig \ref{fig:critical_pts_res}c, certain conditions, such as $1$, $5$, and $6$, demonstrate a relatively high incidence of saddle point patterns, while others, notably $16$ and $17$, display none. This heterogeneity suggests that the occurrence of saddle points is, to a limited extent, condition-dependent and may be modulated by other underlying experimental variables or intrinsic tissue properties. Further investigations, which may require integrating additional experimental variables, are needed to clarify the drivers of this heterogeneity and its biological implications.

\begin{table}[h]
\begin{center}
\caption{\label{tbl:cpa_summ}Summary of critical point analysis results for all $670$ tissue samples, detailing the type and frequency of observed critical points. The analysis reveals an equal division between examples exhibiting saddle point patterns ($333$) and those with no detected critical points ($333$).}
\begin{tabular}{lccccccc}
\hline
\textbf{type}      & \begin{tabular}[c]{@{}c@{}}saddle\\ point\end{tabular} & \begin{tabular}[c]{@{}c@{}}stable\\ node\end{tabular} & \begin{tabular}[c]{@{}c@{}}unstable\\ node\end{tabular} & \begin{tabular}[c]{@{}c@{}}stable\\ focus\end{tabular} & \begin{tabular}[c]{@{}c@{}}unstable\\ focus\end{tabular} & \begin{tabular}[c]{@{}c@{}}center\\ point\end{tabular} & \begin{tabular}[c]{@{}c@{}}absence of\\ critical point\end{tabular} \\ \hline
\textbf{frequency} & 333                                                    & 3                                                     & 0                                                       & 1                                                      & 0                                                        & 0                                                      & 333                                                                 \\ \hline
\end{tabular}
\end{center}
\end{table}

While PCA highlights dominant modes of contraction, the critical point analysis uncovers local spatial heterogeneities and complex deformation patterns that are not captured by variance-based dimensionality reduction alone. This underscores the potential value of integrating topological analysis with PCA to better characterize spatiotemporal contraction dynamics.
A promising direction for future critical point analysis involves tracking the temporal dynamics of these features, enabling us to observe how critical points arise, migrate, or vanish throughout the tissue contraction cycles. Such an approach could shed light on the mechanical evolution and stability of tissue regions during active deformation. Additionally, a thorough investigation of the spatial organization of critical points would be valuable-—for instance, assessing whether their distribution is random or exhibits clustering within specific tissue domains. This could reveal underlying mechanical microenvironments and potential hotspots of activity. Most intriguingly, integrating critical point information with high-resolution immunostained imaging would allow for direct correlation of critical point locations with cellular architecture and tissue composition. This multimodal analysis could uncover important mechanistic links between topological features in displacement fields and the biological makeup of the tissue, paving the way for new insights into tissue structure–function relationships.

\begin{figure}[p]
\begin{center}
\includegraphics[width=1\textwidth]{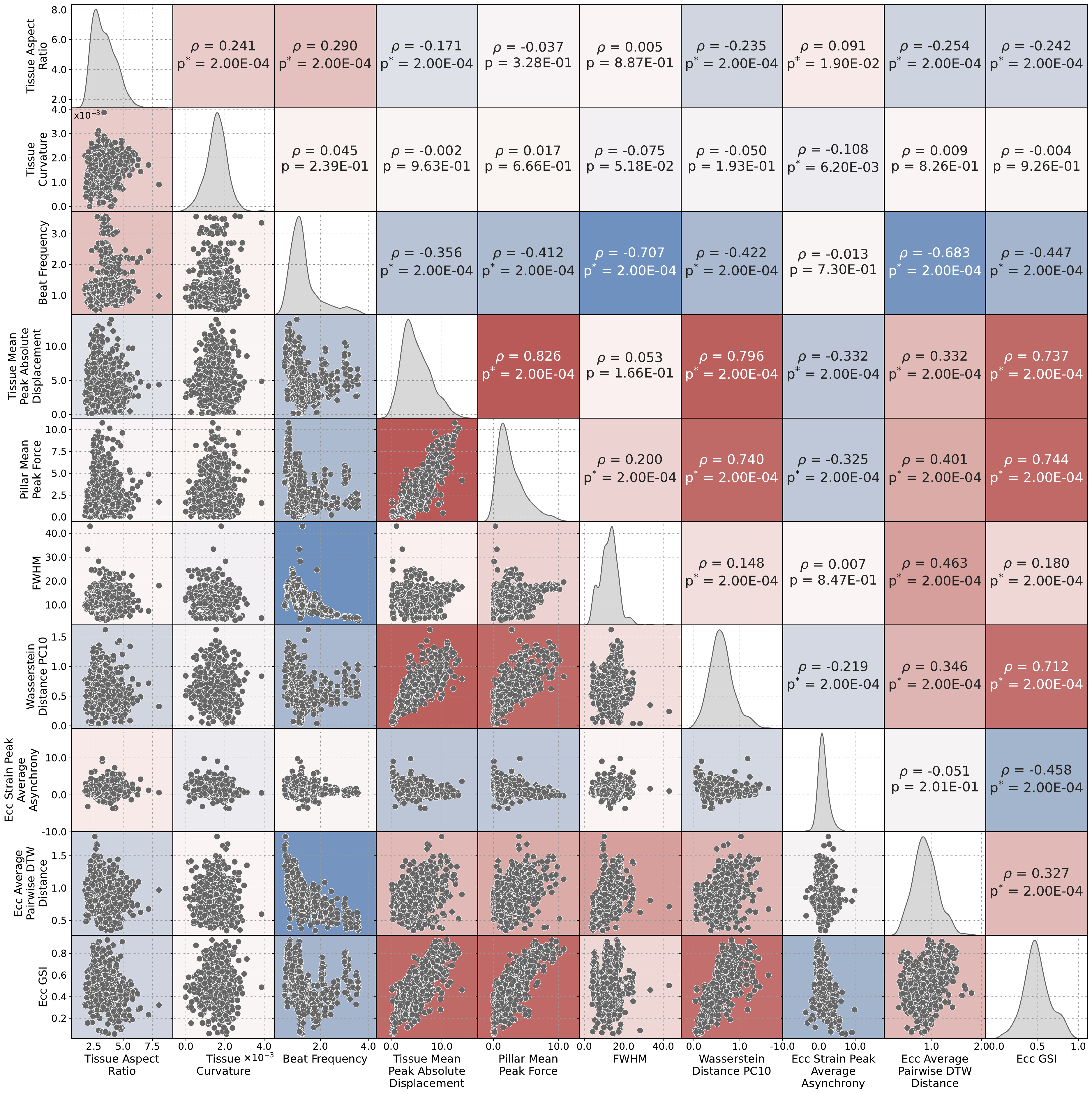}
\caption{\label{fig:pairplot_corr}Comprehensive visualization of pairwise relationships among metrics. The upper triangle displays Spearman’s correlation coefficients ($\rho$) along with the corresponding p-values, providing both the strength and statistical significance of monotonic associations between metrics. The lower triangle features pairplots that visualize the joint distribution of each metric pair. Kernel Density Estimate (KDE) plots along the diagonal illustrate the distribution of individual metrics, highlighting variation in data spread.}
\end{center}
\end{figure}

\subsection{Feature correlations reflect both redundancy and novel information} 
\label{subsec:correlations}

\subsubsection{Interrelationships and redundancy among extracted metrics}

We next investigated the relationships among the extracted metrics in order to assess their informational value and identify potential redundancy. Our goal was to determine whether all metrics are essential, or if some exhibit sufficiently high inter-correlation to warrant exclusion. To this end, we evaluated multicollinearity among the metrics using two complementary measures: the condition number, which provides an overall assessment of collinearity, and the variance inflation factor (VIF), which pinpoints the specific sources. \citep{belsley2005regression}.

The condition number is derived as the maximum of the condition indices, which themselves are computed from the eigenvalues ($\lambda$) of the scaled correlation (or covariance) matrix of the remaining metrics in a regression model as $\text{condition index}_i = \sqrt{(\lambda_{\max}/\lambda_{i})}$ where $\lambda_i$ are the eigenvalues of the correlation matrix of the predictors, and $\lambda_{\max}$ is the largest eigenvalue. Then for each metric, VIF is calculated as $\text{VIF}_i = 1/(1 - R_i^2)$, where $R_i^2$ is the coefficient of multiple determination from regressing metric $i$ against the remaining metrics.

Commonly accepted thresholds for weak collinearity are a condition number less than $10$ and VIF values below $5$ \citep{belsley2005regression}. Table \ref{tbl:multicol_summ} summarizes the multicollinearity analysis. The results demonstrate that, while the complete set of $16$ metrics displays weak global collinearity (condition number below $10$), certain strain-derived metrics exhibit elevated VIF values. When metrics related to $E_{cr}$ and $E_{rr}$ are excluded, both the condition number and variance inflation factors for the remaining metrics drop well below critical thresholds--except for ``Tissue Mean Peak Absolute Displacement'' and ``Pillar Mean Peak Force''--suggesting minimal collinearity. Based on this assessment, we restrict our downstream analyses to the $10$ key metrics identified in Table \ref{tbl:multicol_summ}.
 
\begin{table}[h]
\begin{center}
\caption{\label{tbl:multicol_summ}Multicollinearity was assessed using condition number and variance inflation factor. Collinearity is generally low, with condition numbers for both the full set of $16$ metrics ($8.04$) and the $10$ selected metrics ($5.67$) well below the threshold of $10$. The $10$ selected metrics are highlighted in bold font. Notable multicollinearity is confined to strain-derived metrics; excluding those linked to $E_{cr}$ and $E_{rr}$ yields consistently low VIF values for the remaining metrics, except for ``Tissue Mean Peak Absolute Displacement'' and ``Pillar Mean Peak Force,`` which are highly inter-correlated.}
\begin{tabular}{l >{\centering\arraybackslash}p{0.1\linewidth}>{\centering\arraybackslash}p{0.1\linewidth}}\toprule
\textbf{Metric}                        & \multicolumn{2}{c}{\textbf{Variance Inflation Factor}} \\\midrule 
\textbf{Tissue Aspect Ratio }                   &    1.40            & 1.32            \\ 
\textbf{Tissue Curvature }                      &    1.11            & 1.10            \\ 
\textbf{Beat Frequency}                         &    3.63            & 2.91            \\ 
\textbf{Tissue Mean Peak Absolute Displacement} &    5.11            & 4.91            \\ 
\textbf{Pillar Mean Peak Force}                &    7.28            & 5.0             \\ 
\textbf{Full Width at Half Maximum}                                  &    2.38            & 2.16            \\ 
\textbf{Wasserstein Distance PC10}              &    3.02            & 2.70            \\ 
\textbf{Ecc Strain Peak Average Asynchrony}     &    19.05           & 1.11            \\ 
Ecr Strain Peak Average Asynchrony     &    22.93           & --              \\ 
Err Strain Peak Average Asynchrony     &    15.04           & --              \\ 
\textbf{Ecc Average Pairwise DTW Distance}      &    3.07            & 2.01            \\ 
Ecr Average Pairwise DTW Distance      &    3.90            & --              \\ 
Err Average Pairwise DTW Distance      &    2.84            & --              \\ 
\textbf{Ecc GSI }                               &    7.54            & 3.12            \\ 
Ecr GSI                                &    3.87            & --              \\ 
Err GSI                                &    2.80            & --              \\ \toprule
\textbf{Condition number}              &    \textbf{8.04}   & \textbf{5.67}   \\ \bottomrule 
\end{tabular}
\end{center}
\end{table}

As a next step, we investigated pairwise relationships among all metrics. We first calculated the Spearman's rank correlation coefficient ($\rho$) \citep{spearman1961proof}, which captures monotonic relationships regardless of linearity, for every metric pair. Recognizing, as Anscombe's quartet \citep{anscombe1973graphs} exemplifies, that summary statistics alone can mask nuanced patterns in the data, we complemented the correlation analysis with pairplots for each metric combination to visually examine their associations. In addition, we included kernel density estimate (KDE) plots to assess the distribution and variability of each metric. The width of the KDE curve reflects the degree of variability: wider curves indicate greater variance, while narrow curves suggest more closely grouped observations. Moreover, skewness in the KDE curves may indicate data asymmetry or the presence of outliers.

To streamline interpretation and avoid redundancy inherent in symmetric correlation matrices, we condensed the results in Fig. \ref{fig:pairplot_corr}, focusing on the $10$ representative metrics with minimal collinearity identified in Table \ref{tbl:multicol_summ}. The figure is organized as a $10\times10$ matrix: the diagonal elements feature KDE plots illustrating each metric’s distribution, the upper triangle displays Spearman's correlation coefficients, and the lower triangle contains pairplots visualizing the relationships. Comprehensive results for all $16$ metrics are similarly presented in Fig. \ref{fig:corr_all}.

Surveying the matrix, the upper triangle reveals that most metric pairs show negligible ($|\rho| < 0.3$) or weak ($0.3 \leq |\rho| < 0.5$) monotonic associations \citep{mukaka2012guide, pakay2023foundations}, suggesting that the metrics largely capture distinct aspects of tissue behavior. Moderate ($0.5 \leq |\rho| < 0.7$) and strong ($0.7 \leq |\rho| < 0.9$) correlations are relatively rare and limited to several metric clusters. Noteworthy examples of strong positive correlations include ``Tissue Mean Peak Absolute Displacement,`` ``Pillar Mean Peak Force,`` ``Wasserstein Distance PC10,`` and ``$E_{cc}$ GSI.`` In contrast, ``Beat Frequency,`` ``Full Width at Half Maximum,`` and ``$E_{cc}$ Average Pairwise DTW Distance`` form a cluster of negative correlations. Nevertheless, perfect correlations are absent; the strongest observed Spearman coefficient is between ``Tissue Mean Peak Absolute Displacement'' and ``Pillar Mean Peak Force'' ($\rho=0.826$), underscoring that each metric retains unique informational value.

The KDE plots along the diagonal confirm that metrics such as ``Tissue Curvature'' and ``$E_{cc}$ Strain Peak Average Asynchrony`` are relatively consistent across conditions, while others like ``Pillar Mean Peak Force'' and ``Full Width at Half Maximum`` display greater variability and occasional outliers. The pairplots in the lower triangle further validate the absence of pronounced non-monotonic or anomalous relationships, providing visual confirmation that the data structure is well-represented by the computed Spearman correlations.

\begin{figure}[h]
\begin{center}
\includegraphics[width=1\textwidth]{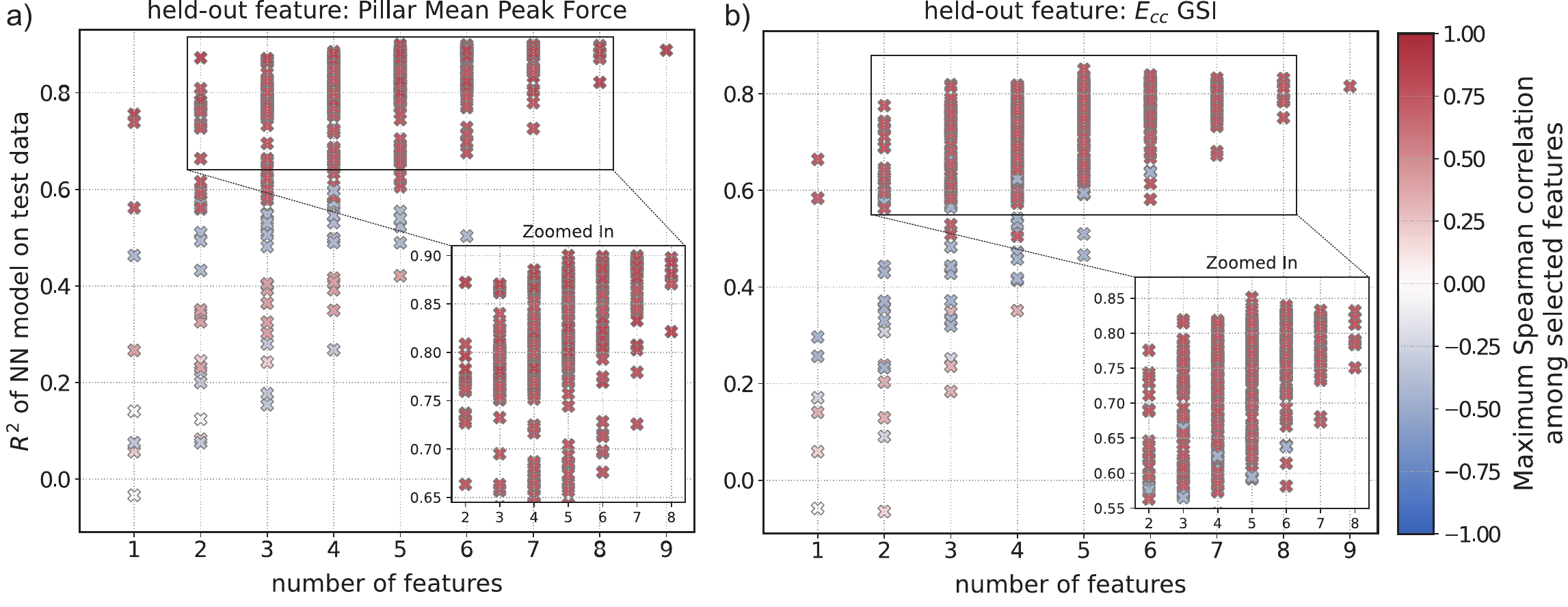}
\caption{\label{fig:nn_train_v2}Summary of prediction performance for all multilayer perceptron (MLP) models trained using different numbers and combinations of input features, with (a) ``Pillar Mean Peak Force'' and (b) ``$E_{cc}$ GSI'' held out as target variables. For each case, the $R^2$ value of the neural network on test data is shown as a function of the number of input features. Results are color-coded according to the maximum Spearman’s correlation coefficient observed among the selected input features, highlighting the impact of feature inter-correlation on model performance. Insets provide a closer view of regions with higher $R^2$.}
\end{center}
\end{figure}

\subsubsection{Prediction-based assessment of metric redundancy}

To provide an alternative perspective on metric redundancy, we designed an exhaustive prediction-based strategy using machine learning methods. From the set of $10$ metrics, we selected one as a held-out target and used the remaining $9$ metrics as input features for prediction. For these $9$ metrics, we considered all possible feature combinations of size $k$ ($k \in [1,9]$), resulting in a total of $511$ unique subsets ($\sum_{k=1}^{9}$$_9\rm C_k$). For each subset, we trained a multilayer perceptron (MLP) using the scikit-learn Python library (v$1.7.1$) \citep{scikit-learn}, employing a parameter grid search to optimize model depth (up to three layers) and the hyperparameters \verb|alpha| and \verb|learning_rate_init|, selecting the configuration with the lowest validation loss. Fixed common hyperparameters include \verb|activation=`relu`|, \verb|solver=`adam`|, \verb|num_epochs=1000|, \verb|patience=15|, \verb|lr_patience=8|, \verb|decay_factor=0.5|, and \verb|random_state=42|. The dataset was split into $60\%$, $20\%$, and $20\%$ for training, validation, and test, respectively. This analysis was performed for two representative metrics with notably high inter-correlation and elevated VIF values: ``Pillar Mean Peak Force'' and ``$E_{cc}$ GSI.`` 

The outcomes of this predictive analysis are summarized in Fig \ref{fig:nn_train_v2}, where we evaluated the reconstruction accuracy using the $R^2$ score on the test dataset. Each outcome is color-coded based on the highest Spearman’s correlation coefficient identified among the input features. In general, we observed that the predictive performance improves as the number of input features increases, particularly when those features include variables with strong Spearman correlation to the held-out metric. This trend demonstrates that access to more relevant features enhances the model's ability to capture the underlying structure of the data. However, after a certain point, the addition of more features yields only marginal gains, and the $R^2$ curve plateaus. This plateauing effect indicates that once the most influential variables have been incorporated into the model, further expansion of the input set provides diminishing returns, highlighting redundancy among certain features. For example, a model trained with $2$ metrics, with one being ``Tissue Mean Peak Absolute Displacement,'' can predict ``Pillar Mean Peak Force'' with a similar accuracy to one trained on all $9$ features.   

This analysis reveals intricate, non-linear relationships between metrics that transcend the scope of standard correlation measures. Importantly, subsets of features with moderate maximum Spearman correlation ($0.5 < |\rho_{max}| < 0.7$) can still achieve substantial predictive accuracy. Future studies employing larger datasets, rigorously defined metrics, and advanced machine learning techniques will be pivotal in disentangling the connections among diverse tissue characteristics, such as the links between structural properties and heterogeneous contractile function. Such approaches can also help determine the minimum set of metrics required for comprehensive characterization of tissue dynamics. Ultimately, these strategies will provide critical insights for experimental design and support more informed and targeted investigations.

\begin{figure}[p]
\begin{center}
\includegraphics[angle=0,width=1\textwidth]{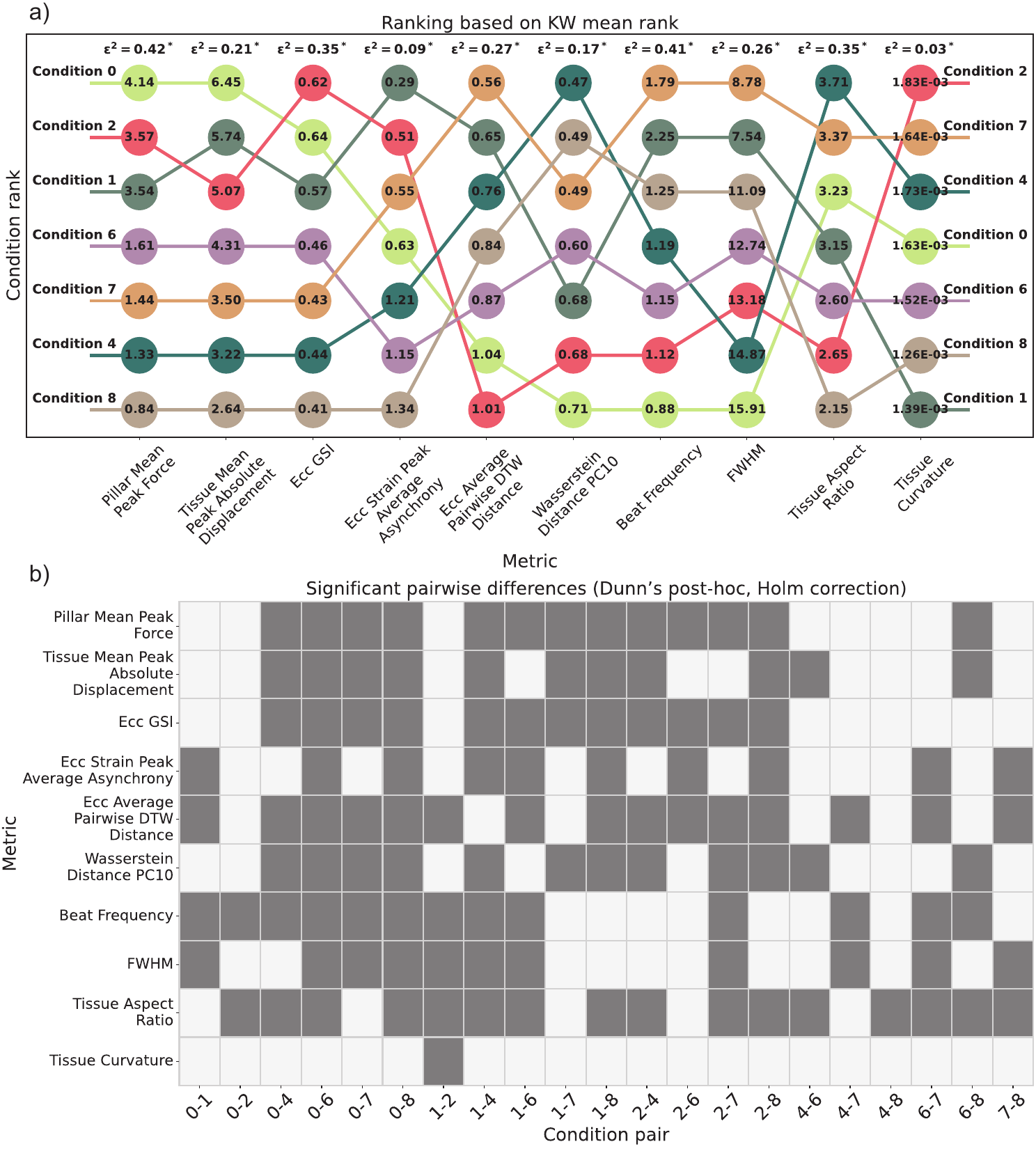}
\caption{\label{fig:feat_imp}Statistical comparison of selected features across experimental conditions: (a) bump chart showing condition rankings based on Kruskal-Wallis mean rank with feature median values displayed within the circle markers. Kruskal–Wallis effect sizes ($\epsilon^2$) are summarized above each feature, with asterisks indicating statistical significance. Ranking order (ascending or descending) is based on each metric's favorable direction; (b) binary matrix of significant pairwise differences between condition pairs identified by Dunn’s post-hoc test with Holm–Bonferroni correction, where dark grey cells indicate comparisons with p $<0.05$.}
\end{center}
\end{figure}

\subsection{Condition-level comparisons depend on the selected feature} 
\label{subsec:statistics}
In this Section, we investigate whether the choice of metric influences the analysis outcome. Our approach builds on the correlation and redundancy analyses presented in Section \ref{subsec:correlations}, which motivated the use of the reduced set of $10$ metrics. From the $20$ experimental conditions, we retained only $7$ with more than $25$ samples to ensure adequate statistical power and meaningful comparisons for this component of our analysis.

\begin{figure}[p]
\begin{center}
\includegraphics[width=1\textwidth]{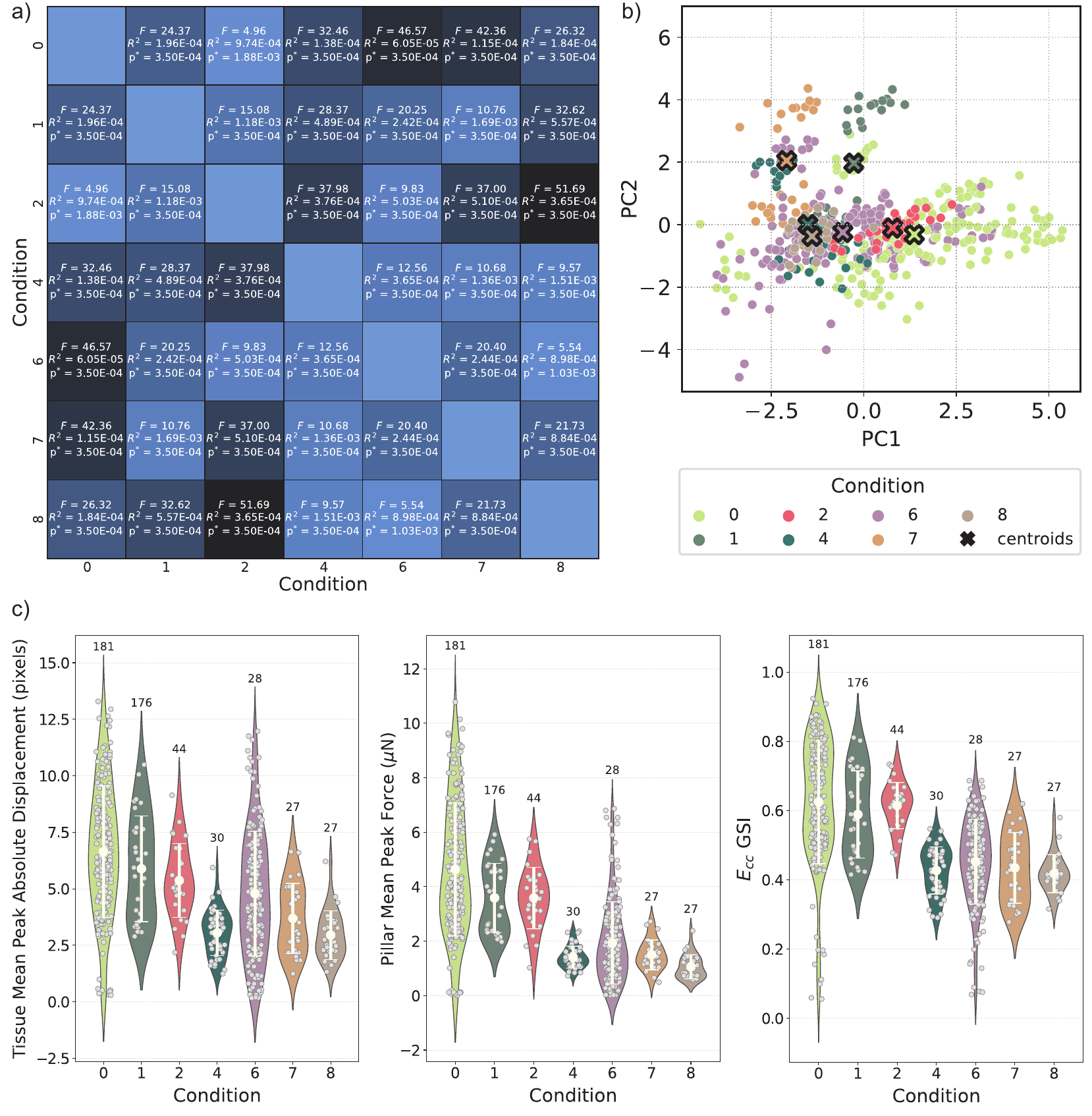}
\caption{\label{fig:condition_variance}Multivariate statistical analysis across experimental conditions. (a) Annotated heatmap of PERMANOVA results, with darker blue shades indicating higher pseudo-$F$ statistics; $R^2$ and p-values are also displayed for each condition pair with an asterisk denoting statistical significance. (b) Principal component projection (PC1 vs. PC2) of the dataset comprising the $10$ metrics and $7$ conditions, with data points color-coded by condition and centroids marked for each group. (c) Violin plots of $3$ representative metrics, ``Tissue Mean Peak Absolute Displacement,'' ``Pillar Mean Peak Force,'' and ``$E_{cc}$ Average Pairwise DTW Distance,'' showing mean ± standard deviation for each condition.}
\end{center}
\end{figure}

To select a suitable statistical framework, we considered several key properties of our data: metric distributions are generally non-normal, group variances are unequal, and sample sizes vary across conditions. Additionally, our analyses require comparison across multiple groups for each metric. Given these characteristics, the Kruskal–Wallis H test \citep{kruskal1952use} is well-suited for our needs. As a nonparametric, rank-based method, it serves as an alternative to one-way ANOVA for comparing $3$ or more independent samples. The interpretation of the Kruskal–Wallis test \citep{kruskal1952use} depends heavily on the underlying distributions of the observations \citep{dinno2015nonparametric}: when group distributions differ markedly, the test assesses stochastic dominance; if distributions are identical, it tests for differences in medians; and with symmetric distributions, it tests for mean differences. In our context, the metric distributions across the $7$ conditions (Figs \ref{fig:condition_variance}c and \ref{fig:condition_variance_app}b) are neither identical nor symmetric; hence, we interpret the Kruskal–Wallis test results as reflecting differences in stochastic dominance among conditions.

To complement the assessment of statistical significance (p < $0.05$), we also report an effect size $\epsilon^2$ \citep{kelley1935unbiased}, defined as $\epsilon^2 = (H - k + 1)/(N - 1)$, where $H$ is the Kruskal–Wallis statistic, $k$ is the number of conditions ($k = 7$), and $N$ is the total number of tissue samples ($N = 513$). Following the guidelines in \citep{field2024discovering}, values of $0.01 \le \epsilon^2 < 0.06$ indicate a small effect, $0.06 \le \epsilon^2 < 0.14$ a medium effect, and $\epsilon^2 > 0.14$ a large effect.

Because a significant Kruskal–Wallis result does not identify which specific groups differ, we follow up with Dunn’s test which compares mean ranks for stochastic dominance \citep{dunn1964multiple} among multiple pairwise post-hoc comparisons. To control for type I errors arising from multiple testing, we apply the Holm–Bonferroni correction \citep{holm1979simple}.

We present the results of this analysis in Fig. \ref{fig:feat_imp}. Fig \ref{fig:feat_imp}a summarizes the condition Kruskal-Wallis mean rankings for each metric, where favorable ranks are positioned at the top. For clarity of interpretation, the metrics are ordered such that higher-ranked values correspond to more desirable outcomes. For instance, distance-based metrics are arranged in ascending order, where smaller distances are favored, reflecting an assumed preference for homogeneous tissue, whereas force-related metrics are ordered in descending fashion, reflecting the desirability of higher force generation.  

At the top of the chart, Kruskal–Wallis effect sizes ($\epsilon^2$) are displayed for each metric, with statistically significant results marked by an asterisk. Most metrics exhibit large effect sizes, with the exception of ``$E_{cc}$ Strain Peak Average Asynchrony,`` which shows a medium effect, and ``Tissue Curvature,`` which shows a small effect. Each circular marker (Fig \ref{fig:feat_imp}a) denotes the median metric value for that condition. Generally, rankings by median and Kruskal-Wallis mean rank are in agreement; however, any discrepancies highlight underlying non-standard distributions--such as multimodality, skewness, or outliers--where the mean rank offers a more accurate comparative ordering. For example, although condition $0$ has a higher median $E_{cc}$ GSI than condition $2$, its rank is lower, which aligns with the relatively bimodal and skewed distribution seen in $E_{cc}$ GSI for condition $0$ (see Fig \ref{fig:condition_variance_app}b).

When comparing how conditions rank across individual metrics (Fig \ref{fig:feat_imp}a), we observe notable consistency among the first $3$, ``Tissue Mean Peak Absolute Displacement,'' ``Pillar Mean Peak Force,'' and ``$E_{cc}$ GSI,'' for which the relative ordering of conditions remains largely stable. This alignment suggests that conditions producing greater tissue displacement or contraction also tend to generate higher forces and exhibit more globally synchronized beating.
In contrast, the rankings diverge considerably for the remaining metrics. For instance, conditions $0$ and $2$, which appear highly favorable according to the first $3$ metrics, move to the bottom of the ranking when evaluated using ``$E_{cc}$ Average Pairwise DTW Distance``, ``Wasserstein Distance PC10``, ``Beat Frequency``, and ``Full Width at Half Maximum.`` This shift indicates that, despite exhibiting strong temporal synchrony, tissues under these conditions display more heterogeneous strain time series profiles across spatial regions, produce displacement vector fields that deviate more from the low-dimensional reconstruction based on the first $10$ principal components, beat at higher frequencies, and exhibit broader beat profiles. Thus, the choice of metric can substantially alter the outcome of the comparison, at times leading to contrasting interpretations of condition favorability.                                

These observations must be interpreted in conjunction with the statistical significance of pairwise comparisons obtained from the Holm-corrected Dunn’s test \citep{dunn1964multiple, holm1979simple} (Fig. \ref{fig:feat_imp}b). The results reveal that certain condition pairs exhibit virtually no significant differences across any metric, for example, pairs $0$–$2$ and $4$–$8$, whereas others, such as $0$–$6$ and $0$–$8$, differ significantly on nearly all metrics. In this context, the apparent ranking of conditions $0$, $1$, and $2$ for ``Tissue Mean Peak Absolute Displacement,'' ``Pillar Mean Peak Force,'' and ``$E_{cc}$ GSI'' does not hold statistical support: the pairwise differences among these conditions are not significant.

\subsubsection{Note on the dangers of analytical flexibility}

A critical implication of our statistical analysis is the risk of inadvertently engaging in data dredging, or p-hacking \citep{simonsohn2014p}. This occurs when researchers explore a wide array of analytical choices such as varying the selection of variables, data subsets, or preprocessing decisions, and selectively report only those results that achieve statistical significance. In our dataset, this risk may take the form of highlighting only significant findings, repeatedly modifying data-cleaning criteria (for example, removing or retaining outliers), or performing multiple subgroup analyses but disclosing only those yielding p-values below 0.05. Notably, these behaviors may arise inadvertently rather than from intentional misconduct. Nevertheless, they undermine statistical validity and inflate the likelihood of false positives, presenting spurious outcomes as genuine discoveries. 

In the present work, our aim is not to advance specific biological conclusions but to illustrate, in a conclusion-agnostic manner, the potential range of analytical comparisons that can be performed with these types of data, and highlight the methodological considerations that accompany them. As larger and more complex datasets become increasingly common, we encourage researchers to carefully document and justify key analytical decisions, consider pre-specifying their primary analysis strategy (for example, through preregistration), and transparently report relevant analytical alternatives where practical. Such practices need not require exhaustive reporting of every possible analysis, but fostering transparency in the analytical workflow can help minimize false discoveries and strengthen the reproducibility and robustness of scientific findings \citep{forstmeier2017detecting, weston2019recommendations, lakens2024benefits}.

\subsection{Multivariate analysis reveals distinct condition centroids amid broad dispersion} 
\label{subsec:multivariate}
As a final example of what is possible with these additional metrics, we examined how the combined features differ across experimental conditions using Permutational Multivariate Analysis of Variance (PERMANOVA) \citep{anderson2014permutational}. PERMANOVA is a nonparametric test that evaluates whether multivariate observations differ between groups based on a chosen distance matrix. Unlike MANOVA \citep{weinfurt1995multivariate}, it does not require multivariate normality and is widely used with skewed, sparse, or zero-inflated data. The method partitions the total sum of squares in the distance matrix and assesses group differences by permuting group labels to construct a null distribution for the pseudo-F statistic.

However, PERMANOVA is sensitive to differences in both group centroids and group dispersions. To evaluate whether groups differ in their multivariate dispersion, defined as the average distance of observations to their group centroid, we also applied PERMDISP (Permutational Analysis of Multivariate Dispersions) \citep{anderson2014permutational}. Interpreting PERMANOVA together with PERMDISP provides a more nuanced understanding of the multivariate structure: significant PERMANOVA results in the absence of dispersion differences suggest genuine centroid shifts, whereas significant dispersion differences indicate that variation in group spread may contribute to or even drive the observed group separation.

For this analysis, we focused on the reduced set of $10$ metrics across the $7$ selected experimental conditions defined in Section \ref{subsec:statistics}. PERMANOVA was implemented in Python using the \texttt{scikit-bio} library \citep{skbio}, with Euclidean distances computed after standardizing the data. To ensure robust statistical inference, we used $59{,}999$ permutations with a fixed random seed of $123$. Pairwise PERMANOVA tests were conducted for all condition combinations, and the Holm-Bonferroni method \citep{holm1979simple} was applied to correct for multiple comparisons. The results are summarized in Figs.~\ref{fig:condition_variance} and \ref{fig:condition_variance_app}.

Figure~\ref{fig:condition_variance}a displays the PERMANOVA outputs, including the pseudo-$F$ statistic (the ratio of between-group to within-group mean squares obtained by partitioning sums of squares derived from the distance matrix), the explained variation $R^2$ (the proportion of distance-based variation explained by group membership), and the associated $p$-values. All condition pairs exhibit statistically significant differences in their multivariate distance structures, with pairs $2$-$8$, $0$-$6$, $0$-$7$, and $2$-$4$ exhibiting the highest pseudo-$F$ values. However, the $R^2$ values are extremely small (on the order of $10^{-3}$ to $10^{-4}$). These findings indicate that although group differences are statistically significant, the effect sizes are minimal, meaning that condition explains only a very small fraction of the total multivariate structure. Thus, the observed differences are statistically detectable but very subtle.

To further interpret the PERMANOVA findings, we supplemented our analysis with PERMDISP (Permutational Analysis of Multivariate Dispersions), also implemented using \texttt{scikit-bio} \citep{skbio} (Fig. \ref{fig:condition_variance_app}a). PERMDISP tests whether groups differ in their multivariate dispersion, defined as the average distance of observations to their respective group centroids. Combining the results of PERMANOVA and PERMDISP leads to two primary scenarios:

\begin{enumerate}
    \item \textbf{Significant PERMANOVA, extremely small $R^2$, and non-significant PERMDISP} (pairs: $0$-$1$, $0$-$6$, $0$-$7$, $1$-$4$, $1$-$6$, $1$-$7$, $2$-$8$, $4$-$7$, $6$-$8$):\\
    In these cases, dispersions do not differ across groups, and the significant PERMANOVA result is therefore most consistent with a subtle shift in group centroids rather than dispersion-driven effects. These pairs likely reflect genuine but very small differences in multivariate location.

    \item \textbf{Significant PERMANOVA, extremely small $R^2$, and significant PERMDISP} (pairs: $0$-$2$, $0$-$4$, $0$-$8$, $1$-$2$, $1$-$8$, $2$-$4$, $2$-$6$, $2$-$7$, $4$-$6$, $4$-$8$, $6$-$7$, $7$-$8$):\\
    In these pairs, groups differ in dispersion, meaning that the significant PERMANOVA result may be partly or entirely driven by heterogeneous spread rather than differences in centroid location. Consequently, interpretations of group separation should be made with caution, because dispersion differences can inflate or mimic significance in PERMANOVA.
\end{enumerate}

To visually explore these patterns, we performed a principal component analysis (PCA) on the tissue examples using the $10$ metrics. Figure \ref{fig:condition_variance}b shows the data projected onto the first two principal components, color-coded by condition. This visualization shows separated condition centroids (with the exceptions of pairs $0$–$2$ and $4$–$8$, which are closer in space), but extensive overlap and large dispersion within each condition exists, underscoring the effects observed in the PERMANOVA and PERMDISP results.

Additionally, Fig.~\ref{fig:condition_variance}c displays violin plots for three representative metrics: ``Tissue Mean Peak Absolute Displacement,'' ``Pillar Mean Peak Force,'' and ``$E_{cc}$ GSI.'' These visualizations, together with the complementary violin plots for the remaining seven metrics in Fig.~\ref{fig:condition_variance_app}b, help contextualize the multivariate statistical findings. For example, the condition pair $0$-$2$ ($F = 4.96$, PERMANOVA; $F = 43.82$, PERMDISP, both statistically significant) exhibits a large difference in within-group dispersion and only modest separation in group centroids in multivariate space. In contrast, the pair $2$-$8$ ($F = 51.69$, PERMANOVA, statistically significant; $F = 0.17$, PERMDISP, not significant) shows clear separation between group centroids with relatively similar within-group dispersions.

\section{Conclusion} 
\label{sec:concl}
In this work, we present a computational pipeline for quantifying dynamic behavior in brightfield videos of beating cardiac microbundles, leveraging and expanding our open-source tools ``MicroBundleCompute'' \citep{kobeissi2024microbundlecompute} and ``MicroBundlePillarTrack'' \citep{kobeissi2024microbundlepillartrack}. We introduce $16$ metrics that capture heterogeneous spatiotemporal contractility by integrating structural and functional features with hybrid spatial and temporal descriptors, and we apply them to a dataset of $670$ cardiac tissues spanning $20$ experimental conditions on the fibroTUG platform \citep{kobeissi2024fibroTUG, depalma2024matrix}. Through statistical and machine learning analyses, we assess the relevance, redundancy, and necessity of each metric and identify a core subset required to effectively characterize tissue behavior. Dimensionality reduction methods (PCA, UMAP, t-SNE) show that no combination of metrics perfectly separates experimental conditions, underscoring the inherent biological complexity of this system. Analysis of denoised displacement fields further reveals that tissue contraction is primarily linear and isotropic along the major axis, with saddle points present in approximately $50\%$ of samples, an observation that warrants deeper investigation into the relationship between contractile patterns and microstructure.

Investigation of metric interrelationships showed that multicollinearity and redundancy are limited. After reducing the strain-derived metrics by retaining only the $E_{cc}$ component, a refined set of 10 metrics captured the majority of the informational value. Correlations among metrics were mostly weak to moderate, with the strongest Spearman coefficient ($\rho = 0.826$) observed between ``Tissue Mean Peak Absolute Displacement'' and ``Pillar Mean Peak Force'' and the weakest between ``Tissue Curvature'' and ``$E_{cc}$ GSI'' ($\rho = -0.004$). Machine learning models further supported this conclusion by demonstrating that predictive accuracy improves when correlated features are included, but only up to a point, indicating that redundancy among metrics is modest. These results also highlight that quantitative interpretations can vary depending on the specific metrics used, particularly when comparing across experimental conditions. Multivariate analyses revealed that within-condition dispersion often exceeds between-condition separation, underscoring the need for cautious interpretation when drawing biological conclusions. Collectively, our analytical framework provides a robust and reproducible approach for the comprehensive study of cardiac microtissue contractility across diverse experimental scenarios, and the full Python implementation is openly available on GitHub (\href{https://github.com/HibaKob/MicroBundleAnalysis}{https://github.com/HibaKob/MicroBundleAnalysis}) to support broad adoption and continued development.

While our framework was applied to a single cardiac tissue platform, the methodology is readily extensible to time-lapse imaging of cardiac tissues across a variety of experimental setups and modalities where approximating full-field deformation is feasible \citep{boudou2012microfabricated, ewoldt2024hypertrophic, zhao2019platform, karakan2024geometry, tsan2021physiologic}. Moreover, these approaches can be adapted to other engineered tissue types, such as the actuated two-dimensional muscle sheets described in \citep{rios2023mechanically}. Moving forward, analyses leveraging this comprehensive suite of metrics can guide the design of more rigorous and comparable experiments, enabling the identification of optimal culture conditions and configurations that enhance hiPSC-CM tissue maturation. This, in turn, will facilitate reproducible extraction of mechanical phenotypes across diverse testbeds and directly inform the development of improved computational models of engineered tissues.

In particular, a promising direction for future work is to integrate these dynamic metrics with detailed structural descriptors obtained from fluorescent staining and complementary imaging modalities, such as calcium imaging. This integrated approach could address important questions, such as how much predictive power can be gained from easily obtained structural metrics derived from still images, or how structural organization, ranging from sarcomere geometry and alignment to the broader architecture of the extracellular matrix and cell arrangement, influences contractile function. Ultimately, these investigations may help determine whether brightfield imaging alone captures sufficient information to characterize tissue behavior, or if complementary modalities, such as calcium imaging, are necessary. In particular, it will be valuable to evaluate if synchrony metrics like the GSI can provide robust functional insights using only brightfield data. While this remains a challenging question, it presents a compelling avenue for advancing tissue engineering research.

In summary, the pipeline and metrics developed in this work contribute a versatile and extensible toolkit for the quantitative analysis of engineered tissue dynamics. By making these methods openly available, we hope to accelerate discovery, facilitate cross-platform comparisons, and inspire innovative experiments in the field of cardiac tissue engineering and beyond.

\section{Acknowledgements}
This work was supported by the CELL-MET Engineering Research Center (NSF ERC ECC-1647837) and the National Science Foundation (Grant CMMI-2311640). We thank Boston University Research Computing Services for providing the computational infrastructure and tools that enabled the analyses reported here. We are also grateful to Professor Paul Barbone for his insightful feedback and guidance on principal component analysis.

\appendix 

\label{sec:appendix}


\section{Supplementary Figures}
\label{appendix-a}

\renewcommand{\thefigure}{S\arabic{figure}}
\setcounter{figure}{0}

\begin{figure}[!ht]
\begin{center}
\includegraphics[width=0.8\textwidth]{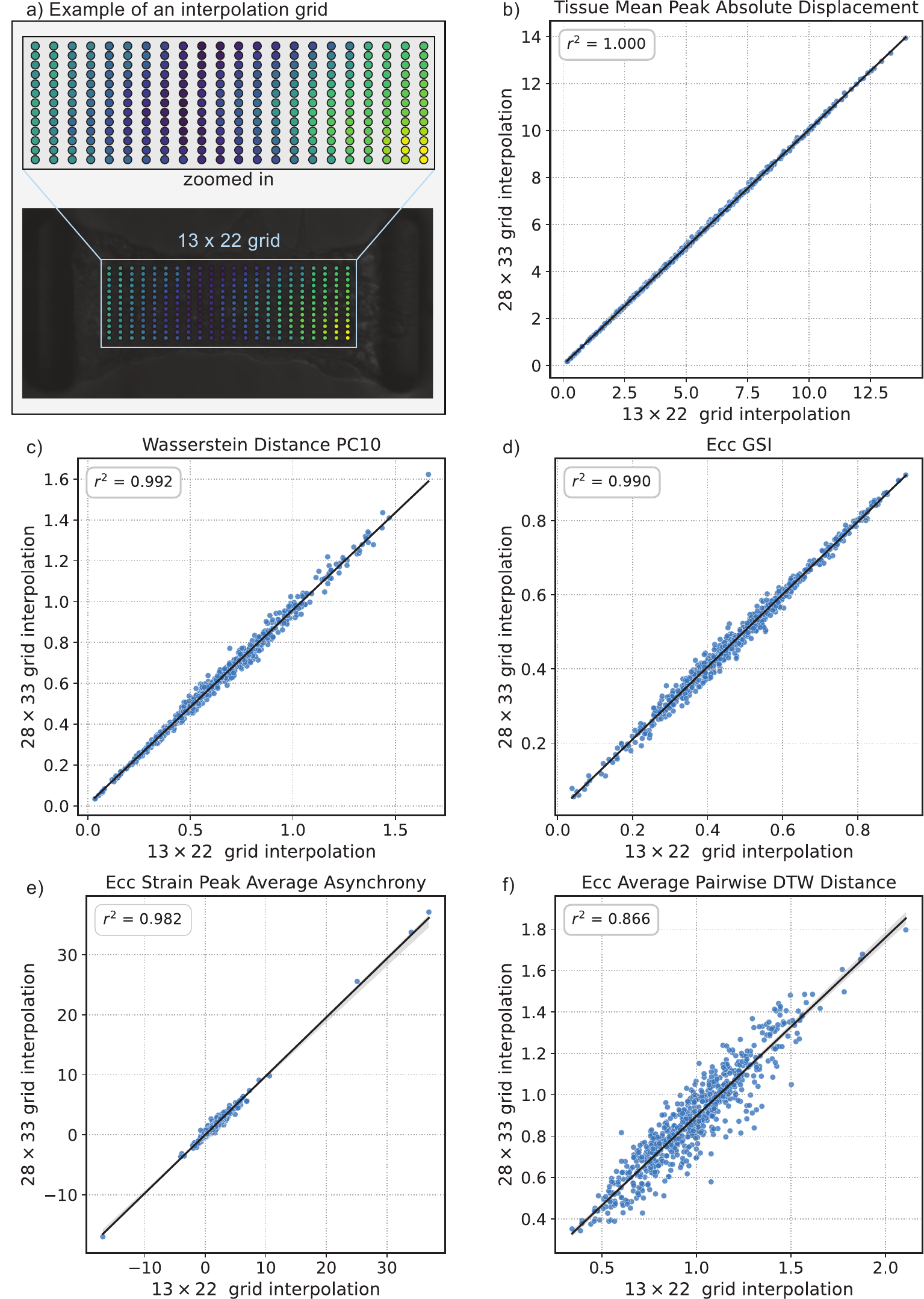}
\caption{\label{fig:grid_sens}Grid sensitivity analysis for all grid-dependent metrics. (a) Example of a tissue-specific $13\times22$ interpolation grid used for displacement and strain calculations, with a zoomed-in view. For our main analysis, a finer $28\times33$ grid is applied. Subplots (b–f) compare metric values computed on both grid resolutions, demonstrating convergence with increasing grid size. While most metrics show strong agreement (high $r^2$ values) between grids, ``$E_{cc}$ Average Pairwise DTW Distance'' exhibits greater sensitivity to grid resolution. Based on these results, the $28\times33$ grid provides sufficient resolution for robust metric calculations.}
\end{center}
\end{figure}

\begin{figure}[p]
\begin{center}
\includegraphics[width=1\textwidth]{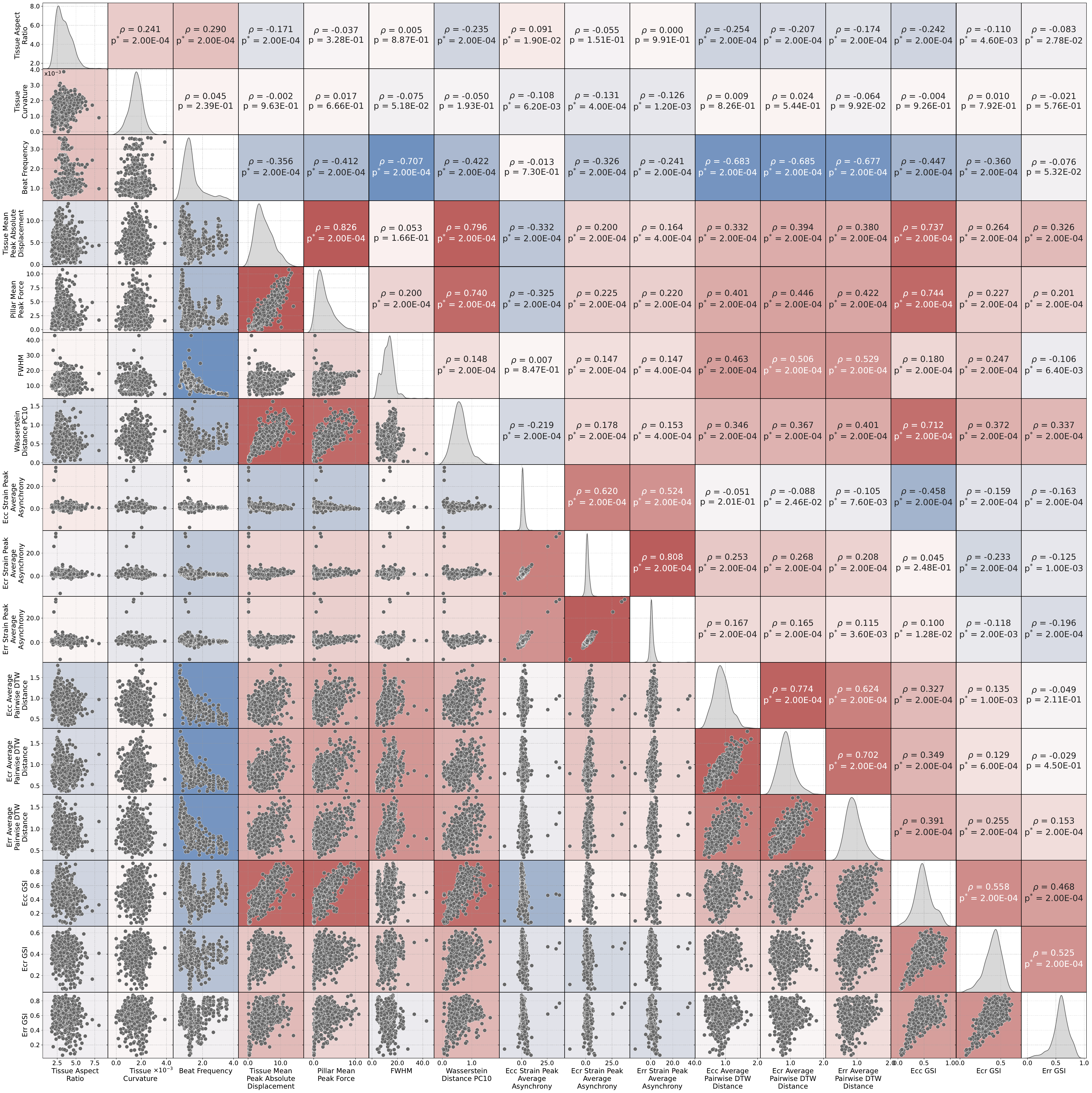}
\caption{\label{fig:corr_all}Expanded version of Fig \ref{fig:pairplot_corr} displaying all $16$ extracted metrics. The upper triangle presents the Spearman’s correlation coefficients ($\rho$) for each metric pair, accompanied by their respective p-values. The lower triangle contains pairplots depicting the joint distributions for each pair of metrics, providing further support for the monotonicity assumption underlying the Spearman correlation analysis. Along the diagonal, Kernel Density Estimate (KDE) plots illustrate the individual distribution of each metric, highlighting differences in data spread and underlying variability across metrics.}
\end{center}
\end{figure}

\begin{figure}[p]
\begin{center}
\includegraphics[width=0.88\textwidth]{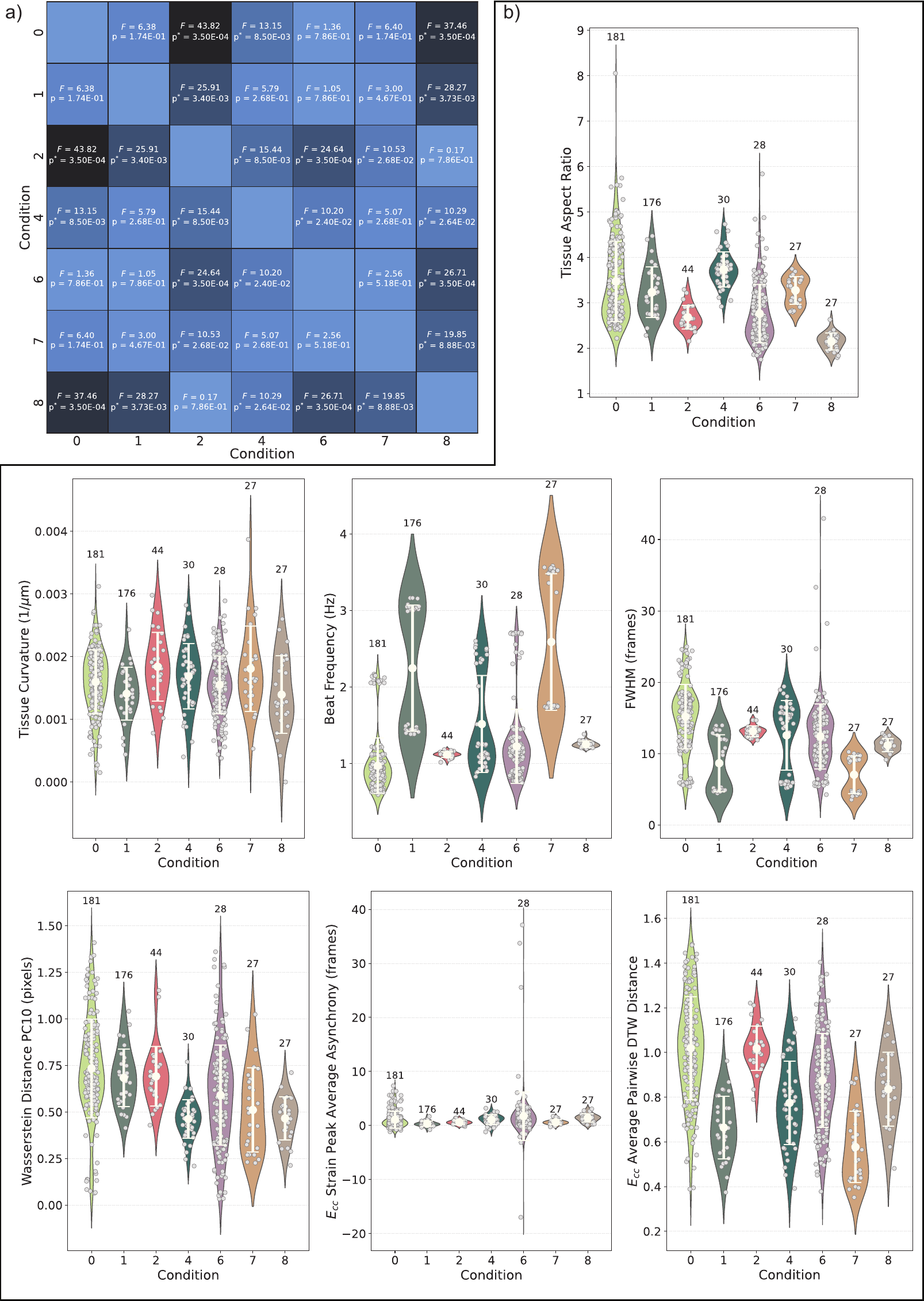}
\caption{\label{fig:condition_variance_app}Multivariate statistical analysis across experimental conditions. (a) Annotated heatmap of PERMDISP results, where deeper blue tones represent higher pseudo-$F$ statistics. P-values are shown for each condition pair, with an asterisk indicating statistical significance. (b) Violin plots for the remaining $7$ metrics from Fig. \ref{fig:condition_variance}, depicting the distribution, mean, and standard deviation for each condition.}
\end{center}
\end{figure}

\FloatBarrier
\newpage
\bibliographystyle{vancouver}
\typeout{}
\bibliography{references}  

@article{de2024tissue,
  title={The tissue engineering revolution: from bench research to clinical reality},
  author={De Chiara, Francesco and Ferret-Mi{\~n}ana, Ainhoa and Fern{\'a}ndez-Costa, Juan M and Ram{\'o}n-Azc{\'o}n, Javier},
  journal={Biomedicines},
  volume={12},
  number={2},
  pages={453},
  year={2024},
  publisher={MDPI},
  doi={10.3390/biomedicines12020453}
}

@article{hoang2025tissue,
  title={Tissue Engineering and Regenerative Medicine: Perspectives and Challenges},
  author={Hoang, Van T and Nguyen, Quyen Thi and Phan, Trang Thi Kieu and Pham, Trang H and Dinh, Nhung Thi Hong and Anh, Le Phuong Hoang and Dao, Lan Thi Mai and Bui, Van Dat and Dao, Hong-Nhung and Le, Duc Son and others},
  journal={MedComm},
  volume={6},
  number={5},
  pages={e70192},
  year={2025},
  publisher={Wiley Online Library},
  doi={10.1002/mco2.70192}
}

@article{de2021scaffold,
  title={Scaffold-free cell-based tissue engineering therapies: advances, shortfalls and forecast},
  author={De Pieri, Andrea and Rochev, Yury and Zeugolis, Dimitrios I},
  journal={NPJ Regenerative medicine},
  volume={6},
  number={1},
  pages={18},
  year={2021},
  doi={10.1038/s41536-021-00133-3},
  publisher={Nature Publishing Group UK London}
}

@article{kalkunte2024review,
  title={A review on machine learning approaches in cardiac tissue engineering},
  author={Kalkunte, Nikhith and Cisneros, Jorge and Castillo, Edward and Zoldan, Janet},
  journal={Frontiers in Biomaterials Science},
  volume={3},
  pages={1358508},
  year={2024},
  doi={10.3389/fbiom.2024.1358508},
  publisher={Frontiers Media SA}
}

@article{luo2024current,
  title={Current challenges in imaging the mechanical properties of tissue engineered grafts},
  author={Luo, Lu and Okur, Kerime Ebrar and Bagnaninchi, Pierre O and El Haj, Alicia J},
  journal={Frontiers in Biomaterials Science},
  volume={3},
  pages={1323763},
  year={2024},
  doi={10.3389/fbiom.2024.1323763},
  publisher={Frontiers Media SA}
}

@article{cho2022challenges,
  title={Challenges and opportunities for the next generation of cardiovascular tissue engineering},
  author={Cho, Sangkyun and Discher, Dennis E and Leong, Kam W and Vunjak-Novakovic, Gordana and Wu, Joseph C},
  journal={Nature Methods},
  volume={19},
  number={9},
  pages={1064--1071},
  year={2022},
  doi={10.1038/s41592-022-01591-3},
  publisher={Nature Publishing Group US New York}
}

@article{eldeeb2022biomaterials,
  title={Biomaterials for tissue engineering applications and current updates in the field: a comprehensive review},
  author={Eldeeb, Alaa Emad and Salah, Salwa and Elkasabgy, Nermeen A},
  journal={Aaps Pharmscitech},
  volume={23},
  number={7},
  pages={267},
  year={2022},
  doi={10.1208/s12249-022-02419-1},
  publisher={Springer}
}

@article{ouyang2017imaging,
  title={The imaging tsunami: computational opportunities and challenges},
  author={Ouyang, Wei and Zimmer, Christophe},
  journal={Current Opinion in Systems Biology},
  volume={4},
  pages={105--113},
  year={2017},
  doi={10.1016/j.coisb.2017.07.011},
  publisher={Elsevier}
}

@article{kemmer2023building,
  title={Building a FAIR image data ecosystem for microscopy communities},
  author={Kemmer, Isabel and Keppler, Antje and Serrano-Solano, Beatriz and Rybina, Arina and {\"O}zdemir, Bu{\u{g}}ra and Bischof, Johanna and El Ghadraoui, Ayoub and Eriksson, John E and Mathur, Aastha},
  journal={Histochemistry and Cell Biology},
  volume={160},
  number={3},
  pages={199--209},
  year={2023},
  doi={10.1007/s00418-023-02203-7},
  publisher={Springer}
}

@article{dou2022microengineered,
  title={Microengineered platforms for characterizing the contractile function of in vitro cardiac models},
  author={Dou, Wenkun and Malhi, Manpreet and Zhao, Qili and Wang, Li and Huang, Zongjie and Law, Junhui and Liu, Na and Simmons, Craig A and Maynes, Jason T and Sun, Yu},
  journal={Microsystems \& Nanoengineering},
  volume={8},
  number={1},
  pages={26},
  year={2022},
  doi={10.1038/s41378-021-00344-0},
  publisher={Nature Publishing Group UK London}
}

@article{bertram2023open,
  title={Open science},
  author={Bertram, Michael G and Sundin, Josefin and Roche, Dominique G and S{\'a}nchez-T{\'o}jar, Alfredo and Thor{\'e}, Eli SJ and Brodin, Tomas},
  journal={Current biology},
  volume={33},
  number={15},
  pages={R792--R797},
  year={2023},
  doi={10.1016/j.cub.2023.05.036 External Link}, 
  publisher={Elsevier}
}

@article{wilkinson2016fair,
  title={The FAIR Guiding Principles for scientific data management and stewardship},
  author={Wilkinson, Mark D and Dumontier, Michel and Aalbersberg, IJsbrand Jan and Appleton, Gabrielle and Axton, Myles and Baak, Arie and Blomberg, Niklas and Boiten, Jan-Willem and da Silva Santos, Luiz Bonino and Bourne, Philip E and others},
  journal={Scientific data},
  volume={3},
  number={1},
  pages={1--9},
  year={2016},
  doi={https://doi.org/10.1038/sdata.2016.18},
  publisher={Nature Publishing Group}
}

@article{ashammakhi2022highlights,
  title={Highlights on advancing frontiers in tissue engineering},
  author={Ashammakhi, Nureddin and GhavamiNejad, Amin and Tutar, Rumeysa and Fricker, Annabelle and Roy, Ipsita and Chatzistavrou, Xanthippi and Hoque Apu, Ehsanul and Nguyen, Kim-Lien and Ahsan, Taby and Pountos, Ippokratis and others},
  journal={Tissue Engineering Part B: Reviews},
  volume={28},
  number={3},
  pages={633--664},
  year={2022},
  doi={10.1089/ten.teb.2021.00},
  publisher={Mary Ann Liebert, Inc., publishers 140 Huguenot Street, 3rd Floor New~…}
}

@article{national2018open,
  title={Open science by design: Realizing a vision for 21st century research},
  author={{National Academies of Sciences and Medicine and Global Affairs and Board on Research Data and Committee on Toward an Open Science Enterprise}},
  year={2018},
  doi={doi.org/10.17226/25116},
  journal={The National Academies Press},
  publisher={National Academies Press}
}

@article{du2023advances,
  title={Advances in spatial transcriptomics and related data analysis strategies},
  author={Du, Jun and Yang, Yu-Chen and An, Zhi-Jie and Zhang, Ming-Hui and Fu, Xue-Hang and Huang, Zou-Fang and Yuan, Ye and Hou, Jian},
  journal={Journal of translational medicine},
  volume={21},
  number={1},
  pages={330},
  year={2023},
  doi={10.1186/s12967-023-04150-2},
  publisher={Springer}
}

@article{grases2025practical,
  title={A practical guide to spatial transcriptomics: lessons from over 1000 samples},
  author={Grases, Daniela and Porta-Pardo, Eduard},
  journal={Trends in Biotechnology},
  year={2025},
  doi={10.1016/j.tibtech.2025.08.020},
  publisher={Elsevier}
}

@article{hirt2014cardiac,
  title={Cardiac tissue engineering: state of the art},
  author={Hirt, Marc N and Hansen, Arne and Eschenhagen, Thomas},
  journal={Circulation research},
  volume={114},
  number={2},
  pages={354--367},
  year={2014},
  doi={10.1161/CIRCRESAHA.114.300522},
  publisher={Lippincott Williams \& Wilkins Hagerstown, MD}
}

@article{ewoldt2025induced,
  title={Induced pluripotent stem cell-derived cardiomyocyte in vitro models: benchmarking progress and ongoing challenges},
  author={Ewoldt, Jourdan K and DePalma, Samuel J and Jewett, Maggie E and Karakan, M {\c{C}}a{\u{g}}atay and Lin, Yih-Mei and Mir Hashemian, Paria and Gao, Xining and Lou, Lihua and McLellan, Micheal A and Tabares, Jonathan and others},
  journal={Nature methods},
  volume={22},
  number={1},
  pages={24--40},
  year={2025},
  doi={10.1038/s41592-024-02480-7},
  publisher={Nature Publishing Group US New York}
}

@article{zhuang2022opportunities,
  title={Opportunities and challenges in cardiac tissue engineering from an analysis of two decades of advances},
  author={Zhuang, Richard Z and Lock, Roberta and Liu, Bohao and Vunjak-Novakovic, Gordana},
  journal={Nature Biomedical Engineering},
  volume={6},
  number={4},
  pages={327--338},
  year={2022},
  doi={10.1038/s41551-022-00885-3},
  publisher={Nature Publishing Group UK London}
}

@article{ketabat2024cardiac,
  title={Cardiac Tissue Engineering: A Journey from Scaffold Fabrication to In Vitro Characterization},
  author={Ketabat, Farinaz and Alcorn, Jane and Kelly, Michael E and Badea, Ildiko and Chen, Xiongbiao},
  journal={Small Science},
  volume={4},
  number={9},
  pages={2400079},
  year={2024},
  doi={10.1002/smsc.202400079},
  publisher={Wiley Online Library}
}

@article{rivera2023automated,
  title={Automated assessment of human engineered heart tissues using deep learning and template matching for segmentation and tracking},
  author={Rivera-Arbel{\'a}ez, Jos{\'e} M and Keekstra, Danjel and Cofi{\~n}o-Fabres, Carla and Boonen, Tom and Dostanic, Milica and Ten Den, Simone A and Vermeul, Kim and Mastrangeli, Massimo and van den Berg, Albert and Segerink, Loes I and others},
  journal={Bioengineering \& Translational Medicine},
  volume={8},
  number={3},
  pages={e10513},
  year={2023},
  doi={10.1002/btm2.10513},
  publisher={Wiley Online Library}
}

@article{rivera2025forcetracker,
  title={FORCETRACKER: A versatile tool for standardized assessment of tissue contractile properties in 3D Heart-on-Chip platforms},
  author={Rivera-Arbel{\'a}ez, Jos{\'e} M and Dostani{\'c}, Milica and Windt, Laura M and Stein, Jeroen M and Cofi{\~n}o-Fabres, Carla and Boonen, Tom and Wiendels, Maury and Van Den Berg, Albert and Segerink, Loes I and Mummery, Christine L and others},
  journal={PloS one},
  volume={20},
  number={2},
  pages={e0314985},
  year={2025},
  doi={10.1371/journal.pone.0314985},
  publisher={Public Library of Science San Francisco, CA USA}
}

@article{sala2018musclemotion,
  title={MUSCLEMOTION: a versatile open software tool to quantify cardiomyocyte and cardiac muscle contraction in vitro and in vivo},
  author={Sala, Luca and Van Meer, Berend J and Tertoolen, Leon GJ and Bakkers, Jeroen and Bellin, Milena and Davis, Richard P and Denning, Chris and Dieben, Michel AE and Eschenhagen, Thomas and Giacomelli, Elisa and others},
  journal={Circulation research},
  volume={122},
  number={3},
  pages={e5--e16},
  year={2018},
  publisher={Am Heart Assoc},
  doi = {10.1161/CIRCRESAHA.117.312067}
}

@article{tsan2021physiologic,
  title={Physiologic biomechanics enhance reproducible contractile development in a stem cell derived cardiac muscle platform},
  author={Tsan, Yao-Chang and DePalma, Samuel J and Zhao, Yan-Ting and Capilnasiu, Adela and Wu, Yu-Wei and Elder, Brynn and Panse, Isabella and Ufford, Kathryn and Matera, Daniel L and Friedline, Sabrina and others},
  journal={Nature Communications},
  volume={12},
  number={1},
  pages={6167},
  year={2021},
  publisher={Nature Publishing Group UK London},
  doi = {10.1038/s41467-021-26496-1}
}

@article{tamargo2021millipillar,
  title={milliPillar: a platform for the generation and real-time assessment of human engineered cardiac tissues},
  author={Tamargo, Manuel Alejandro and Nash, Trevor Ray and Fleischer, Sharon and Kim, Youngbin and Vila, Olaia Fernandez and Yeager, Keith and Summers, Max and Zhao, Yimu and Lock, Roberta and Chavez, Miguel and others},
  journal={ACS biomaterials science \& engineering},
  volume={7},
  number={11},
  pages={5215--5229},
  year={2021},
  publisher={ACS Publications},
  doi = {10.1021/acsbiomaterials.1c01006}
}

@article{ronaldson2019engineering,
  title={Engineering of human cardiac muscle electromechanically matured to an adult-like phenotype},
  author={Ronaldson-Bouchard, Kacey and Yeager, Keith and Teles, Diogo and Chen, Timothy and Ma, Stephen and Song, LouJin and Morikawa, Kumi and Wobma, Holly M and Vasciaveo, Alessandro and Ruiz, Edward C and others},
  journal={Nature protocols},
  volume={14},
  number={10},
  pages={2781--2817},
  year={2019},
  publisher={Nature Publishing Group UK London},
  doi = {10.1038/s41596-019-0189-8}
}

@article{psaras2021caltrack,
  title={CalTrack: high-throughput automated calcium transient analysis in cardiomyocytes},
  author={Psaras, Yiangos and Margara, Francesca and Cicconet, Marcelo and Sparrow, Alexander J and Repetti, Giuliana G and Schmid, Manuel and Steeples, Violetta and Wilcox, Jonathan AL and Bueno-Orovio, Alfonso and Redwood, Charles S and others},
  journal={Circulation research},
  volume={129},
  number={2},
  pages={326--341},
  year={2021},
  doi={10.1161/CIRCRESAHA.121.31886},
  publisher={Lippincott Williams \& Wilkins Hagerstown, MD}
}

@article{lian2012robust,
  title={Robust cardiomyocyte differentiation from human pluripotent stem cells via temporal modulation of canonical Wnt signaling},
  author={Lian, Xiaojun and Hsiao, Cheston and Wilson, Gisela and Zhu, Kexian and Hazeltine, Laurie B and Azarin, Samira M and Raval, Kunil K and Zhang, Jianhua and Kamp, Timothy J and Palecek, Sean P},
  journal={Proceedings of the National Academy of Sciences},
  volume={109},
  number={27},
  pages={E1848--E1857},
  year={2012},
  doi={10.1073/pnas.1200250109},
  publisher={National Academy of Sciences}
}

@article{burridge2014chemically,
  title={Chemically defined generation of human cardiomyocytes},
  author={Burridge, Paul W and Matsa, Elena and Shukla, Praveen and Lin, Ziliang C and Churko, Jared M and Ebert, Antje D and Lan, Feng and Diecke, Sebastian and Huber, Bruno and Mordwinkin, Nicholas M and others},
  journal={Nature methods},
  volume={11},
  number={8},
  pages={855--860},
  year={2014},
  doi={10.1038/nmeth.2999},
  publisher={Nature Publishing Group US New York}
}

@article{sarkans2021rembi,
  title={REMBI: Recommended Metadata for Biological Images—enabling reuse of microscopy data in biology},
  author={Sarkans, Ugis and Chiu, Wah and Collinson, Lucy and Darrow, Michele C and Ellenberg, Jan and Grunwald, David and H{\'e}rich{\'e}, Jean-Karim and Iudin, Andrii and Martins, Gabriel G and Meehan, Terry and others},
  journal={Nature methods},
  volume={18},
  number={12},
  pages={1418--1422},
  year={2021},
  doi={10.1038/s41592-021-01166-8},
  publisher={Nature Publishing Group US New York}
}

@article{schapiro2022miti,
  title={MITI minimum information guidelines for highly multiplexed tissue images},
  author={Schapiro, Denis and Yapp, Clarence and Sokolov, Artem and Reynolds, Sheila M and Chen, Yu-An and Sudar, Damir and Xie, Yubin and Muhlich, Jeremy and Arias-Camison, Raquel and Arena, Sarah and others},
  journal={Nature methods},
  volume={19},
  number={3},
  pages={262--267},
  year={2022},
  doi={10.1038/s41592-022-01415-4},
  publisher={Nature Publishing Group US New York}
}

@article{hosseini2023fair,
  title={FAIR high content screening in bioimaging},
  author={Hosseini, Rohola and Vlasveld, Matthijs and Willemse, Joost and van de Water, Bob and Le D{\'e}v{\'e}dec, Sylvia E and Wolstencroft, Katherine J},
  journal={Scientific Data},
  volume={10},
  number={1},
  pages={462},
  year={2023},
  doi={10.1038/s41597-023-02367-w},
  publisher={Nature Publishing Group UK London}
}

@misc{Mohammadzadeh2025dataset,
author = {Mohammadzadeh, Saeed and Tsan, Yao-chang and Kobeissi, Hiba and Lejeune, Emma and Helms, Adam},
publisher = {Harvard Dataverse},
title = {{SarcGraph - 2D Cardiac Muscle Bundle}},
UNF = {UNF:6:FeYENWUsTWi22Z6OilrWgQ==},
year = {2025},
version = {V1},
doi = {10.7910/DVN/GHMKWJ},
url = {https://doi.org/10.7910/DVN/GHMKWJ}
}

@article{iudin2023empiar,
  title={EMPIAR: the electron microscopy public image archive},
  author={Iudin, Andrii and Korir, Paul K and Somasundharam, Sriram and Weyand, Simone and Cattavitello, Cesare and Fonseca, Neli and Salih, Osman and Kleywegt, Gerard J and Patwardhan, Ardan},
  journal={Nucleic Acids Research},
  volume={51},
  number={D1},
  pages={D1503--D1511},
  year={2023},
  doi={10.1093/nar/gkac1062},
  publisher={Oxford University Press}
}

@article{mohammadzadeh2025quantifying,
  title={Quantifying HiPSC-CM structural organization at scale with deep learning-enhanced SarcGraph},
  author={Mohammadzadeh, Saeed and Lejeune, Emma},
  journal={PLOS Computational Biology},
  volume={21},
  number={10},
  pages={e1013436},
  year={2025},
  doi={10.1371/journal.pcbi.1013436},
  publisher={Public Library of Science San Francisco, CA USA}
}

@article{mohammadzadeh2025sarcgraph,
  title={SarcGraph for High-Throughput Regional Analysis of Sarcomere Organization and Contractile Function in 2D Cardiac Muscle Bundles},
  author={Mohammadzadeh, Saeed and Tsan, Yao-Chang and Renberg, Aaron and Kobeissi, Hiba and Helms, Adam and Lejeune, Emma},
  journal={arXiv preprint arXiv:2511.11913},
  doi={10.48550/arXiv.2511.11913},
  year={2025}
}

@article{toepfer2019sarctrack,
  title={SarcTrack: an adaptable software tool for efficient large-scale analysis of sarcomere function in hiPSC-cardiomyocytes},
  author={Toepfer, Christopher N and Sharma, Arun and Cicconet, Marcelo and Garfinkel, Amanda C and M{\"u}cke, Michael and Neyazi, Meraj and Willcox, Jon AL and Agarwal, Radhika and Schmid, Manuel and Rao, Jyoti and others},
  journal={Circulation research},
  volume={124},
  number={8},
  pages={1172--1183},
  year={2019},
  doi={10.1161/CIRCRESAHA.118.314505},
  publisher={Lippincott Williams \& Wilkins Hagerstown, MD}
}

@article{davidson2020myofibroblast,
  title={Myofibroblast activation in synthetic fibrous matrices composed of dextran vinyl sulfone},
  author={Davidson, Christopher D and Jayco, Danica Kristen P and Matera, Daniel L and DePalma, Samuel J and Hiraki, Harrison L and Wang, William Y and Baker, Brendon M},
  journal={Acta biomaterialia},
  volume={105},
  pages={78--86},
  year={2020},
  doi={10.1016/j.actbio.2020.01.009},
  note={PMID:31945504},
  publisher={Elsevier}
}

@article{depalma2021microenvironmental,
  title={Microenvironmental determinants of organized iPSC-cardiomyocyte tissues on synthetic fibrous matrices},
  author={DePalma, Samuel J and Davidson, Christopher D and Stis, Austin E and Helms, Adam S and Baker, Brendon M},
  journal={Biomaterials science},
  volume={9},
  number={1},
  pages={93--107},
  year={2021},
  doi={10.1039/d0bm01247e.},
  note={PMID:33325920},
  publisher={Royal Society of Chemistry}
}

@article{depalma2024matrix,
  title={Matrix architecture and mechanics regulate myofibril organization, costamere assembly, and contractility in engineered myocardial microtissues},
  author={DePalma, Samuel J and Jilberto, Javiera and Stis, Austin E and Huang, Darcy D and Lo, Jason and Davidson, Christopher D and Chowdhury, Aamilah and Kent III, Robert N and Jewett, Maggie E and Kobeissi, Hiba and others},
  journal={Advanced Science},
  volume={11},
  number={47},
  pages={2309740},
  year={2024},
  doi={10.1002/advs.202309740},
  note={PMID: 39558513},
  publisher={Wiley Online Library}
}

@article{kobeissi2024microbundlecompute,
  title={MicroBundleCompute: Automated segmentation, tracking, and analysis of subdomain deformation in cardiac microbundles},
  author={Kobeissi, Hiba and Jilberto, Javiera and Karakan, M {\c{C}}a{\u{g}}atay and Gao, Xining and DePalma, Samuel J and Das, Shoshana L and Quach, Lani and Urquia, Jonathan and Baker, Brendon M and Chen, Christopher S and others},
  journal={Plos one},
  volume={19},
  number={3},
  pages={e0298863},
  year={2024},
  doi={10.1371/journal.pone.0298863},
  note={PMID: 38530829},
  publisher={Public Library of Science San Francisco, CA USA}
}

@article{kobeissi2024microbundlepillartrack,
  title={MicroBundlePillarTrack: A Python package for automated segmentation, tracking, and analysis of pillar deflection in cardiac microbundles},
  author={Kobeissi, Hiba and Gao, Xining and DePalma, Samuel J and Ewoldt, Jourdan K and Wang, Miranda C and Das, Shoshana L and Jilberto, Javiera and Nordsletten, David and Baker, Brendon M and Chen, Christopher S and others},
  journal={Micropublication Biology},
  volume={2024},
  doi={10.17912/micropub.biology.001231},
  note={PMID: 39114859},
  year={2024}
}

@misc{kobeissi2024fibroTUG,
  author = {Kobeissi, Hiba and Gao, Xining and DePalma, Samuel J and Ewoldt, Jourdan K and Wang, Miranda C and Das, Shoshana L and Jilberto, Javiera and Nordsletten, David and Baker, Brendon M and Chen, Christopher S and others},
  title = {FibroTUG platforms: Time-lapse microscopy dataset of engineered cardiac microbundles},
  year = {2024},
  doi={10.5061/dryad.3r2280gqd},
  url = {10.5061/dryad.3r2280gqd}
    }

@misc{kobeissi2024strain,
  author = {Kobeissi, Hiba and Gao, Xining and DePalma, Samuel J and Ewoldt, Jourdan K and Wang, Miranda C and Das, Shoshana L and Jilberto, Javiera and Nordsletten, David and Baker, Brendon M and Chen, Christopher S and others},
  title = {Strain gauge platforms: Time-lapse microscopy dataset of engineered cardiac microbundles},
  year = {2024},
  doi={10.5061/dryad.sqv9s4nbg},
  url = {10.5061/dryad.sqv9s4nbg}
    }

@inproceedings{shi1994good,
  title={Good features to track},
  author={Shi, Jianbo and others},
  booktitle={1994 Proceedings of IEEE conference on computer vision and pattern recognition},
  pages={593--600},
  year={1994},
  organization={IEEE},
  doi = {10.1109/CVPR.1994.323794}
}

@inproceedings{lucas1981iterative,
author = {Lucas, Bruce D. and Kanade, Takeo},
title = {An Iterative Image Registration Technique with an Application to Stereo Vision},
year = {1981},
publisher = {Morgan Kaufmann Publishers Inc.},
address = {San Francisco, CA, USA},
booktitle = {Proceedings of the 7th International Joint Conference on Artificial Intelligence - Volume 2},
pages = {674–679},
numpages = {6},
location = {Vancouver, BC, Canada},
series = {IJCAI'81}
}

@article{bouguet2001pyramidal,
  title={Pyramidal implementation of the affine lucas kanade feature tracker description of the algorithm},
  author={Bouguet, Jean-Yves and others},
  journal={Intel corporation},
  volume={5},
  number={1-10},
  pages={4},
  year={2001}
}

@article{zimmerman2009deformation,
  title={Deformation gradients for continuum mechanical analysis of atomistic simulations},
  author={Zimmerman, Jonathan A and Bammann, Douglas J and Gao, Huajian},
  journal={International Journal of Solids and Structures},
  volume={46},
  number={2},
  pages={238--253},
  year={2009},
  doi={10.1016/j.ijsolstr.2008.08.036},
  publisher={Elsevier}
}

@article{benkley2023estimation,
  title={Estimation of the deformation gradient tensor by particle tracking near a free boundary with quantified error},
  author={Benkley, T and Li, C and Kolinski, J},
  journal={Experimental Mechanics},
  volume={63},
  number={7},
  pages={1255--1270},
  year={2023},
  doi= {10.1007/s11340-023-00981-8},
  publisher={Springer}
}

@article{pearson1901onlines,
  title={On lines and planes of closest fit to systems of points in space},
  author={Pearson, Karl},
  journal={The London, Edinburgh, and Dublin philosophical magazine and journal of science},
  volume={2},
  number={11},
  pages={559--572},
  year={1901},
  doi={10.1080/14786440109462720},
  publisher={Taylor \& Francis}
}

@incollection{hotelling1992relations,
  title={Relations between two sets of variates},
  author={Hotelling, Harold},
  booktitle={Breakthroughs in statistics: methodology and distribution},
  pages={162--190},
  year={1992},
  doi={10.2307/2333955},
  publisher={Springer}
}

@article{grama2014computation,
  title={Computation of full-field strains using principal component analysis},
  author={Grama, SN and Subramanian, SJ},
  journal={Experimental Mechanics},
  volume={54},
  number={6},
  pages={913--933},
  year={2014},
  doi={10.1007/s11340-013-9800-z},
  publisher={Springer}
}

@article{tyagi2017improving,
  title={Improving three-dimensional mechanical imaging of breast lesions with principal component analysis},
  author={Tyagi, Mohit and Wang, Yuqi and Hall, Timothy J and Barbone, Paul E and Oberai, Assad A},
  journal={Medical physics},
  volume={44},
  number={8},
  pages={4194--4203},
  year={2017},
  doi={10.1002/mp.12372},
  publisher={Wiley Online Library}
}

@article{abney2011principal,
  title={Principal component analysis of dynamic relative displacement fields estimated from MR images},
  author={Abney, Teresa M and Feng, Yuan and Pless, Robert and Okamoto, Ruth J and Genin, Guy M and Bayly, Philip V},
  journal={PLoS One},
  volume={6},
  number={7},
  pages={e22063},
  year={2011},
  doi={10.1371/journal.pone.0022063},
  publisher={Public Library of Science San Francisco, USA}
}

@article{arzani2021data,
  title={Data-driven cardiovascular flow modelling: examples and opportunities},
  author={Arzani, Amirhossein and Dawson, Scott TM},
  journal={Journal of the Royal Society Interface},
  volume={18},
  number={175},
  pages={20200802},
  year={2021},
  doi={10.1098/rsif.2020.0802},
  publisher={The Royal Society}
}

@article{lu2008mpca,
  title={MPCA: Multilinear principal component analysis of tensor objects},
  author={Lu, Haiping and Plataniotis, Konstantinos N and Venetsanopoulos, Anastasios N},
  journal={IEEE transactions on Neural Networks},
  volume={19},
  number={1},
  pages={18--39},
  year={2008},
  doi={10.1109/TNN.2007.901277},
  publisher={IEEE}
}

@article{lu2011survey,
  title={A survey of multilinear subspace learning for tensor data},
  author={Lu, Haiping and Plataniotis, Konstantinos N and Venetsanopoulos, Anastasios N},
  journal={Pattern Recognition},
  volume={44},
  number={7},
  pages={1540--1551},
  year={2011},
  doi={10.1016/j.patcog.2011.01.004},
  publisher={Elsevier}
}

@article{virtanen2020scipy,
  title={SciPy 1.0: fundamental algorithms for scientific computing in Python},
  author={Virtanen, Pauli and Gommers, Ralf and Oliphant, Travis E and Haberland, Matt and Reddy, Tyler and Cournapeau, David and Burovski, Evgeni and Peterson, Pearu and Weckesser, Warren and Bright, Jonathan and others},
  journal={Nature methods},
  volume={17},
  number={3},
  pages={261--272},
  year={2020},
  doi={10.1038/s41592-019-0686-2},
  publisher={Nature Publishing Group US New York}
}

@article{scikit-learn,
  title={Scikit-learn: Machine Learning in {P}ython},
  author={Pedregosa, F. and Varoquaux, G. and Gramfort, A. and Michel, V.
          and Thirion, B. and Grisel, O. and Blondel, M. and Prettenhofer, P.
          and Weiss, R. and Dubourg, V. and Vanderplas, J. and Passos, A. and
          Cournapeau, D. and Brucher, M. and Perrot, M. and Duchesnay, E.},
  journal={Journal of Machine Learning Research},
  volume={12},
  pages={2825--2830},
  year={2011}
}

@Inbook{wall2003singular,
author={Wall, Michael E. and Rechtsteiner, Andreas and Rocha, Luis M.},
editor={Berrar, Daniel P. and Dubitzky, Werner and Granzow, Martin},
title={Singular Value Decomposition and Principal Component Analysis},
bookTitle={A Practical Approach to Microarray Data Analysis},
year={2003},
publisher={Springer US},
address={Boston, MA},
pages={91--109},
isbn={978-0-306-47815-4},
doi={10.1007/0-306-47815-3_5},
url={https://doi.org/10.1007/0-306-47815-3_5}
}

@misc{scipy_rbf_doc,
  author = {{SciPy Developers}},
  title = {{scipy.interpolate.RBFInterpolator} --- {SciPy} v1.13.1 Manual},
  url = {https://docs.scipy.org/doc/scipy-1.13.1/reference/generated/scipy.interpolate.RBFInterpolator.html#scipy.interpolate.RBFInterpolator},
  year = {2024},
  note = {Accessed on 2025-11-19}
}

@misc{scipy_wd,
  author = {{SciPy Developers}},
  title = {{scipy.stats.wasserstein\_distance\_nd} --- {SciPy} v1.13.1 Manual},
  url = {https://docs.scipy.org/doc/scipy-1.13.1/reference/generated/scipy.stats.wasserstein_distance_nd.html},
  year = {2024},
  note = {Accessed on 2025-11-23}
}

@article{ramdas2017wasserstein,
  title={On wasserstein two-sample testing and related families of nonparametric tests},
  author={Ramdas, Aaditya and Garc{\'\i}a Trillos, Nicol{\'a}s and Cuturi, Marco},
  journal={Entropy},
  volume={19},
  number={2},
  pages={47},
  year={2017},
  publisher={MDPI},
  doi={doi.org/10.3390/e19020047}
}

@incollection{perry1975critical,
  title={Critical points in flow patterns},
  author={Perry, Anthony E and Fairlie, BD},
  booktitle={Advances in geophysics},
  volume={18},
  pages={299--315},
  year={1975},
  publisher={Elsevier}
}

@article{perry1987description,
  title={A description of eddying motions and flow patterns using critical-point concepts},
  author={Perry, Anthony E and Chong, Min S},
  journal={Annual Review of Fluid Mechanics},
  volume={19},
  pages={125--155},
  doi={10.1146/annurev.fl.19.010187.001013},
  year={1987}
}

@article{helman1989representation,
  title={Representation and display of vector field topology in fluid flow data sets},
  author={Helman, James and Hesselink, Lanbertus},
  journal={Computer},
  volume={22},
  number={08},
  pages={27--36},
  year={1989},
  doi= {10.1109/2.35197},
  publisher={IEEE Computer Society}
}

@article{helman1991visualizing,
  title={Visualizing vector field topology in fluid flows},
  author={Helman, James L and Hesselink, Lambertus},
  journal={IEEE Computer Graphics and Applications},
  volume={11},
  number={3},
  pages={36--46},
  year={1991},
  doi={},
}

@article{lee2001flow,
  title={Flow field analysis of a turbulent boundary layer over a riblet surface},
  author={Lee, S-J and Lee, S-H},
  journal={Experiments in fluids},
  volume={30},
  number={2},
  pages={153--166},
  year={2001},
  doi={10.1007/s003480000150},
  publisher={Springer}
}

@article{adrian2000analysis,
  title={Analysis and interpretation of instantaneous turbulent velocity fields},
  author={Adrian, Ronald J and Christensen, Kenneth T and Liu, Z-C},
  journal={Experiments in fluids},
  volume={29},
  number={3},
  pages={275--290},
  year={2000},
  doi={10.1007/s003489900087},
  publisher={Springer}
}

@article{effenberger2010finding,
  title={Finding and classifying critical points of 2D vector fields: a cell-oriented approach using group theory},
  author={Effenberger, Felix and Weiskopf, Daniel},
  journal={Computing and Visualization in Science},
  volume={13},
  number={8},
  pages={377--396},
  year={2010},
  doi={10.1007/s00791-011-0152-x},
  publisher={Springer}
}

@article{shu1994vector,
  title={Vector field analysis for oriented patterns},
  author={Shu, Chiao-Fe and Jain, Ramesh C},
  journal={IEEE Transactions on Pattern Analysis and Machine Intelligence},
  volume={16},
  number={9},
  pages={946--950},
  year={1994},
  doi={10.1109/34.310692},
  publisher={IEEE}
}

@article{theisel2005topological,
  title={Topological methods for 2D time-dependent vector fields based on stream lines and path lines},
  author={Theisel, Holger and Weinkauf, Tino and Hege, H-C and Seidel, H-P},
  journal={IEEE Transactions on Visualization and Computer Graphics},
  volume={11},
  number={4},
  pages={383--394},
  year={2005},
  doi={10.1109/TVCG.2005.68},
  publisher={IEEE}
}

@article{liu2024two,
  title={Two-dimensional vector field topology and scalar fields in viscous flows: Reconstruction methods},
  author={Liu, Tianshu and Salazar, David M},
  journal={Physics of Fluids},
  volume={36},
  number={7},
  year={2024},
  doi={10.1063/5.0215393},
  publisher={AIP Publishing}
}

@article{townsend2018detection,
  title={Detection and analysis of spatiotemporal patterns in brain activity},
  author={Townsend, Rory G and Gong, Pulin},
  journal={PLoS computational biology},
  volume={14},
  number={12},
  pages={e1006643},
  year={2018},
  doi={10.1371/journal.pcbi.1006643},
  publisher={Public Library of Science San Francisco, CA USA}
}

@article{hamid2021wave,
  title={Wave-like dopamine dynamics as a mechanism for spatiotemporal credit assignment},
  author={Hamid, Arif A and Frank, Michael J and Moore, Christopher I},
  journal={Cell},
  volume={184},
  number={10},
  pages={2733--2749},
  year={2021},
  doi={10.1016/j.cell.2021.03.046 External Link},
  publisher={Elsevier}
}

@article{xu2023interacting,
  title={Interacting spiral wave patterns underlie complex brain dynamics and are related to cognitive processing},
  author={Xu, Yiben and Long, Xian and Feng, Jianfeng and Gong, Pulin},
  journal={Nature human behaviour},
  volume={7},
  number={7},
  pages={1196--1215},
  year={2023},
  doi={10.1038/s41562-023-01626-5},
  publisher={Nature Publishing Group UK London}
}

@article{qiu2022mapping,
  title={Mapping transcriptomic vector fields of single cells},
  author={Qiu, Xiaojie and Zhang, Yan and Martin-Rufino, Jorge D and Weng, Chen and Hosseinzadeh, Shayan and Yang, Dian and Pogson, Angela N and Hein, Marco Y and Min, Kyung Hoi Joseph and Wang, Li and others},
  journal={Cell},
  volume={185},
  number={4},
  pages={690--711},
  year={2022},
  doi={10.1016/j.cell.2021.12.045},
  publisher={Elsevier}
}

@article{sha2024reconstructing,
  title={Reconstructing growth and dynamic trajectories from single-cell transcriptomics data},
  author={Sha, Yutong and Qiu, Yuchi and Zhou, Peijie and Nie, Qing},
  journal={Nature Machine Intelligence},
  volume={6},
  number={1},
  pages={25--39},
  year={2024},
  doi={doi.org/10.1038/s42256-023-00763-w},
  publisher={Nature Publishing Group UK London}
}

@article{zhu2025quantifying,
  title={Quantifying Landscape and Flux from Single-Cell Omics: Unraveling the Physical Mechanisms of Cell Function},
  author={Zhu, Ligang and Wang, Jin},
  journal={JACS Au},
  volume={5},
  number={8},
  pages={3738--3757},
  year={2025},
  doi={10.1021/jacsau.5c00620},
  publisher={ACS Publications}
}

@article{pancorbo2023vector,
  title={Vector field heterogeneity for the assessment of locally disorganised cardiac electrical propagation wavefronts from high-density multielectrodes},
  author={Pancorbo, Luc{\'\i}a and Ruip{\'e}rez-Campillo, Samuel and Tormos, {\'A}lvaro and Guill, Antonio and Cervig{\'o}n, Raquel and Alberola, Antonio and Chorro, Francisco Javier and Millet, Jos{\'e} and Castells, Francisco},
  journal={IEEE Open Journal of Engineering in Medicine and Biology},
  volume={5},
  pages={32--44},
  year={2023},
  doi={10.1109/OJEMB.2023.3344349},
  publisher={IEEE}
}

@article{dunn1961multiple,
  title={Multiple comparisons among means},
  author={Dunn, Olive Jean},
  journal={Journal of the American statistical association},
  volume={56},
  number={293},
  pages={52--64},
  year={1961},
  doi={10.1080/01621459.1961.10482090},
  publisher={Taylor \& Francis}
}

@article{tonko2025vector,
  title={Vector field heterogeneity as a novel omnipolar mapping metric for functional substrate characterization in scar-related ventricular tachycardias},
  author={Tonko, Johanna B and Ruip{\'e}rez-Campillo, Samuel and Cabero-Vidal, Gema and Cabrera-Borrego, Eva and Roney, Caroline and Jim{\'e}nez-J{\'a}imez, Juan and Millet, Jos{\'e} and Castells, Francisco and Lambiase, Pier D},
  journal={Heart Rhythm},
  volume={22},
  number={5},
  pages={1218--1228},
  year={2025},
  doi={10.1016/j.hrthm.2024.10.066},
  publisher={Elsevier}
}

@inbook{brasselet2009vector,
  title={Vector fields on singular varieties},
  chapter = {1},
  author={Brasselet, Jean-Paul and Seade, Jos{\'e} and Suwa, Tatsuo},
  volume={1987},
  year={2009},
  publisher={Springer Science \& Business Media}
}

@inproceedings{rubner1998metric,
  title={A metric for distributions with applications to image databases},
  author={Rubner, Yossi and Tomasi, Carlo and Guibas, Leonidas J},
  booktitle={Sixth international conference on computer vision (IEEE Cat. No. 98CH36271)},
  pages={59--66},
  year={1998},
  doi={10.1109/ICCV.1998.710701},
  organization={IEEE}
}

@inproceedings{rubner1997earth,
  title={The earth mover’s distance, multi-dimensional scaling, and color-based image retrieval},
  author={Rubner, Yossi and Guibas, Leonidas J and Tomasi, Carlo},
  booktitle={Proceedings of the ARPA image understanding workshop},
  volume={661},
  pages={668},
  year={1997}
}

@inproceedings{lavin1998feature,
  title={Feature comparisons of vector fields using earth mover's distance},
  author={Lavin, Yingmei and Batra, Rajesh and Hesselink, Lambertus},
  booktitle={Proceedings Visualization'98 (Cat. No. 98CB36276)},
  pages={103--109},
  year={1998},
  doi={10.1109/VISUAL.1998.745291},
  organization={IEEE}
}

@article{rimehaug2023uncovering,
  title={Uncovering circuit mechanisms of current sinks and sources with biophysical simulations of primary visual cortex},
  author={Rimehaug, Atle E and Stasik, Alexander J and Hagen, Espen and Billeh, Yazan N and Siegle, Josh H and Dai, Kael and Olsen, Shawn R and Koch, Christof and Einevoll, Gaute T and Arkhipov, Anton},
  journal={elife},
  volume={12},
  pages={e87169},
  year={2023},
  doi={10.7554/eLife.87169},
  publisher={eLife Sciences Publications Limited}
}

@article{lu2025multi,
  title={Multi-Dimensional Wasserstein Distance Implementation in Scipy},
  author={Lu, Zehao},
  journal={arXiv preprint arXiv:2510.23651},
  doi={10.48550/arXiv.2510.23651},
  year={2025}
}

@article{li2007synchronization,
  title={Synchronization measurement of multiple neuronal populations},
  author={Li, Xiaoli and Cui, Dong and Jiruska, Premysl and Fox, John E and Yao, Xin and Jefferys, John GR},
  journal={Journal of neurophysiology},
  volume={98},
  number={6},
  pages={3341--3348},
  year={2007},
  doi = {10.1152/jn.00977.2007},
  note ={PMID: 17913983},
  publisher={American Physiological Society}
}

@article{patel2012dynamic,
  title={Dynamic changes in neural circuit topology following mild mechanical injury in vitro},
  author={Patel, Tapan P and Ventre, Scott C and Meaney, David F},
  journal={Annals of biomedical engineering},
  volume={40},
  number={1},
  pages={23--36},
  year={2012},
  doi={10.1007/s10439-011-0390-6},
  publisher={Springer}
}

@article{Shearme1968some,
  author={Shearme, J. and Leach, P.},
  journal={IEEE Transactions on Audio and Electroacoustics}, 
  title={Some experiments with a simple word recognition system}, 
  year={1968},
  volume={16},
  number={2},
  pages={256-261},
  doi={10.1109/TAU.1968.1161985},
  publisher={IEEE}
}

@article{itakura1975minimum,
  author={Itakura, Fumitada},
  journal={IEEE Transactions on Acoustics, Speech, and Signal Processing}, 
  title={Minimum prediction residual principle applied to speech recognition}, 
  year={1975},
  volume={23},
  number={1},
  pages={67-72},
  doi={10.1109/TASSP.1975.1162641},
  publisher={IEEE}
}

@article{sakoe1978dynamic,
  author={Sakoe, H. and Chiba, S.},
  journal={IEEE Transactions on Acoustics, Speech, and Signal Processing}, 
  title={Dynamic programming algorithm optimization for spoken word recognition}, 
  year={1978},
  volume={26},
  number={1},
  pages={43-49},
  doi={10.1109/TASSP.1978.1163055},
  publisher={IEEE}
  }

@inproceedings{berndt1994using,
  title={Using dynamic time warping to find patterns in time series},
  author={Berndt, Donald J and Clifford, James},
  booktitle={Proceedings of the 3rd international conference on knowledge discovery and data mining},
  pages={359--370},
  year={1994},
  location = {Seattle, WA}, 
  series = {AAAIWS'94},
  publisher = {AAAI Press}
}

@article{jeong2011weighted,
  title={Weighted dynamic time warping for time series classification},
  author={Jeong, Young-Seon and Jeong, Myong K and Omitaomu, Olufemi A},
  journal={Pattern recognition},
  volume={44},
  number={9},
  pages={2231--2240},
  year={2011},
  doi={10.1016/j.patcog.2010.09.022},
  publisher={Elsevier}
}

@article{olivares2019inferring,
  title={On inferring intentions in shared tasks for industrial collaborative robots},
  author={Olivares-Alarcos, Alberto and Foix, Sergi and Alenya, Guillem},
  journal={Electronics},
  volume={8},
  number={11},
  pages={1306},
  year={2019},
  doi={10.3390/electronics8111306},
  publisher={MDPI}
}

@article{miralles2023forecasting,
  title={Forecasting COVID-19 cases using dynamic time warping and incremental machine learning methods},
  author={Miralles-Pechu{\'a}n, Luis and Kumar, Ankit and Su{\'a}rez-Cetrulo, Andr{\'e}s L},
  journal={Expert Systems},
  volume={40},
  number={6},
  pages={e13237},
  year={2023},
  doi={10.1111/exsy.13237},
  publisher={Wiley Online Library}
}

@misc{meert2020DTAIDistance,
author = {Meert, Wannes and Hendrickx, Kilian and Van Craenendonck, Toon and Robberechts, Pieter and Blockeel, Hendrik and Davis, Jesse},
doi = {10.5281/zenodo.3981067},
month = aug,
title = {{DTAIDistance}},
url = {https://github.com/wannesm/dtaidistance},
version = {2},
year = {2020}
}

@article{healy2024uniform,
  title={Uniform manifold approximation and projection},
  author={Healy, John and McInnes, Leland},
  journal={Nature Reviews Methods Primers},
  volume={4},
  number={1},
  pages={82},
  year={2024},
  publisher={Nature Publishing Group UK London}
}

@article{maaten2008visualizing,
  title={Visualizing data using t-SNE},
  author={Maaten, Laurens van der and Hinton, Geoffrey},
  journal={Journal of machine learning research},
  volume={9},
  number={Nov},
  pages={2579--2605},
  year={2008}
}

@article{spearman1961proof,
 author = {C. Spearman},
 journal = {The American Journal of Psychology},
 number = {1},
 pages = {72--101},
 publisher = {University of Illinois Press},
 title = {The Proof and Measurement of Association between Two Things},
 volume = {15},
 doi={10.2307/1412159},
 year = {1904}
}

@article{anscombe1973graphs,
  title={Graphs in statistical analysis},
  author={Anscombe, Francis J},
  journal={The american statistician},
  volume={27},
  number={1},
  pages={17--21},
  year={1973},
  doi={10.1080/00031305.1973.10478966},
  publisher={Taylor \& Francis}
}

@article{mukaka2012guide,
  title={A guide to appropriate use of correlation coefficient in medical research},
  author={Mukaka, Mavuto M},
  journal={Malawi medical journal},
  volume={24},
  number={3},
  pages={69--71},
  doi={},
  note={PMID:23638278},
  year={2012}
}

@inbook{pakay2023foundations,
  title={Foundations of Biomedical Science: Quantitative Literacy: Theory and Problems},
  chapter={11},
  author={Pakay, Julian},
  year={2023},
  publisher={La Trobe eBureau}
}

@book{belsley2005regression,
  title={Regression diagnostics: Identifying influential data and sources of collinearity},
  author={Belsley, David A and Kuh, Edwin and Welsch, Roy E},
  year={2005},
  publisher={John Wiley \& Sons}
}

@article{jilberto2023data,
  title={A data-driven computational model for engineered cardiac microtissues},
  author={Jilberto, Javiera and DePalma, Samuel J and Lo, Jason and Kobeissi, Hiba and Quach, Lani and Lejeune, Emma and Baker, Brendon M and Nordsletten, David},
  journal={Acta biomaterialia},
  volume={172},
  pages={123--134},
  year={2023},
  doi={10.1016/j.actbio.2023.10.025},
  publisher={Elsevier}
}

@misc{jilberto2025data,
  title={Evaluating Constrained and Unconstrained Mixture Frameworks for Predicting Engineered Heart Tissue Mechanics - Under Review},
  author={Jilberto, Javiera and DePalma, Samuel J and Ntim, David and Kobeissi, Hiba and Lejeune, Emma and Helms, Adam and Baker, Brendon M and Nordsletten, David},
  journal={Acta biomaterialia},
  doi={10.2139/ssrn.5958154},
  publisher={Acta biomaterialia}
}

@article{kruskal1952use,
  title={Use of ranks in one-criterion variance analysis},
  author={Kruskal, William H and Wallis, W Allen},
  journal={Journal of the American statistical Association},
  volume={47},
  number={260},
  pages={583--621},
  year={1952},
  doi={10.1080/01621459.1952.10483441},
  publisher={Taylor \& Francis}
}

@article{dunn1964multiple,
  title={Multiple comparisons using rank sums},
  author={Dunn, Olive Jean},
  journal={Technometrics},
  volume={6},
  number={3},
  pages={241--252},
  year={1964},
  doi={10.1080/00401706.1964.10490181},
  publisher={Taylor \& Francis}
}

@article{holm1979simple,
  title={A simple sequentially rejective multiple test procedure},
  author={Holm, Sture},
  journal={Scandinavian journal of statistics},
  pages={65--70},
  year={1979},
  publisher={JSTOR}
}

@article{kelley1935unbiased,
  title={An unbiased correlation ratio measure},
  author={Kelley, Truman L},
  journal={Proceedings of the National Academy of Sciences},
  volume={21},
  number={9},
  pages={554--559},
  year={1935},
  doi={10.1073/pnas.21.9.554}
}

@book{field2024discovering,
  title={Discovering statistics using IBM SPSS statistics},
  author={Field, Andy},
  year={2024},
  publisher={Sage publications limited}
}

@article{simonsohn2014p,
  title={P-curve: a key to the file-drawer.},
  author={Simonsohn, Uri and Nelson, Leif D and Simmons, Joseph P},
  journal={Journal of experimental psychology: General},
  volume={143},
  number={2},
  pages={534},
  year={2014},
  doi={10.1037/a0033242},
  publisher={American Psychological Association}
}

@article{anderson2014permutational,
  title={Permutational multivariate analysis of variance (PERMANOVA)},
  author={Anderson, Marti J},
  journal={Wiley statsref: statistics reference online},
  pages={1--15},
  year={2014},
  doi = {10.1002/9781118445112.stat07841},
  publisher={Wiley Online Library}
}

@misc{skbio,
  title={biocore/scikit-bio: scikit-bio 0.5. 9: Maintenance release},
  author={Rideout, Jai Ram and Caporaso, Greg and Bolyen, Evan and McDonald, Daniel and V{\'a}zquez Baeza, Yoshiki and Ca{\~n}ardo Alastuey, Jorge and Pitman, Anders and Morton, Jamie and Navas, Jose and Gorlick, Kestrel and others},
  howpublished = {10.5281/zenodo.593387},
  year={2023}
}

@inbook{weinfurt1995multivariate,
  title={Reading and understanding multivariate statistics =},
  chapter={Multivariate analysis of variance.},
  author={Weinfurt, Kevin P},
  pages={245–276},
  year={1995},
  publisher={American Psychological Association}
}

@article{rios2023mechanically,
  title={Mechanically programming anisotropy in engineered muscle with actuating extracellular matrices},
  author={Rios, Brandon and Bu, Angel and Sheehan, Tara and Kobeissi, Hiba and Kohli, Sonika and Shah, Karina and Lejeune, Emma and Raman, Ritu},
  journal={Device},
  volume={1},
  number={4},
  year={2023},
  doi={10.1016/j.device.2023.100097},
  publisher={Elsevier}
}

@article{boudou2012microfabricated,
  title={A microfabricated platform to measure and manipulate the mechanics of engineered cardiac microtissues},
  author={Boudou, Thomas and Legant, Wesley R and Mu, Anbin and Borochin, Michael A and Thavandiran, Nimalan and Radisic, Milica and Zandstra, Peter W and Epstein, Jonathan A and Margulies, Kenneth B and Chen, Christopher S},
  journal={Tissue Engineering Part A},
  volume={18},
  number={9-10},
  pages={910--919},
  year={2012},
  publisher={Mary Ann Liebert, Inc. 140 Huguenot Street, 3rd Floor New Rochelle, NY 10801 USA},
  doi = {10.1089/ten.tea.2011.0341}
}

@article{ewoldt2024hypertrophic,
  title={Hypertrophic cardiomyopathy--associated mutations drive stromal activation via EGFR-mediated paracrine signaling},
  author={Ewoldt, Jourdan K and Wang, Miranda C and McLellan, Micheal A and Cloonan, Paige E and Chopra, Anant and Gorham, Joshua and Li, Linqing and DeLaughter, Daniel M and Gao, Xining and Lee, Joshua H and others},
  journal={Science Advances},
  volume={10},
  number={42},
  pages={eadi6927},
  year={2024},
  doi={10.1126/sciadv.adi6927},
  publisher={American Association for the Advancement of Science}
}

@article{karakan2024geometry,
  title={Geometry and length control of 3D engineered heart tissues using direct laser writing},
  author={Karakan, M {\c{C}}a{\u{g}}atay and Ewoldt, Jourdan K and Segarra, Addianette J and Sundaram, Subramanian and Wang, Miranda C and White, Alice E and Chen, Christopher S and Ekinci, Kamil L},
  journal={Lab on a Chip},
  volume={24},
  number={6},
  pages={1685--1701},
  year={2024},
  doi={10.1039/D3LC00752A},
  publisher={Royal Society of Chemistry}
}

@article{zhao2019platform,
  title={A platform for generation of chamber-specific cardiac tissues and disease modeling},
  author={Zhao, Yimu and Rafatian, Naimeh and Feric, Nicole T and Cox, Brian J and Aschar-Sobbi, Roozbeh and Wang, Erika Yan and Aggarwal, Praful and Zhang, Boyang and Conant, Genevieve and Ronaldson-Bouchard, Kacey and others},
  journal={Cell},
  volume={176},
  number={4},
  pages={913--927},
  year={2019},
  publisher={Elsevier}, 
  doi = {https://doi.org/10.1016/j.cell.2018.11.042}
}

@article{dinno2015nonparametric,
  title={Nonparametric pairwise multiple comparisons in independent groups using Dunn's test},
  author={Dinno, Alexis},
  journal={The Stata Journal},
  volume={15},
  number={1},
  pages={292--300},
  year={2015},
  doi={10.1177/1536867X1501500117},
  publisher={SAGE Publications Sage CA: Los Angeles, CA}
}

@article{forstmeier2017detecting,
  title={Detecting and avoiding likely false-positive findings--a practical guide},
  author={Forstmeier, Wolfgang and Wagenmakers, Eric-Jan and Parker, Timothy H},
  journal={Biological Reviews},
  volume={92},
  number={4},
  pages={1941--1968},
  year={2017},
  doi={10.1111/brv.12315},
  publisher={Wiley Online Library}
}

@article{weston2019recommendations,
  title={Recommendations for increasing the transparency of analysis of preexisting data sets},
  author={Weston, Sara J and Ritchie, Stuart J and Rohrer, Julia M and Przybylski, Andrew K},
  journal={Advances in methods and practices in psychological science},
  volume={2},
  number={3},
  pages={214--227},
  year={2019},
  doi={10.1177/2515245919848684},
  publisher={Sage Publications Sage CA: Los Angeles, CA}
}

@article{lakens2024benefits,
  title={The benefits of preregistration and Registered Reports},
  author={Lakens, Dani{\"e}l and Mesquida, Cristian and Rasti, Sajedeh and Ditroilo, Massimiliano},
  journal={Evidence-Based Toxicology},
  volume={2},
  number={1},
  pages={2376046},
  year={2024},
  doi={10.1080/2833373X.2024.2376046},
  publisher={Taylor \& Francis}
}

@article{scalzo2021dense,
  title={Dense optical flow software to quantify cellular contractility},
  author={Scalzo, S{\'e}rgio and Afonso, Marcelo QL and da Fonseca, N{\'e}li J and Jesus, Itamar CG and Alves, Ana Paula and Mendon{\c{c}}a, Carolina ATF and Teixeira, Vanessa P and Biagi, Diogo and Cruvinel, Estela and Santos, Anderson K and others},
  journal={Cell Reports Methods},
  volume={1},
  number={4},
  year={2021},
  doi={10.1016/j.crmeth.2021.100044},
  publisher={Elsevier}
}

@article{huebsch2015automated,
  title={Automated video-based analysis of contractility and calcium flux in human-induced pluripotent stem cell-derived cardiomyocytes cultured over different spatial scales},
  author={Huebsch, Nathaniel and Loskill, Peter and Mandegar, Mohammad A and Marks, Natalie C and Sheehan, Alice S and Ma, Zhen and Mathur, Anurag and Nguyen, Trieu N and Yoo, Jennie C and Judge, Luke M and others},
  journal={Tissue Engineering Part C: Methods},
  volume={21},
  number={5},
  pages={467--479},
  year={2015},
  doi={10.1089/ten.tec.2014.0283},
  publisher={Mary Ann Liebert, Inc. 140 Huguenot Street, 3rd Floor New Rochelle, NY 10801 USA}
}

@article{mery2023light,
  title={Light-driven biological actuators to probe the rheology of 3D microtissues},
  author={M{\'e}ry, Adrien and Ruppel, Artur and Revilloud, Jean and Balland, Martial and Cappello, Giovanni and Boudou, Thomas},
  journal={Nature Communications},
  volume={14},
  number={1},
  pages={717},
  year={2023},
  publisher={Nature Publishing Group UK London}
}

\end{document}